\documentclass[final,1p,times]{elsarticle}

\usepackage{latexsym}
\usepackage{amssymb}
\usepackage{amsmath}
\usepackage{color}
\usepackage{psfrag}
\usepackage{graphicx}

\usepackage{mathrsfs}
\usepackage{txfonts}
\usepackage{tikz,tikz-3dplot} 
\usepackage{bm} 
\usetikzlibrary{calc,patterns,decorations.pathmorphing,decorations.markings}

\usetikzlibrary{calc,arrows}
\usetikzlibrary{decorations.text}
\usetikzlibrary{decorations.pathmorphing}

\mathchardef\Gamma="7100 
\mathchardef\Delta="7101
\mathchardef\Theta="7102
\mathchardef\Lambda="7103
\mathchardef\Xi="7104
\mathchardef\Pi="7105
\mathchardef\Sigma="7106
\mathchardef\Upsilon="7107
\mathchardef\Phi="7108
\mathchardef\Psi="7109
\mathchardef\Omega="710A

\newcommand{\qqeqq}{\qquad \textrm{and} \qquad}
\newcommand{\qeq}{\quad \textrm{and} \quad}

\newcommand{\dts}{\partial_t}
\newcommand{\dxs}{\partial_x}

\newcommand{\mP}{\mathcal{P}}
\newcommand{\mQ}{\mathcal{Q}}
\newcommand{\mM}{\mathcal{M}}

\newcommand{\gint}{g_{\rm I}}
\newcommand{\pint}{p_{\rm I}}
\newcommand{\cint}[1]{c_{{\rm I} #1}}

\newcommand{\rhoint}{\rho_{\scriptstyle\textrm{I}}}

\newcommand{\phider}{\phi}
\newcommand{\zetader}{\zeta}

\newcommand{\nonc}{\sigma}
\newcommand{\nonce}{\Sigma}
\newcommand{\noncep}{\Xi}

\newcommand{\dxsb}[1]{\partial_x \! \left( #1 \right)}

\newcommand{\etapp}{\tilde{\eta}}

\newcommand{\Ei}{\mathcal{E}}
\newcommand{\eint}{\varepsilon}

\newcommand{\trasp}{^{\scriptstyle\textrm{T}}}

\newcommand{\calW}{\mathcal{W}}

\newcommand{\iph}{_{i+1/2}}

\newcommand{\apdqm}{\mathcal{A}^+\Delta Q_{i-1/2}}
\newcommand{\amdqp}{\mathcal{A}^-\Delta Q_{i+1/2}}

\newcommand{\apmdqp}{\mathcal{A}^{\pm}\Delta Q_{i+1/2}}

\newcommand{\cref}{c_\ast}

\newcommand{\tilhFm}{{F}_{i-1/2}^{\rm h}}
\newcommand{\tilhFp}{{F}_{i+1/2}^{\rm h}}

\newcommand{\iM}{\mathscr{M}}

\journal{}
\begin{document}
\begin{frontmatter}

\title{Arbitrary-rate relaxation techniques for the numerical modeling of compressible two-phase
flows with heat and mass transfer}
%
\author[pelanti]{Marica~Pelanti\corref{cor1}}
\ead{marica.pelanti@ensta-paris.fr}

\cortext[cor1]{Corresponding author. Tel.:  +33 1 69 31 98 19; Fax:  +33 1 69 31 99 97.}
\address[pelanti]{IMSIA, UMR 9219 ENSTA-CNRS-EDF-CEA,\\ ENSTA Paris - Institut Polytechnique de Paris,\\ 828,~Boulevard des Mar\'echaux, 91120 Palaiseau, France}

\begin{abstract}
We describe compressible two-phase flows by a single-velocity six-equation flow model, which is
composed of the phasic mass and total energy equations, one volume fraction
equation, and the mixture momentum equation. The model contains relaxation source terms
accounting for volume, heat and mass transfer. The equations are numerically solved via a   fractional step algorithm, where we alternate between the solution of the homogeneous hyperbolic portion of the system via a HLLC-type wave propagation scheme, and the solution of a sequence of three systems of ordinary differential equations for the relaxation source terms driving the flow toward mechanical, thermal and chemical equilibrium. In the literature often numerical relaxation procedures are based on simplifying assumptions, namely
simple equations of state, such as the stiffened gas one, and instantaneous relaxation processes.
These simplifications of the flow physics  might be inadequate for the description
of the thermodynamical processes involved in various flow problems. In the present work we introduce new numerical relaxation techniques  with two significant properties: the capability to describe heat and mass transfer processes of arbitrary relaxation time, and the applicability to a general equation of state. We show the effectiveness of the proposed  methods by presenting several numerical experiments.

\end{abstract}

\end{frontmatter}

\section{Introduction}

\label{sec:1}

The modeling of multifluid and multiphase flows has applications  in numerous fields of science,
 largely in  many sectors of engineering such as aerospace, naval and nuclear technologies. In the present work
we are interested in the simulation of compressible multiphase flows that might involve shocks,
interfaces, and phase transition processes. Examples of flows of interest are 
those occurring in underwater explosions \cite{cole}, nuclear power plants,
and fuel injection systems. We describe these flows by a hyperbolic  single-velocity six-equation  
compressible two-phase flow model that we have first studied in \cite{pelanti-shyue},
and which is a variant of the six-equation model presented in \cite{saurel-PB}. We employ  a 
diffuse-interface approach, cf.~\cite{saurel-pantano}.
The model system is composed of  the phasic mass and total energy equations for the two phases,
one volume fraction equation, and the mixture momentum equation.
The model   contains  mechanical, thermal and chemical relaxation  source terms,
 accounting respectively for volume, heat and mass transfer.
 The considered   model belongs to the class of Baer--Nunziato-type \cite{baer-nunz} multiphase compressible flow models.
 The seven-equation two-phase flow model of Baer--Nunziato \cite{baer-nunz} (and the variant of Saurel--Abgrall \cite{sa-ab:multi})
 is the most   general model able to account for velocity, pressure, temperature and chemical potential disequilibria
 between the phases. From this  full non-equilibrium seven-equation  model endowed with relaxation source terms 
 a hierarchy of relaxed models can be established by considering combinations of infinite-rate relaxation
 processes driving the flow to different levels of equilibrium \cite{gaute-tore}.
 The six-equation model considered in the present work represents the relaxed velocity equilibrium  model obtained from the 
 seven-equation Baer--Nunziato model in the limit of instantaneous kinetic equilibrium. From the  six-equation single-velocity
 model a sub-hierarchy of relaxed models can be then obtained \cite{flatten-lund:rel,lund:rel}. 
 In the limit of instantaneous mechanical relaxation
 we obtain the five-equation pressure equilibrium model
 of Kapila \textit{et al.} \cite{kapila}. In the limit of instantaneous mechanical  and thermal relaxation
 we obtain a four-equation pressure and temperature equilibrium two-phase model \cite{lund_proc,lemart-boil,saurelCF16,saurelCF17,demou-boil}, and in the limit of 
 full instantaneous thermodynamic equilibrium we obtain
 the three-equation Homogeneous Equilibrium Model (HEM) \cite{st-we:2phase}.
 Let us note that the numerical solution method for the six-equation model must be able to approximate solutions of the relaxed models in the hierarchy when appropriate instantaneous relaxation processes are activated.
 We also  recall a different four-equation two-phase flow model of the Baer--Nunziato class, the liquid-vapor 
 Homogeneous Relaxation Model (HRM) with mass transfer of \cite{HRMmodelBK,HRMmodel}, which does not enter 
 in the aforementioned hierarchy. 
 The thermodynamic closure of this
 model consists in the assumption of  mechanical equilibrium and  vapor phase at saturation,
 and the model accounts for thermal disequilibrium.

The considered class of models with relaxation source terms in the literature is classically solved numerically via a fractional
step algorithm where one alternates between the solution of the homogeneous hyperbolic portion of the model system and 
the solution of a sequence of systems of ordinary differential equations for the relaxation source terms
\cite{sa-ab:multi,sa-lem:multi,saurel-PB,saurel-PA,waha-code,zein-HW,lund_proc,pelanti-shyue,
daude-galon-gao-blaud,met-mas-saur-proc,
rodio-abgrall,pelanti-amc,saurel-martelot,lemart-boil,
saurelCF16,saurelCF17,brynge-MCF,SBC-JCPbubble,schmi-ecogen,saurel-sodium}.
We also adopt here this operator splitting approach for the numerical approximation of the
six-equation model, and for the solution  of the homogeneous system we employ
 a second-order accurate finite volume wave propagation scheme \cite{rjlbook,rjl:wavepr} based on the HLLC-type Riemann solver
 that we  have presented in \cite{pelanti-shyue} (and which later we have also re-interpreted as a Suliciu-type Riemann 
 solver \cite{delorenzo-hllc}). 
 
 For the numerical approximation of the mechanical, thermal and chemical relaxation processes often in the literature
 it is assumed that these processes are instantaneous \cite{saurel-PB,saurel-PA,zein-HW, met-mas-saur-proc,rodio-abgrall,
 lemart-boil,saurelCF16,saurelCF17,daude-galon-gao-blaud,schmi-ecogen}, and this assumption was also made in our
 previous six-equation numerical model \cite{pelanti-shyue,pelanti-amc}. This simplifying hypothesis is advantageous
  because in this
 case one does not need to solve the system of ordinary equations that govern the relaxation process,
 but it suffices to impose equilibrium conditions to
 obtain a system of algebraic equations to be solved for the unknown relaxed equilibrium state.
 The assumption of instantaneous mechanical equilibrium can be indeed  considered appropriate for the flows of interest
 (see also for instance the discussion on characteristic relaxation scales in \cite{kapila}).
 On the other hand, the hypothesis  of instantaneous thermo-chemical relaxation might be  inadequate for the 
  description of the thermodynamical processes
involved in several flow problems.  For instance, in some transient phenomena such as fast depressurizations
the delay of vaporization and the appearance of metastable states are key features in the flow evolution, 
and they can be described only by models that account for non-instantaneous  mass transfer,
such as those in \cite{HRMmodel,herard-ducts,lund_proc,delor-laf-pelanti-IJMF}.
Another simplification often considered in the literature is the choice of a simple equation
of state,  the stiffened gas equation of state \cite{saurel-PB,saurel-PA,zein-HW, met-mas-saur-proc,rodio-abgrall,
 lemart-boil,daude-galon-gao-blaud,brynge-MCF,SBC-JCPbubble,schmi-ecogen}, which results from a linearization
 of the more general Mie--Gru\"neisen pressure law \cite{men-plo:rp}. The stiffened gas 
 equation of state is  very convenient for
numerical purposes,  however it might not allow an accurate flow characterization
over a wide temperature range, and in particular for liquid-vapor flows  it might not
provide a precise estimation
of the saturation  conditions  \cite{lemeta-eos}.  
Some more recent multiphase numerical models for liquid-vapor flows adopt a slightly more accurate equation of state,
the Noble--Abel stiffened gas equation of state \cite{saurel-lemet-NASG,saurelCF16,saurelCF17,saurel-sodium}, 
and few models adopt complex and very precise equations of state
such as the IAPWS Industrial Formulation 1997 for Water and Steam \cite{iapws97}, which  we have used in previous work
\cite{delor-laf-pelanti-IJMF,delor-laf-pelanti-JCP19,
delor-laf-pelanti-NED}.

One main objective of the present work is to develop new relaxation techniques for heat and mass transfer
capable to model processes  of any relaxation rate, both instantaneous infinite-rate processes and slow finite-rate
 ones. We are primarily interested in arbitrary-rate mass transfer and the capability to model
 metastable states in vapor-liquid flows with phase transition. Another objective is the design of relaxation techniques efficiently applicable to a general equation of state.  
 A known  difficulty encountered in the numerical solution of a 
system of ordinary equations with a relaxation source term is the stiffness of the problem
in case of  nearly instantaneous  relaxation, which would require computationally expensive implicit time integration techniques. Our idea consists in describing the relaxation processes by systems of ordinary equations
obtained from the governing two-phase equations that admit analytical semi-exact exponential solutions. Similar approaches using exponential solutions
to solve stiff relaxation systems were used for instance in 
\cite{HRMmodel,ch-rjl-gw:deton,pelanti-leveque-dusty,norway-exponential,delor-laf-pelanti-IJMF,delor-laf-pelanti-JCP19}.
Let us remark some differences with respect to our previous work 
\cite{delor-laf-pelanti-JCP19,delor-laf-pelanti-NED} on
relaxation techniques for non-instantaneous heat and mass transfers and general equation of state.
The principal thermal and chemical relaxation procedures proposed in \cite{delor-laf-pelanti-JCP19,delor-laf-pelanti-NED}
 were based on relaxation systems derived from physical principles  solved
 numerically via explicit Runge--Kutta methods with adaptive step size. 
These explicit methods were not suited for stiff problems, and the
employment of implicit solvers was found too computationally expensive, thus the procedures were not adequate for stiff instantaneous or nearly instantaneous processes. To solve
problems with infinite-rate transfers alternative techniques based on exponential solutions were 
briefly proposed in the Appendix of~\cite{delor-laf-pelanti-JCP19}. Nonetheless
 these techniques  were specifically aimed at the limit case of infinitely fast relaxation and 
 built  differently  with respect to the procedures  of the present work.
 In particular it was assumed \textit{a priori} an exponential decay of the pressure, temperature and chemical 
 potential differences, whereas in the present work  the relaxations systems with exponential solution are obtained from the equations of the two-phase parent and relaxed models after assuming some quantities constant during the relaxation process.
  
The relaxation procedures developed here results to be simple, robust  and effective, and by construction they
can be also used for other two-phase models belonging to the hierarchy established from
the Baer--Nunziato model. Moreover, the techniques  guarantee
consistency of the values of the relaxed states with the mixture pressure law, so that
the numerical method is mixture-energy-consistency in the
sense defined in \cite{pelanti-shyue}.

This article is structured as follows. In Section~\ref{sec:sixeq_model}  we present the six-equation two-phase  flow model under study. In Section~\ref{sec:hierarchy} we recall the hierarchy of relaxed  
models established from the parent six-equation model.
Examples of equations of state to close the model systems used in the numerical experiments are reported in
Section~\ref{sec:eos}. 
In Section~\ref{sec:numet} we outline the fractional step method
employed to solve the two-phase equations. In Section~\ref{sec:solhom} we illustrate the HLLC-type scheme
used for the solution of the homogeneous system, and we detail then in Section~\ref{sec:relax-step}
the new relaxation techniques to treat the phase transfer source terms. 
Numerical experiments are finally presented in Section~\ref{sec:numexp}, including  tests
  with shocks, interfaces, evaporation waves and metastable states.

\section{Single-velocity six-equation two-phase compressible flow model}

\label{sec:sixeq_model}

We consider a compressible flow composed of two phases that we assume in kinetic equilibrium
with velocity $\vec{u}$. The volume fraction, density,  pressure, specific internal energy 
of each phase will be denoted by
$\alpha_k$, $\rho_k$, $p_k$, $\varepsilon_k$, $k=1,2$, respectively. 
We will denote  the phasic internal energy per unit volume with $\Ei_k = \rho_k \varepsilon_k$, and the phasic total energy per unit volume with
$E_k=\Ei_k+\rho_k \frac{|\vec{u}|^2}{2}$.
The saturation condition is $\alpha_1+\alpha_2=1$.
The mixture density is $\rho=\sum_{k=1}^2 \alpha_k \rho_k\,$, the mixture 
internal energy per unit volume $\Ei= \sum_{k=1}^2 \alpha_k \Ei_k\,$,  
and the mixture total energy $E=\sum_{k=1}^2 \alpha_k E_k\,$. The nomenclature of the variables is summarized in  Table~\ref{tab_nom}.
We describe the two-phase flow by the following system \cite{pelanti-shyue} consisting of $5+d$ equations, where $d$ denotes the spatial dimension:
\begin{subequations}
  \label{eq:2phasesys}
  \begin{eqnarray}
  \label{eq:2phase_alpha}
  &&  \dts \alpha_{1} +  \vec{u} \cdot  \nabla \alpha_1 = \mP,\\
 [1mm]
  &&\dts(\alpha_1 \rho_1) +\nabla \cdot (\alpha_1 \rho_1 \vec{u}) = \mM,\\
  [1mm]
  &&\dts(\alpha_2 \rho_2) +\nabla \cdot (\alpha_2 \rho_2 \vec{u}) = -\mM,\\
  [1mm]
 && \dts (\rho \vec{u})+\nabla \cdot (\rho \vec{u} \otimes \vec{u} 
 + (\alpha_1 p_1+\alpha_2 p_2)\mathbb{I} )=0,\\
  [1mm]
  && \dts (\alpha_1 E_1) + \nabla \cdot(\alpha_1(E_1+p_1)\vec{u}) +\nonce
  = -\pint\mP+\mQ +\textstyle \left(\gint +\textstyle\frac{|\vec{u}|^2}{2}\right) \mM,\\
  [1mm]
&&\dts (\alpha_2 E_2) +\nabla \cdot(\alpha_2(E_2+p_2)\vec{u}) -\nonce
  = \pint\mP-\mQ - \textstyle \left(\gint +\textstyle\frac{|\vec{u}|^2}{2}\right) \mM,
 \end{eqnarray}
where the non-conservative  term $\nonce$ appearing in 
the phasic total energy equations is given by
\begin{equation}
\label{eq:nonce}
   \nonce =-\vec{u}\cdot \vec{\noncep}\,,\qquad
   \vec{\noncep} = Y_2\nabla(\alpha_1 p_1) -Y_1\nabla(\alpha_2 p_2)\,.
\end{equation}
\end{subequations}
Here $Y_k=\frac{\alpha_k \rho_k}{\rho}$ is the mass fraction of phase $k$.
Above we have denoted with $\mP$, $\mQ$ and $\mM$ the volume,  heat and mass   transfer terms between
the two phases. These transfer terms are expressed as relaxation terms:
\begin{equation}
\label{eq:relterms}
\mP =\mu(p_1-p_2), \quad \mQ=\vartheta(T_2-T_1), \quad \mM=\nu(g_2-g_1),
\end{equation}
where $T_k$ denotes the phasic temperature, and $g_k$  the phasic chemical potential. $\mu$, $\vartheta$, and $\nu$ are parameters or more generally
functions expressing the rate of 
mechanical, thermal and chemical relaxation, respectively. 
Here we are interested in modeling flows in mechanical equilibrium, hence we will always consider that mechanical relaxation is an instantaneous process, 
thus we assume $\mu\rightarrow +\infty$. 
Indeed, following the same idea of \cite{saurel-PA,saurel-PB,pelanti-shyue}, the parent non-equilibrium two-phase flow model with instantaneous pressure relaxation (\ref{eq:2phasesys})
is used to approximate solutions to the limiting pressure-equilibrium flow model (see model (\ref{eq:sisp}) in section \ref{sec:5eqm}), which is the physical flow model of interest. Concerning thermal and chemical relaxation, in contrast to \cite{saurel-PA,pelanti-shyue},
no specific assumption is made for the heat 
and mass transfer rate, and hence for $\vartheta$ and $\nu$.
The quantity $\pint$ is an interface pressure and $\gint$ is an interface chemical potential.
The definition of the relaxation parameters or functions and of the interface  quantities needs to be consistent with the second law on thermodynamics,
namely the entropy production for the mixture must be positive. 
Sufficient conditions are (see proof in \cite{flatten-lund:rel}):
\begin{equation}
\mu \geq 0, \quad \vartheta\geq 0, \quad \nu \geq 0, 
\end{equation}
and
\begin{equation}
\label{eq:interfcons}
\pint\in \{\min(p_1,p_2),\max(p_1,p_2)\}, \quad \gint \in \{\min(g_1,g_2),\max(g_1,g_2)\}.
\end{equation}
Hence, it suffices to define the interface  quantities $\pint$ and $\gint$ as convex combinations
of the respective phasic quantities. Concerning $\pint$, for our numerical tests we have used the 
definition proposed in \cite{saurel-gav-ren,saurel-PB}, which  we already used in \cite{pelanti-shyue}:
$\pint=\frac{Z^a_1 p_1+Z^a_2p_2}{Z^a_1+Z_2^a}$, where $Z^a_k = \rho_k c_k$ is the
acoustic impedance of phase $k$. Other definitions are possible,
for instance the one suggested in \cite{sa-ab:multi}, $\pint=\alpha_1 p_1+\alpha_2p_2$. 
Concerning $\gint$, we will see that we do not need to  define it in our numerical scheme.
To close the model system an equation of state for each phase must be provided,
for instance through the specification of the pressure relations $p_k(\Ei_k,\rho_k)$
and the temperature relations $T_k(p_k,\rho_k)$. If thermo-chemical transfer terms are not considered,
then the specification of the pressure laws $p_k(\Ei_k,\rho_k)$ (incomplete equation of state) suffices
to solve the model system.

The two-phase model above is hyperbolic and the eigenvalues associated to the direction $\vec{n}$
are given by $\lambda_{1,5+d}  = \vec{u}\cdot \vec{n}\mp c_{\rm f}\,$, 
$\,\lambda_{l} =\vec{u}\cdot \vec{n}\,$, for $l=2,\ldots,4+d$ (eigenvalue of multiplicity $3+d$). Here
 $c_{\rm f}$ is the non-equilibrium (frozen) speed of sound, defined by
\begin{equation}
c_{\rm f}^2 = \left(\frac{\partial p_{\rm m}}{\partial \rho}\right)_{s_k,Y_k,\alpha_k,\, k=1,2},
\end{equation}
where $s_k$ denotes the  entropy of phase $k$, and where we have introduced the mixture pressure\break 
$p_{\rm m}(\rho,s_1,s_2,Y_1,\alpha_1)$ =  
$\sum_{k=1}^2 \alpha_k p_k\left(s_k,\rho \frac{Y_k}{\alpha_k}\right)$.
From this we obtain:
\begin{equation}
\label{eq:cfroz}
c_{\rm f} = \displaystyle \sqrt{Y_1 c_1^2 + Y_2 c_2^2}\,. 
\end{equation}
Here $c_k$ is the speed of sound of phase $k$, defined by
$c_k^2 = \left(\frac{\partial p_k}{\partial \rho_k}\right)_{\!s_k}$, which  can be expressed as:
\begin{equation}
c_k=
  \sqrt{\Gamma_k h_k + \chi_k}\,, 
\end{equation}
where $h_k = (\mathcal{E}_k +p_k)/\rho_k$ is the specific enthalpy
of phase $k$, and 
\begin{equation}
\label{eq:gruncoef}
\Gamma_k = \left(\frac{\partial p_k}{\partial \Ei_k}\right)_{\!\rho_k},
\qquad
\chi_k = \left(\frac{\partial p_k}{\partial \rho_k}\right)_{\!\Ei_k}.
\end{equation}
 Note that the sum of the phasic total energy equations recovers a conservation law for the
mixture total energy $E=\sum_{k=1}^2 \alpha_k E_k$:
\begin{equation}
\label{eq:mixEt}
\dts E + \nabla \cdot (E\vec{u} + (\alpha_1p_1+\alpha_2p_2) \vec{u}) = 0\,.
\end{equation}
For later use, let us also write here the equations for the phasic pressures:
\begin{equation}
\label{eq:2phase_pres}
\dts p_k +\vec{u} \cdot \nabla p_k+\rho_k c_k^2 \nabla \cdot \vec{u} 
=\textstyle \frac{\Gamma_k}{\alpha_k} \left[-\left(
\pint -\rho_k^2 \textstyle\left(\frac{\partial \varepsilon_k}{\partial \rho_k}\right)_{p_k}
\right)\mP+  \mQ +\gint \mM\right](-1)^{k-1},\quad k=1,2.
\end{equation}



\begin{table}[h!]
\centering
\begin{tabular}{|l|}
\hline 
$\rho_k$ = phasic density\\
[1mm]
$\alpha_k$ = volume fraction of phase $k$ ($\alpha_1+\alpha_2 =1$)\\
[1mm]
$\vec{u}$ = velocity vector\\
[1mm]
$\varepsilon_k$ = phasic specific internal energy\\
[1mm]
$\Ei_k = \rho_k \, \varepsilon_k $ = phasic internal energy per unit volume\\
[1mm]
$E_k =  \Ei_k + \rho_k \, \frac{|\vec{u}|^2}{2}$ =  phasic total energy per unit volume\\
[1mm]
$p_k$ = phasic pressure\\
[1mm]
$p_{\rm m} = \alpha_1 p_1+\alpha_2p_2$\\
[1mm]
$p$ = mixture equilibrium pressure\\
[1mm]
$\pint$ = interface pressure\\
[1mm]
$\rho = \alpha_1 \rho_1 +\alpha_2 \rho_2$ = mixture density\\
[1mm]
$Y_k = \frac{\alpha_k \, \rho_k}{\rho}$ = mass fraction of phase $k$
($Y_1+Y_2=1$)\\
[1mm]
$\varepsilon =Y_1 \varepsilon_1 + Y_2 \varepsilon_2 =$
mixture specific internal energy\\
[1mm]
$\Ei = \rho \varepsilon = \alpha_1 \Ei_1 +\alpha_2 \Ei_2$ = mixture internal energy per unit volume\\
[1mm]
$E = \Ei + \frac{1}{2} \rho |\vec{u}|^2 = \alpha_1 E_1 +\alpha_2 E_2$ = 
mixture total energy per unit volume\\
[1mm]
$h_k =\frac{\Ei_k +p_k}{\rho_k}$ = phasic specific enthalpy\\
[1mm]
$h =Y_1 h_1 +Y_2 h_2$ = mixture specific enthalpy\\
[1mm]
$c_k = \sqrt{\left(\frac{\partial p_k}{\partial \rho_k}\right)_{s_k}}$ = sound speed of phase $k$\\
[1mm]
$c_{\rm f} = \sqrt{Y_1 c_1^2+Y_2c_2^2}$ = non-equilibrium (frozen) mixture sound speed\\
[1mm]
$T_k$ = phasic temperature\\
[1mm]
$T$ = mixture equilibrium temperature\\ 
[1mm]
$s_k$ = phasic entropy\\
[1mm]
$s = Y_1s_1+Y_2s_2$ =  mixture entropy\\
[1mm]
$g_k$ = phasic chemical potential\\
[1mm]
$\gint$ = interface chemical potential\\
[1mm]
$\Gamma_k = \left(\frac{\partial p_k}{\partial \Ei_k}\right)_{\rho_k}$ = Gr\"uneisen coefficient of phase $k$\\
[1mm]
$\chi_k = \left(\frac{\partial p_k}{\partial \rho_k}\right)_{\Ei_k}$\\
[1mm]
$\phi_k = \left(\frac{\partial \rho_k}{\partial T_k}\right)_{p_k} = -\rho_k \beta_k$,
$\quad \beta_k$  = phasic coefficient of thermal expansion\\
[1mm]
$\zeta_k = \left(\frac{\partial \rho_k}{\partial p_k}\right)_{T_k} = \rho_k \mathcal{K}_{Tk}$, 
$\quad \mathcal{K}_{Tk}$ = phasic isothermal compressibility\\
[1mm]
$\mathcal{K}_{Sk} = \frac{1}{\rho_k c_k^2}$ = phasic isentropic compressibility\\
[1mm]
$\kappa_{pk} = T_k\left(\frac{\partial s_k}{\partial T_k}\right)_{p_k}$  = 
$\left(\frac{\partial h_k}{\partial T_k}\right)_{p_k}$
=specific heat capacity at constant pressure\\ 
[1mm]
$\kappa_{vk} = T_k\left(\frac{\partial s_k}{\partial T_k}\right)_{\rho_k}$ =
$\left(\frac{\partial \varepsilon_k}{\partial T_k}\right)_{\rho_k}$ =specific 
heat capacity at constant volume\\
[1mm]
$C_{pk} = \alpha_k \rho_k \kappa_{pk}$ = 
phasic extensive heat capacity at constant pressure\\
\hline
\end{tabular}
\caption{Nomenclature of variables.}
\label{tab_nom}
\end{table}


\section{Hierarchy of single-velocity relaxed  two-phase flow models}

\label{sec:hierarchy}

From the parent six-equation non-equilibrium model (\ref{eq:2phasesys})
presented in the previous section we can establish 
a hierarchy of hyperbolic relaxed single-phase two-phase flow models by considering
the limit of combinations of instantaneous relaxation processes, see \cite{flatten-lund:rel,lund:rel}.
The $p$-relaxed and $pT$-relaxed model equations recalled below will be used in the
construction of the relaxation procedures in Section \ref{sec:arbraterel}.

\subsection{Five-equation $p$-relaxed two-phase flow model}
\label{sec:5eqm}

We assume that 
the flow is driven instantaneously to mechanical equilibrium, $p_1=p_2=p$,
hence we consider  $\mu \rightarrow +\infty$.
The $p$-relaxed (pressure equilibrium) model, corresponding to the well known Kapila 
\textit{et al.}~model \cite{kapila} (see also \cite{gui-mur}), consists of $4 +d$ equations:
\begin{subequations}
\label{eq:sisp}
\begin{eqnarray}
\label{eq:sisp_alpha}
\hspace*{-9mm}  &&  \dts \alpha_{1} + \vec{u} \cdot \nabla \alpha_1 - 
\textstyle\frac{\alpha_1 \alpha_2}{D} (\rho_2 c_2^2 -\rho_1c_1^2)
\nabla \cdot \vec{u} = \textstyle\frac{1}{D} \left(\alpha_2 \Gamma_1+ \alpha_1\Gamma_2\right)\!\mQ
+\textstyle\frac{1}{D} \left(\alpha_2 \cint{g1}^2 + \alpha_1\cint{g2}^2\right)\!\mM\,,
\\
  [1mm]
\hspace*{-9mm}  &&\dts(\alpha_1 \rho_1) +\nabla \cdot (\alpha_1 \rho_1 \vec{u}) = \mM\,,\\
  [1mm]
\hspace*{-9mm}  &&\dts(\alpha_2 \rho_2) +\nabla \cdot (\alpha_2 \rho_2 \vec{u}) = -\mM\,,\\
 [1mm]
\hspace*{-9mm} && \dts (\rho \vec{u})+\nabla \cdot (\rho \vec{u} \otimes \vec{u} + p\mathbb{I}) =0 \,,\\
  [1mm]
 \hspace*{-9mm} && \dts E +\nabla \cdot ((E+p)\vec{u}) =0\,,
\end{eqnarray}
\end{subequations}
where
\begin{equation}
\label{eq:Dpar}
D = \alpha_1 \rho_2 c_2^2 + \alpha_2 \rho_1 c_1^2\,
\end{equation}
and
\begin{equation}
\cint{gk}^2 = \Gamma_k(\gint-h_k)+c_k^2 = \Gamma_k\gint + \chi_k\,, \quad k=1,2\,.
\end{equation}
The derivation of the above $p$-relaxed system from the parent system (\ref{eq:2phasesys}) 
is detailed in Appendix \ref{sec_redmod}, and it has been also
illustrated  in our work~\cite{pelanti-shyue-3p} for a more general $N$-phase model.   
Given the phasic energy laws $\Ei_k(p_k,\rho_k)$, the mixture pressure law $p=p(\Ei,\rho_1,\rho_2,\alpha_1)$  for this  model
is determined by the mixture energy relation
\begin{equation}
\label{eq:mixpresimpl}
\Ei = \alpha_1\Ei_1(p,\rho_1)+\alpha_2\Ei_2(p,\rho_2)\,,
\end{equation}
where we have used the isobaric condition $p_1=p_2=p$.
The speed of sound associated to the model is defined by
\begin{equation}
c_p^2 = \left(\frac{\partial p}{\partial \rho}\right)_{s_k,Y_k,\,k=1,2}\,,
\end{equation} 
which gives the well known Wood's speed of sound
\begin{equation}
c_p = \left( \rho \sum_{k=1}^2 \frac{\alpha_k}{\rho_k c_k^2}\right)^{-\frac{1}{2}}\,.
\end{equation}
Note that the term $D$ (\ref{eq:Dpar}) can be written in terms of $c_p$, 
$\frac{1}{D} = \frac{\rho c_p^2}{\rho_1 c_1^2 \rho_2 c_2^2}$. 
The pressure equation is:
\begin{equation}
\dts p +\vec{u}\cdot \nabla p+\rho c_p^2\nabla \cdot \vec{u}  = 
\textstyle \frac{1}{D}
\left[ ( \Gamma_1\rho_2c_2^2-\Gamma_2\rho_1c_1^2 )\mQ+
(\rho_2c_2^2\cint{g1}^2-\rho_1c_1^2\cint{g2}^2)\mM
\right]\,.
\end{equation}
Let us now write the equations for the phasic temperatures $T_k$, $k=1,2$,
which we will use in the following:
\begin{equation}
\label{eq:sisp_temp}
\begin{split}
\dts T_k &+\vec{u}\cdot \nabla T_k +\frac{\rho c_p^2}{\phider_k}
\left(-\zetader_k+\frac{1}{c_k^2}\right)\nabla \cdot \vec{u}\\
& = \frac{1}{\phider_k D}
\left[ (-1)^k\frac{\rho_k}{\alpha_k}\left( \alpha_2\Gamma_1+\alpha_1\Gamma_2\right) 
-\zetader_k (\Gamma_1\rho_2c_2^2-\Gamma_2\rho_1c_1^2)\right]\mQ\\
&+  \frac{1}{\phider_k}\left\{ \frac{(-1)^{k-1}}{\alpha_k}+ 
\frac{1}{D}
\left[(-1)^{k}\frac{\rho_k}{\alpha_k} (\alpha_2 \cint{g1}^2+\alpha_1\cint{g2}^2)
-\zetader_k(\rho_2c_2^2\cint{g1}^2-\rho_1c_1^2\cint{g2}^2)\right]\right\}\mM\,,
\end{split}
\end{equation}
where we have introduced the derivatives
\begin{equation}
\label{eq:phizeta}
\phider_k= \left(\frac{\partial\rho_k}{\partial T_k}\right)_{p_k} =-\rho_k\beta_k
\qeq \zetader_k= \left(\frac{\partial \rho_k}{\partial p_k}\right)_{T_k} =
\rho_k \mathcal{K}_{Tk}\,,
\end{equation}
where $\beta_k$ denotes the coefficient of thermal expansion and
$\mathcal{K}_{Tk}$ the isothermal compressibility.
Note also that we have the relations:
\begin{equation}
\beta_k=\frac{\Gamma_k \kappa_{pk}}{c_k^2} = \frac{\Gamma_k C_{pk}}{c_k^2 \alpha_k \rho_k}
\qeq  \mathcal{K}_{Tk} = \mathcal{K}_{Sk}+\frac{\beta_k^2 T_k}{\rho_k \kappa_{pk}} = 
\frac{1}{\rho_k c_k^2}+\frac{\beta_k^2 T_k \alpha_k}{C_{pk}}\,,
\end{equation}
where $\mathcal{K}_{Sk}=\frac{1}{\rho_k c_k^2}$ is the isentropic compressibility, and where 
$C_{pk}=\alpha_k \rho_k \kappa_{pk}$  and 
$\kappa_{pk}= \frac{\partial h_k}{\partial T_k}\bigl|_{p_k}= T_k \frac{\partial s_k}{\partial T_k}\bigl|_{p_k}$.

\vspace{2mm}

\noindent
\textbf{Remark.}  In our previous work \cite{pelanti-shyue} an additional source term of the form 
$\mM/\rhoint$ was written in the equation for the volume fraction $\alpha_1$ of the above six-equation two-phase model (\ref{eq:2phasesys}),  with $\rhoint$ representing an interface density.
Similar to \cite{flatten-lund:rel}, this term is not included in the present  model. 
 The purpose of the term $\mM/\rhoint$ in \cite{pelanti-shyue}
was to indicate the influence of the mass transfer process on the evolution  of the volume fraction.
Nonetheless, the rigorous derivation of the pressure-relaxed model (\ref{eq:sisp}) from the system (\ref{eq:2phasesys}) (see Appendix~\ref{sec_redmod}) reveals that indeed 
mass transfer terms  affect $\alpha_k$ via the pressure relaxation process, as we observe
from the contribution of $\mM$ appearing in 
(\ref{eq:sisp_alpha}).
Note  that  the presence of the term $\mM/\rhoint$ eventually  does not affect the numerical model and the numerical results
presented in \cite{pelanti-shyue} since there $\nu$ = 0 or $\nu \rightarrow +\infty$, and the  procedure for treating instantaneous chemical
relaxation consists in imposing directly algebraic thermodynamic equilibrium conditions.



\subsection{Four-equation $pT$-relaxed two-phase flow model}

We now assume that the flow is driven instantaneously to both mechanical and thermal equilibrium, $p_1=p_2=p$, $T_1=T_2=T$. Hence we consider the limit $\mu \rightarrow +\infty$ and $\vartheta \rightarrow +\infty$.
We obtain the following reduced model composed of $3+d$ equations (used for instance in \cite{lund_proc,lemart-boil,saurelCF16,saurelCF17,demou-boil}):
\begin{subequations}
\label{eq:sysT}
\begin{eqnarray}
  &&\dts(\alpha_1 \rho_1) +\nabla \cdot (\alpha_1 \rho_1 \vec{u}) = \mM\,,\\
  [1mm]
 &&\dts(\alpha_2 \rho_2) +\nabla \cdot  (\alpha_2 \rho_2 \vec{u}) = -\mM\,,\\
 [1mm]
 && \dts (\rho \vec{u})+\nabla \cdot (\rho \vec{u}\otimes \vec{u} + p\mathbb{I}) =0 \,,\\
  [1mm]
 && \dts E +\nabla \cdot ((E+p)\vec{u}) =0.
\end{eqnarray}
\end{subequations}
The mixture pressure law $p=p(\Ei,\rho_1,\rho_2)$ is determined by the 
energy relation (\ref{eq:mixpresimpl}), together with
the isothermal condition $T_1(p,\rho_1)=T_2(p,\rho_2)$.
The  speed of sound for this model is defined by
\begin{equation}
c_{pT}^2 = \left(  \frac{\partial p}{\partial \rho}\right)_{s,Y_1,Y_2},
\end{equation}
where $s$ is the mixture specific entropy $s=Y_1s_1+Y_2s_2$. This gives
\begin{equation}
\frac{1}{c_{pT}^2} = \frac{1}{c_p^2} + \frac{\rho T C_{p1}C_{p2}}{C_{p1}+C_{p2}} 
\left(\frac{\Gamma_2}{\rho_2 c_2^2} -\frac{\Gamma_1}{\rho_1 c_1^2} \right)^2,
\end{equation}
where we recall $C_{pk}=\alpha_k \rho_k \kappa_{pk}$ (extensive heat capacities).
Let us finally write also the equations for the volume fraction $\alpha_1$, the
temperature $T$ and the pressure $p$:
\begin{eqnarray}
\phantom{a}\hspace*{-9mm}&&\dts \alpha_1 +\vec{u}\cdot \nabla \alpha_1+\rho c_{pT}^2
\left[\alpha_1\alpha_2 \left(\frac{1}{\rho_2c_2^2}-\frac{1}{\rho_1c_1^2}\right) +
\frac{T C_{p1}C_{p2}}{C_{p1}+C_{p2}} \left(\frac{\Gamma_2}{\rho_2 c_2^2} 
-\frac{\Gamma_1}{\rho_1 c_1^2}  \right)
\left( \frac{\alpha_1 \Gamma_2}{\rho_2 c_2^2} +  \frac{\alpha_2 \Gamma_1}{\rho_1 c_1^2} \right)\right]
\nabla \cdot \vec{u}
\nonumber \\
\phantom{a}\hspace*{-9mm}&& \phantom{a} \hspace{11cm} = 
\mathcal{M}\mathcal{S}_{\alpha}\,,
\\
[2mm]
\phantom{a}\hspace*{-9mm}&& \dts T +\vec{u}\cdot \nabla T+\frac{\rho c_{pT}^2 T}{C_{p1}+C_{p2}}
\left(\frac{C_{p1}\Gamma_1}{\rho_1 c_1^2} + \frac{C_{p2}\Gamma_2}{\rho_2 c_2^2}\right)  \nabla \cdot \vec{u} 
 = \mathcal{M}\mathcal{S}_{T}\,,\\
[2mm]
\phantom{a}\hspace*{-9mm}&&\dts p +\vec{u}\cdot \nabla p+\rho c_{pT}^2\nabla \cdot \vec{u}  = 
\mathcal{M}\mathcal{S}_{p}\,,
\end{eqnarray}
where 
\begin{subequations}
\label{eq:sourcep2}
\begin{eqnarray}
\hspace*{-14mm}&&\mathcal{S}_{\alpha}=\frac{1}{D_T}\left[\left(\frac{\alpha_1}{\Gamma_1} \!+\!\frac{\alpha_2}{\Gamma_2}\right)
(\alpha_1 \phider_1 +\alpha_2\phider_2 )+\alpha_1\alpha_2\left(\frac{\chi_1}{\Gamma_1}-\frac{\chi_2}{\Gamma_2}
\right)
(\phider_1 \zetader_2- \phider_2 \zetader_1)\right],\\
[1mm]
\hspace*{-14mm}&&\mathcal{S}_{T}=\frac{1}{D_T}\left[\left(\!\frac{\chi_2}{\Gamma_2}
\!-\!\frac{\chi_1}{\Gamma_1}\!\right)
(\alpha_1 \zetader_1 \rho_2\!+\! \alpha_2 \zetader_2 \rho_1)+
\left(\!\frac{\rho_1 c_1^2}{\Gamma_1}\!-\!\frac{\rho_2 c_2^2}{\Gamma_2}\!\right)
(\alpha_1 \zetader_1 + \alpha_2 \zetader_2)\!+\!
  \left(\!\frac{\alpha_1}{\Gamma_1} \!+\!\frac{\alpha_2}{\Gamma_2}\!\right) (\rho_2\!-\!\rho_1)\right]\!,\\
[1mm]
\hspace*{-14mm}&&\mathcal{S}_{p}=\frac{1}{D_T}\left[\left(\frac{\chi_1}{\Gamma_1}\!-\!\frac{\chi_2}{\Gamma_2}
\right)
(\alpha_1 \phider_1 \rho_2+ \alpha_2 \phider_2 \rho_1)+
\left(\frac{\rho_2 c_2^2}{\Gamma_2}\!-\!\frac{\rho_1 c_1^2}{\Gamma_1}\right)
(\alpha_1 \phider_1 + \alpha_2 \phider_2)\right],
\end{eqnarray}
with
\begin{equation}
\label{eq:DTsource}
D_T = \alpha_1\alpha_2\left( \frac{\rho_1 c_1^2}{\Gamma_1}- \frac{\rho_2 c_2^2}{\Gamma_2} \right)
(\phider_1\zetader_2-\phider_2 \zetader_1) +
\left(\frac{\alpha_1}{\Gamma_1} +\frac{\alpha_2}{\Gamma_2}\right)
(\alpha_1 \phider_1 \rho_2+\alpha_2\phider_2 \rho_1)\,.
\end{equation}
\end{subequations}
The derivation of these expressions of $\mathcal{S}_{\alpha}$, $\mathcal{S}_{T}$, $\mathcal{S}_{p}$ 
will be illustrated in  Appendix~\ref{sec:source4eq}.


\subsection{Three-equation $pTg$-relaxed two-phase flow model}

For completeness, we also recall the relaxed model obtained by 
assuming full thermodynamic equilibrium, $p_1=p_2=p$, $T_1=T_2=T$, 
and $g_1$= $g_2$. Hence we consider the limit
 $\mu \rightarrow +\infty$, $\vartheta \rightarrow +\infty$, and
$\nu \rightarrow +\infty$. 
We obtain the homogeneous equilibrium model (HEM) 
composed of $2+d$ equations
(see e.g.~\cite{st-we:2phase,clerc,delor-laf-pelanti-IJMF,munkejord-3pJCP}):
\begin{subequations}
\label{eq:pTgHEM}
\begin{eqnarray}
&&\dts \rho +\nabla \cdot (\rho \vec{u}) = 0\,,\\
 [1mm]
 && \dts (\rho \vec{u})+\nabla \cdot (\rho \vec{u}\otimes \vec{u} + p\mathbb{I}) =0 \,,\\
  [1mm]
 && \dts E +\nabla \cdot ((E+p)\vec{u}) =0\,.
\end{eqnarray}
\end{subequations}
The mixture pressure law $p=p(\Ei,\rho)$ is determined by the
energy relation (\ref{eq:mixpresimpl}), the isothermal
condition $T_1(p,\rho_1) = T_2(p,\rho_2)$, and the equilibrium condition
$g_1(p,T) = g_2(p,T)$.
The  speed of sound is defined by:
\begin{equation}
c_{pTg}^2 = \left(  \frac{\partial p}{\partial \rho}\right)_{s},
\end{equation}
which gives (see for instance the systematic  derivation of the 
speeds of sound of the various  models in the hierarchy in \cite{pelanti-shyue-3p})
\begin{equation}
\frac{1}{c_{pTg}^2} = \frac{1}{c_{pT}^2} + \frac{\rho T}{C_{p1}+C_{p2}} 
\left[\frac{\Gamma_1 C_{p1}}{\rho_1 c_1^2} +\frac{\Gamma_2 C_{p2}}{\rho_2 c_2^2} 
-\frac{1}{T} \left(\frac{dT}{dp} \right)_{\rm sat}(C_{p1}+C_{p2})\right]^2.
\end{equation}
We remark that subcharacteristic conditions 
hold for the speeds of sound of the two-phase flow models in the hierarchy
\cite{flatten-lund:rel}:
\begin{equation}
\label{eq:subchar}
c_{pTg} \leq c_{pT}\leq c_{p} \leq c_{\rm f}\,.
\end{equation}
As expected, the speed of sound is reduced 
whenever an additional equilibrium assumption is introduced.

\section{Equation of State (EOS)}

\label{sec:eos}

The numerical techniques that we will present in the following sections can be 
employed for any choice of the equations of state. 
Nonetheless, for the numerical experiments we will consider
two particular equations of state, which can both be written
in the form of the Mie--Gr\"uneisen equation of state
recalled hereafter.

\subsection{Mie--Gr\"uneisen equation of state}

The incomplete Mie--Gr\"uneisen equation of state has the form (see e.g.~\cite{men-plo:rp}):
\begin{equation}
\label{miegrun_eos}
p(\Ei, \rho)=\Gamma(\rho)(\Ei-\rho \varepsilon_{\rm r}(\rho)) +p_{\rm r}(\rho)\,,
\end{equation}
where $\Gamma(\rho)$ is the Gr\"uneisen coefficient defined for a general EOS as
in (\ref{eq:gruncoef}), and $\varepsilon_{\rm r}(\rho)$, $p_{\rm r}(\rho)$ are  reference 
specific energy  and pressure functions, respectively.
An extension of this incomplete EOS to a complete one can be found in \cite{Mie-grun-complete}.
Many equations of state can be written in the form (\ref{miegrun_eos}),
including the JWL and NASG equations of state reported below.

For two-phase flows in mechanical equilibrium where each phase is governed by an equation of state with the form of
the Mie-Gr\"uneisen EOS, it is possible to 
obtain an explicit expression for the mixture pressure law~(\ref{eq:mixpresimpl}):
\begin{equation}
p(\Ei,\rho_1,\rho_2,\alpha_1) = \frac{\Ei-
\left(\alpha_1\rho_1 \varepsilon_{{\rm r}1}(\rho_1) + \alpha_2\rho_2\varepsilon_{{\rm r}2} (\rho_2) \right)
+\left(\alpha_1 \frac{p_{{\rm r}1}(\rho_1)}{\Gamma_1(\rho_1)} +\alpha_2 \frac{p_{{\rm r}2}(\rho_2)}{\Gamma_2(\rho_2)}\right)}{
\frac{\alpha_1}{\Gamma_1(\rho_1)} + \frac{\alpha_2}{\Gamma_2(\rho_2)} }.
\end{equation}
This is an important advantage from the numerical point of view, since solving an implicit equation for the pressure
can be computationally expensive.

\subsection{Jones--Wilkins--Lee (JWL) Equation of State}

The Jones--Wilkins--Lee (JWL) EOS \cite{lee-h-k} has been extensively used to model
gaseous or solid explosives, and it has the form (\ref{miegrun_eos}) with:
\begin{subequations}
\label{jwl_eos}
\begin{eqnarray}
&&\Gamma(\rho) = \Gamma_0\,,\\
&&\eint_{\rm r}(\rho) = \frac{a}{r_1 \rho_0}
\textstyle 
 {\rm e}^{-r_1\frac{\rho_0}{\rho}}
 + \frac{b}{r_2 \rho_0} \textstyle
  {\rm e}^{-r_2\frac{\rho_0}{\rho}} -\eint_0\,,\\
 &&p_{\rm r}(\rho) = a \textstyle 
 \,{\rm e}^{-r_1\frac{\rho_0}{\rho}} 
 +b  {\rm e}^{-r_2\frac{\rho_0}{\rho}}\,.
 \end{eqnarray}
\end{subequations}

\subsection{Noble--Abel Stiffened Gas (NASG) Equation of State}

The Noble--Abel Stiffened Gas (NASG) Equation of State introduced in \cite{saurel-lemet-NASG}
combines the stiffened gas EOS \cite{men-plo:rp} and the Noble--Abel EOS.
It has the form:
\begin{subequations}
\label{eq:nasg}
\begin{eqnarray}
\label{eq:nasgpres}
&&p(\Ei,\rho)  = \frac{\gamma-1}{1-\rho b}(\Ei-\eta \rho) -\gamma \varpi\,,\\
[1mm]
\label{eq:nasgtemp}
&&T(p,\rho)  = \frac{(1-\rho b)(p+\varpi)}{\kappa_v \rho (\gamma-1)}.
\end{eqnarray}
\end{subequations}
Here $\gamma$, $\varpi$, $\eta$, $b$, $\kappa_v$ are material-dependent constant 
parameters. The coefficient $b$ represents the covolume of the fluid and
the choice $b=0$ gives  the classical stiffened gas equation of state.
We can observe that the pressure law in (\ref{eq:nasgpres}) has the form 
(\ref{miegrun_eos}) with
\begin{equation}
\Gamma(\rho) = \textstyle \frac{\gamma-1}{1-\rho \,b}\,,\quad
\varepsilon_{\rm r} =\eta\,,\quad p_{\rm r}  = -\gamma \varpi. 
\end{equation}
Let us also write the expression of the specific entropy $s$, the specific enthalpy $h$,  and 
 the chemical potential (equal for a pure constituent to its specific Gibbs
free energy) $g=h-Ts$:
\begin{subequations}
\begin{eqnarray}
&& s(p,T) = \kappa_v \log \frac{T^\gamma}{(p+\varpi)^{\gamma-1}} +\etapp\,,\\
&& h(T,p) = \kappa_p T +b p +\eta\,,\\  
&&g(p,T) = (\gamma \kappa_v-\etapp)T 
-\kappa_v T\log \frac{T^\gamma}{(p+\varpi)^{\gamma-1}} +\eta +bp,
\end{eqnarray}
\end{subequations}
where $\kappa_p =\gamma \kappa_v$ (specific heat capacity at constant pressure)
 and $\etapp$ are constant parameters.
Let us note that the speed of sound can be written:
\begin{equation}
c = \sqrt{\gamma\frac{p+\varpi}{\rho(1-\rho b)}}\,.
\end{equation}
Finally, we can write the expressions for the derivatives in (\ref{eq:phizeta}): 
\begin{equation}
\phider = -\frac{\rho}{T}(1-\rho \,b) \qqeqq \zetader = \frac{1}{T(\gamma-1)\kappa_v +b(p+\varpi)}(1-\rho \, b). 
\end{equation}

\subsubsection{Saturation curves}
  
For applications to two-phase flows with liquid-vapor transition, given the equation
of state for each phase, the theoretical pressure-temperature saturation curve 
is determined  by the equilibrium conditions $p_1=p_2=p$, $T_1=T_2=T$, $g_1(p,T) = g_2(p,T)$.
Assuming here  each phase   governed by a NASG EOS, the equilibrium  relations give the following equation:
\begin{equation}
A_s+\frac{B_s}{T}  +C_s\log T +D_s\log(p+\varpi_1)-\log(p+\varpi_2) +\frac{p\,E_s}{T}=0,
\end{equation} 
where
\begin{equation}
A_s= \frac{\kappa_{p1}-\kappa_{p2}- \etapp_1+\etapp_2}{\kappa_{p2}-\kappa_{v2}},\,\,
B_s=\frac{\eta_1-\eta_2}{\kappa_{p2}-\kappa_{v2}},\,\,
C_s = \frac{\kappa_{p2}-\kappa_{p1}}{\kappa_{p2}-\kappa_{v2}},\,\,
D_s = \frac{\kappa_{p1}-\kappa_{v1}}{\kappa_{p2}-\kappa_{v2}},\,\,
E_s = \frac{b_1-b_2}{\kappa_{p2}-\kappa_{v2}}.
\end{equation}
The constant parameters in the NASG equations of state of the two phases 
 are determined so that the associated
theoretical saturation curves match the experimental saturation curves 
for the considered material, at least in a certain temperature range, see \cite{saurel-lemet-NASG}. 
The Tables~\ref{tab_water}-\ref{tab_water2} reported in Section~\ref{sec:numexp}  contain sets of parameters determined
 in \cite{saurel-lemet-NASG} for water, Table~\ref{tab_dode} contains a slightly modified set of parameters
 for dodecane taken from \cite{saurel-lemet-NASG}.

\section{Numerical method}

\label{sec:numet}

We now consider the numerical solution of the six-equation model (\ref{eq:2phasesys}), which we rewrite here
in compact vectorial form, denoting with $q \in \mathbb{R}^{5+d}$ the vector of the unknowns: 
\begin{subequations}
\label{eq:sysvect}
\begin{equation}
\label{eq:sysvect1}
\dts q +\nabla \cdot f(q) +\sigma(q,\nabla q) =\psi_{ \mu}(q) + 
\psi_{\vartheta}(q) + \psi_{ \nu}(q)\,,
\end{equation}
\begin{equation}
\label{feq}
q=
\left[
\begin{array}{c}
\alpha_1\\
\alpha_1 \rho_1\\
\alpha_2 \rho_2\\
\rho \vec{u}\\
\alpha_1 E_1\\
\alpha_2 E_2
\end{array}
\right ], \quad
f(q) =\left[
\begin{array}{c}
0\\
\alpha_1 \rho_1 \vec{u}\\
\alpha_2 \rho_2 \vec{u}\\
\rho \vec{u}  \otimes \vec{u} + 
\left(\alpha_1 p_1+\alpha_2p_2 \right)\mathbb{I}\\
\alpha_1 \left (E_1+p_1 \right ) \vec{u} \\
\alpha_2 \left (E_2+p_2 \right ) \vec{u}
\end{array}
\right],\quad
\nonc\left (q,\nabla q \right ) = \left[
\begin{array}{c}
\vec{u} \cdot  \nabla \alpha_1 \\
0\\
0\\
0\\
\nonce \\
-\nonce
\end{array}
\right], 
\end{equation}
\begin{equation}
\psi_{\mu}(q) = \left[
\begin{array}{c}
\mathcal{P}\\
0\\
0\\
0\\
-\pint\mathcal{P}\\
 \pint\mathcal{P}
\end{array}
\right],\quad
\psi_{\vartheta}(q) = \left[
\begin{array}{c}
0\\
0\\
0\\
0\\
\mathcal{Q}\\
-\mathcal{Q}
\end{array}
\right],\quad
\psi_{\nu}(q)= \left[
\begin{array}{c}
0\\
\mathcal{M}\\
-\mathcal{M}\\
0\\
\left(\gint +\frac{|\vec{u}|^2}{2}\right) \mM\\
-\left(\gint +\frac{|\vec{u}|^2}{2}\right) \mM
\end{array}
\right],
\end{equation}
\end{subequations}
with $\nonce(q,\nabla q)$ defined in (\ref{eq:nonce}).
Above  we have put into evidence the conservative 
portion of the spatial derivative contributions in the system as
$\nabla \cdot f(q)$, and we have indicated the non-conservative 
term as $\nonc(q,\nabla q)$. The source terms $\psi_{\mu}(q)$, 
$\psi_{\vartheta}(q)$,
$\psi_{\nu}(q)$ contain  mechanical, thermal and chemical
relaxation terms, respectively, as expressed in (\ref{eq:relterms}).

To numerically solve this system we use a classical fractional step method, where we alternate
between the solution of the homogeneous hyperbolic portion of the system 
via a wave-propagation finite volume scheme and the solution of a sequence 
of ordinary differential equations accounting for the relaxation source terms. 
Denoting with $\tau_{\mu}$, $\tau_{\vartheta}$, $\tau_{\nu}$ the characteristic times
for mechanical, thermal, and chemical relaxation, respectively,
let us note that the underlying assumption here is $\tau_{\mu} \ll \tau_{\vartheta} \ll \tau_{\nu}$
(cf.\ for instance \cite{kapila}).
The algorithm consists of the following steps:

\begin{itemize}
\item[1.] Solution of the homogeneous hyperbolic system 
\begin{equation}
\label{eq:homsys}
\dts q +\nabla \cdot f(q) +\sigma(q,\nabla q) =0\,.
\end{equation}
In the following we will denote with the superscript~$0$ the quantities computed in this step.
\item[2.] Relaxation steps
\begin{itemize}
\item[2(a)] Instantaneous mechanical relaxation. We solve in the limit $\mu \rightarrow +\infty$
the system of ODEs
\begin{equation}
\label{eq:oderelp}
\dts q = \psi_{\mu}(q).
\end{equation}
This step drives instantaneously the flow to pressure equilibrium.
We will denote with superscript~$*$ the quantities computed in this step.
\item[2(b)] Thermal relaxation. We solve 
\begin{equation}
\label{eq:oderelT}
\dts q = \psi_{\mu}(q)+\psi_{\vartheta}(q),
\end{equation}
with  $\mu \rightarrow +\infty$. This step drives the phases
towards thermal equilibrium, while maintaining
pressure equilibrium. We will denote with superscript~$**$ the quantities computed in this step.
\item[2(c)] Chemical relaxation. We solve 
\begin{equation}
\label{eq:oderelg}
\dts q = \psi_{\mu}(q)+\psi_{\vartheta}(q)+\psi_{\nu}(q),
\end{equation}
with $\mu \rightarrow +\infty$. 
 This step drives the phases
towards full thermodynamical equilibrium, while maintaining
pressure equilibrium. We will denote with superscript~$\otimes$ the quantities computed in this step.
\end{itemize}
\end{itemize}
Let us first observe that the step 2(a) is always activated since we model
flows in mechanical equilibrium. 
The steps 2(b) and 2(c) might be activated or not depending on the problem of interest, 
and, moreover, they might be activated only at selected locations, typically at
interfaces, identified by $\min_k{\alpha_k}>\epsilon$, where $\epsilon$ is a given
tolerance (e.g.~$10^{-6}$). If thermal and chemical relaxation are activated unconditionally then
the numerical model approximates solutions to the $pTg$-relaxed model (\ref{eq:pTgHEM}).

\subsection{Mixture-energy-consistency}

In the design of the fractional step method indicated above it is important to ensure
\textit{mixture-energy-consistency}, in the sense defined in \cite{pelanti-shyue}.
Let us denote with  superscript $\#$ the quantities computed in any of the relaxation steps of the
above algorithm, $\# = *,**,\otimes$. 
 Let us then denote with $E^{0,C}$ discrete values
of the mixture total energy that come from a conservative approximation of the conservation law for $E$ in
(\ref{eq:mixEt}).
We say that the numerical scheme based on the fractional step algorithm above is mixture-energy-consistent
if the following two properties are satisfied:
\begin{itemize}
\item[(i)] Mixture total energy conservation consistency, i.e.\ conservation at the discrete level of the mixture total energy:
\begin{equation}
\label{eq:mixencons1}
E^0=E^{0,C}\,,
\end{equation}
where $E^0=(\alpha_1 E_1)^{0} + (\alpha_2 E_2)^{0}$.
\item[(ii)] Relaxed pressure consistency, i.e.\ consistency of the values of the relaxed states with the mixture pressure law for  pressure-equilibrium flows (\ref{eq:mixpresimpl}):
\begin{equation}
\label{eq:mixencons2}
\Ei^{0} = \alpha_1^{\#}\Ei_1(p^{\#},\rho_1^{\#})+\alpha_2^{\#}\Ei_2(p^{\#}, \rho_2^{\#})\,,
\end{equation}
where $\Ei^{0}=E^{0}-\frac{(\rho \vec{u})^{0} \cdot (\rho \vec{u})^{0}}{2\rho^0}$.
\end{itemize}

\section{Solution of the homogeneous system}

\label{sec:solhom}

 To solve the hyperbolic homogeneous  portion of (\ref{eq:sysvect}) 
 we employ the 
wave-propagation algorithms of \cite{rjlbook,rjl:wavepr},
which are a class of Godunov-type finite volume 
methods  to approximate hyperbolic 
systems of partial differential equations. 
We shall consider  here for simplicity the one-dimensional case in the $x$ direction ($d=1$),
and we refer the reader to \cite{rjlbook} for a comprehensive 
presentation of these numerical schemes. Hence we consider here  the solution 
of the one dimensional system  $\dts q+\dxs f(q) +\nonc (q,\dxs q)=0$, $q\in \mathbb{R}^{6}$
(as obtained by setting $\vec{u}=u$ and $\nabla =\partial_x$ in (\ref{eq:sysvect})).
We assume  a   grid with cells of uniform size $\Delta x$, and  we denote
with $Q_{i}^n$ the approximate solution of the system at the $i$th cell 
and at time $t^n$, $i \in \mathbb{Z}$,
 $n \in \mathbb{N}$.
The second-order wave propagation algorithm  has the form
\begin{equation}
\label{eq_wp_1d}
Q_{i}^{n+1}= Q_{i}^n  - \frac{\Delta t}{\Delta x}
         ( \apdqm + \amdqp) 
      - \frac{\Delta t}{\Delta x}
      ( \tilhFp - \tilhFm )\,.        
\end{equation}
Here $\mathcal{A}^{\mp} \Delta Q_{i+1/2}$ 
 are the so-called fluctuations arising from 
 Riemann problems at cell interfaces $(i+1/2)$ between adjacent cells $i$ 
 and $(i+1)$, and $\tilhFp$ 
 are correction terms for (formal) second-order accuracy.
To define the fluctuations, a Riemann solver (cf.~\cite{godrav,toro,rjlbook})
must be provided.
 The solution structure defined by a given solver for a Riemann problem with
 left and right  data $q_{\ell}$ and $q_r$
 can be expressed in general  by a set of $\mathscr{M}$ waves 
$\mathcal{W}^l$ and corresponding speeds $s^l$, $\mathscr{M} \gtreqless 3N$.
For the HLLC-type solver described below $\mathscr{M} = 3$. 
The sum of the waves must be equal to
the initial jump in the vector $q$ of  the system variables:
\begin{equation}
\Delta q \equiv q_r -q_{\ell} = \sum_{l=1}^{\iM} \mathcal{W}^l.
\end{equation}
Moreover, for any variable of the model system governed by a 
conservative equation
the initial  jump in the associated flux function must 
be recovered by the sum of waves multiplied by the corresponding speeds.
In the considered model the conserved quantities are 
$\alpha_k \rho_k$, $k=1,2$, and  $\rho u$, therefore
in order to guarantee conservation we need:
\begin{equation}
 \label{eq-deltaf}
\Delta f^{(\xi)}\equiv f^{(\xi)}(q_{r})-f^{(\xi)}(q_{\ell}) = 
\sum_{l=1}^{\iM}  s^l\mathcal{W}^{l(\xi)}
\end{equation}
for $\xi =2,3,4$, 
where  $f^{(\xi)}$ is the $\xi$th component of the flux vector $f$,
and $\mathcal{W}^{l(\xi)}$  denotes the $\xi$th component of the $l$th wave,
$l=1,\ldots, \iM$. It is clear that conservation of the partial densities  ensures conservation
of the mixture density $\rho = \sum_{k=1}^2\alpha_k\rho_k$.
In addition, we must ensure conservation of the mixture total energy, 
\begin{equation}
 \label{eq-consE}
\Delta f_E\equiv f_E(q_{r})-f_E(q_{\ell}) = 
\sum_{l=1}^{\iM} s^l (\mathcal{W}^{l(5)} +\mathcal{W}^{l(6)}),
\end{equation} 
where $f_{E}=u(E+\sum_{k=1}^2 \alpha_k p_k)$ is the flux function associated 
to the mixture total energy $E$. The relation (\ref{eq-consE}) ensures
the fulfillment of the property (\ref{eq:mixencons1}), and it is necessary for  mixture-energy-consistency
(but not sufficient).
Once the Riemann solution structure 
$\{\mathcal{W}^l\iph,s\iph^l\}_{l=1,\ldots,\iM}$ 
arising at each cell edge $x\iph$ is
defined through a Riemann solver, the fluctuations $\mathcal{A}^{\mp} \Delta Q_{i+1/2}$ 
and the higher-order (second-order) correction fluxes $\tilhFp$
in (\ref{eq_wp_1d}) are computed as
\begin{equation}
   \apmdqp = \sum_{l=1}^\iM
             (s_{\iph}^l)^{\pm}\calW^l \iph\,,
\end{equation}      
where we have used the notation $s^+  = \max(s,0)$, $s^- = \min(s,0)$, and 
\begin{equation}
\tilhFp
= \frac{1}{2} \sum_{l=1}^\iM
      \bigl \vert s\iph^l \bigr \vert  
         \left( 1 - \frac{\Delta t}{\Delta x} \,
	    \bigl\vert s\iph^l \bigr\vert \right) 
	      {\mathcal{W}}^{l\,{\rm h}} \iph\,,
\end{equation}	      
where ${\mathcal{W}}^{l\,{\rm h}}\iph$ are a 
modified version of $\mathcal{W}^{l}\iph$
obtained by applying to $\mathcal{W}^{l}\iph$ a limiter function
(cf.~\cite{rjlbook}).  

\subsection{A simple HLLC-type solver}

\label{sec:HLLC}

In the wave  propagation scheme (\ref{eq_wp_1d}) we use  a simple  HLLC-type Riemann solver, which   we first presented in 
 \cite{pelanti-shyue}. Here we give more details on the derivation since
 the illustration of the derivation in \cite{pelanti-shyue} contained some imprecision
 (although the final formulas were correct).
 
 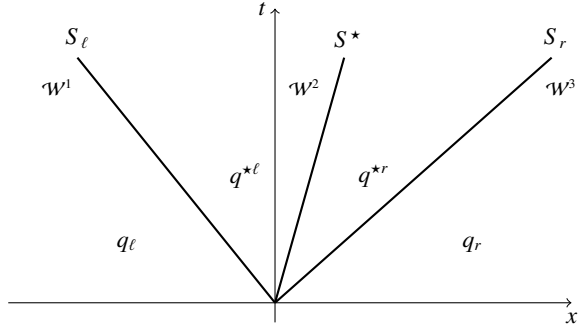
\begin{figure}[h!]
\centering

\begin{tikzpicture}[scale =1.3]


\draw[->] (5.3,0) -- (11,0) node[below]{\small $x$}; 
\draw[->] (8,-0.2) -- (8,3) node[left]{\small $t$};

 \draw[-,thick] (8,0) -- (6.,2.5) node[above]{\small $S_{\ell}$} ;
  
   \draw[-,thick] (8,0) -- (10.8,2.5) node[above]{\small $\,\,S_r$} ;
   \draw[-,thick] (8,0) -- (8.7,2.5) node[above]{\small $\,\,S^\star$} ; 
  
\node[black] at (6.5,0.6) {\small $q_{\ell}$}; 
\node[black] at (10.,0.6) {\small $q_{r}$}; 

\node[black] at (9.,1.3) {\small $q^{\star r}$};
\node[black] at (7.7,1.3) {\small $q^{\star \ell}$};

\node[black] at (5.8,2.2) {\scriptsize $\mathcal{W}^1$}; 
\node[black] at (8.3,2.2) {\scriptsize $\mathcal{W}^2$};
\node[black] at (10.9,2.2) {\scriptsize $\mathcal{W}^3$};

\end{tikzpicture}

\caption{Solution structure of the HLLC-type solver.}
\label{fig:riemannHLLC}
\end{figure}

The Riemann solution structure of the solver is similar to the classical HLLC solver for the
Euler equations \cite{toro-hllc,toro}, and it
consists of three waves $\mathcal{W}^l$, $l=1,2,3$, moving at speeds 
\begin{equation}
\label{eq:Hspeed}
s^1=S_{\ell}\,,\quad s^2=S^\star\,, \qeq s^3=S_r\,,
\end{equation}
which separate four constant states $q_{\ell}$, $q^{\star \ell}$, $q^{\star r}$ and $q_r$ 
(see Figure~\ref{fig:riemannHLLC}). 
In the following we will
indicate with $(\cdot)_{\ell}$ and $(\cdot)_{r}$ quantities corresponding to the states 
 $q_{\ell}$ and $q_r$, respectively. Moreover, we will indicate with 
$(\cdot)^{\star \ell}$ and $(\cdot)^{\star r}$ quantities corresponding to the states $q^{\star \ell}$ and  
$q^{\star r}$ adjacent, respectively on the left and on the right,  to the middle wave propagating at speed $S^\star$.
With this notation, the waves of the HLLC solver are
\begin{equation}
\label{eq:Hwaves}
\mathcal{W}^1=q^{\star \ell} - q_{\ell},\quad
\mathcal{W}^2=q^{\star r} - q^{\star \ell}, \qeq
\mathcal{W}^3=q_{r} - q^{\star r}.
\end{equation}
Invariance conditions for the normal velocity $u$ and for the effective pressure $p_{\rm m}=\alpha_1p_1+\alpha_2p_2$, 
which characterize the exact Riemann solution, are imposed across the middle wave: 
\begin{equation}
\label{invup}
u^{\star \ell}= u^{\star r}\equiv S^{\star} \qeq  p_{\rm m}^{\star \ell}= p_{\rm m}^{\star r}\equiv p^{\star}\,.
\end{equation}
Let us stress however  that the single terms $\alpha_k p_k$ in general vary across this middle wave.
The middle states $q^{\star \ell}$, $q^{\star r}$ are determined by imposing Rankine--Hugoniot 
conditions across the external waves moving at speeds $S_{\ell}$ and $S_r$,
 based on the equations for the partial densities $\alpha_k \rho_k$
 and the conservative portion of the equations governing the phasic momenta
 $\alpha_k \rho_k u$ and the phasic total energies $\alpha_k \rho_k E_k\,$, $\,k=1,2$.
 Note that the homogeneous equations governing the phasic momenta 
 $q^u_k =\alpha_k \rho_k u$, $k=1,2$, have the non-conservative form
 (here in one dimension):
  \begin{equation}
 \dts q^u_k + \dxs f^u_k
 +(-1)^{k-1}\noncep(q,\dxs q) =0\,, \quad k=1,2\,,
 \end{equation} 
 where $f^u_k = \alpha_k \rho_k u^2+\alpha_k p_k$ and  $\noncep(q,\dxs q)=Y_{1} \dxsb{\alpha_2 p_2}-Y_2 \dxsb{\alpha_1 p_1}$ =
   $Y_{1} \dxs p_{\rm m}-\dxsb{\alpha_1 p_1}$ = 
   $-Y_{2} \dxs p_{\rm m}+\dxsb{\alpha_2 p_2}$, as defined in (\ref{eq:nonce}).
Hence we impose:
\begin{subequations}
\label{cond-rh-hllc}
\begin{eqnarray}
\label{rheqr}
 &&f^{(\xi)}(q_{r})-f^{(\xi)}(q^{\star r})   = 
    S_r(q_{r}^{(\xi)}-q^{\star r(\xi)})\,,\quad \xi=2,3,5,6\,,\\    
    \label{rhmomr}
  && f^{u}_k(q_{r})-f^{u}_k(q^{\star r})  = 
    S_r(q_{k\,r}^u-q_k^{u\,\star r} )\,, \\
 [2mm] 
 \label{rheql}  
 &&f^{(\xi)}(q^{\star \ell})-f^{(\xi)}(q_{\ell})  =
    S_{\ell}(q^{\star \ell(\xi)}-q_{\ell}^{(\xi)})\,,\quad \xi=2,3,5,6\,,\\
    \label{rhmoml}
  &&  f^u_k(q^{\star \ell})-f^u_k(q_{\ell})  =
    S_{\ell}(q_k^{u \,\star\ell}-q_{k\,\ell}^u)\,,
\end{eqnarray}
\end{subequations}
where $q^{(\xi)}$ and  $f^{(\xi)}$ are the components of $q$ and  $f(q)$ in (\ref{feq})
in the $d=1$ case.
Let us remark that in writing the conditions above (\ref{cond-rh-hllc}) we 
neglect the contribution of the non-conservative 
terms $\Xi$ appearing in the phasic momentum and energy equations.
We observe however that  the conditions  (\ref{cond-rh-hllc})
imply correct Rankine--Hugoniot conditions
for the conservative equations for the mixture momentum $\rho u$ and the mixture total energy~$E$
(and hence (\ref{eq:mixencons1}) is satisfied).

The  Rankine--Hugoniot conditions  for the partial densities (\ref{rheqr})-(\ref{rheql})
($\xi=2,3$)  determine the
intermediate partial densities. Then  the  conditions for
the conservative portion of the phasic momentum equations (\ref{rhmomr})-(\ref{rhmoml})  determine the intermediate partial pressures:
\begin{equation}
\label{eq_alphap}
(\alpha_k p_k)^{\star \iota} = (\alpha_k p_k)_{\iota} +(\alpha_k p_k)_{\iota}
(S_\iota-u_\iota)(S^{\star}-u_\iota)\,,\quad \iota=\ell,r\,,
\end{equation}
where we have also used the invariance relation for the velocity in (\ref{invup}).
Then, using these expression in the invariance relations for the effective pressure in (\ref{invup})
we obtain  the expression for the speed $S^\star$:
\begin{equation}
\label{Sspeed}
S^\star = \frac{p_r-p_{\ell} +\rho_{\ell} u_{\ell}(S_{\ell}-u_{\ell}) -\rho_r u_r(S_r-u_r)}{
\rho_{\ell}(S_{\ell}-u_{\ell}) -\rho_r(S_r-u_r)}\,,
\end{equation}
where we have used $p_{{\rm m}\ell}=p_{\ell}$ and $p_{{\rm m}r} = p_r$ since
initial Riemann states are characterized by pressure equilibrium.
A definition for the wave speeds must be provided, see e.g.\ \cite{toro,hllc-speeds}. One classical 
 and  simple definition proposed by Davis~\cite{davis-hll} is
\begin{equation}
\label{eqw-davis}
S_{\ell} =  \min(u_{\ell} -c_{{\rm f}\ell}\,,\,u_r-c_{{\rm f}r}) 
\qqeqq S_{r} = \max(u_{\ell} +c_{{\rm f}\ell}\,, \,u_r+c_{{\rm f}r})\,,
\end{equation}
where $c_{\rm f}$ is defined in (\ref{eq:cfroz}).
Another more robust definition has been proposed for instance by Bouchut \cite{bouchut:book} 
(see  \cite{delorenzo-hllc}).
The Rankine--Hugoniot conditions for the conservative portion of the total energy equations 
(\ref{rheqr})-(\ref{rheql}) ($\xi =5,6)$ together with (\ref{eq_alphap}) determine then the 
intermediate states for the total energies.
Finally, we observe that the exact Riemann solution is characterized by the invariance
of the volume fraction across the external waves (the volume fraction is simply advected):
\begin{equation}
\alpha^{\star \ell}_k=\alpha_{k\ell}  \qqeqq    \alpha^{\star r}_k =  \alpha_{kr}\,,\quad k=1,2\,.
\end{equation}
Hence the solution structure for the  volume fractions $\alpha_k$ simply consists of 
single jumps $\alpha_{k,r}-\alpha_{k,\ell}$ across the 2-wave moving at speed $S^\star$.
As we reported in \cite{pelanti-shyue}, the expressions for the middle states are:
\begin{equation}
\label{eq:mstate}
q^{\star \iota} =\left(
\begin{matrix}
\alpha_{1,\iota}\\
(\alpha_1 \rho_1)_\iota\frac{S_{\iota}-u_{\iota}}{S_{\iota}-S^{\star}}\\
(\alpha_2 \rho_2)_\iota\frac{S_{\iota}-u_{\iota}}{S_{\iota}-S^{\star}}\\
\rho_{\iota}\frac{S_{\iota}-u_{\iota}}{S_{\iota}-S^{\star}}S^{\star}\\
(\alpha_1 \rho_1)_\iota\frac{S_{\iota}-u_{\iota}}{S_{\iota}-S^{\star}}
\left(\frac{E_{1,\iota}}{\rho_{1,\iota}}+(S^{\star} -u_{\iota})
\left(S^{\star} + \frac{p_{1,\iota}}{\rho_{1,\iota}(S_{\iota}-u_{\iota})}\right)\right)\\
(\alpha_2 \rho_2)_\iota\frac{S_{\iota}-u_{\iota}}{S_{\iota}-S^{\star}}
\left(\frac{E_{2,\iota}}{\rho_{2,\iota}}+(S^{\star} -u_{\iota})
\left(S^{\star} + \frac{p_{2,\iota}}{\rho_{2,\iota}(S_{\iota}-u_{\iota})}\right)\right)
\end{matrix}
\right), 
\end{equation}
$\iota = \ell,r$. Note that in the above formulas $p_{k,\iota} =   p_{\iota}$,
$k=1,2$, since initial Riemann states satisfy pressure equilibrium conditions.
As seen above the  Rankine--Hugoniot conditions are satisfied by construction
for all the physically  conserved  quantities across the external waves.
We now also observe that Rankine--Hugoniot conditions are satisfied
for the  conserved  quantities
 across the middle wave:
\begin{subequations}
\label{cond-rh-hllc-middle}
\begin{eqnarray}
\label{rheqm}
 &&f^{(\xi)}(q^{\star r}) - f^{(\xi)}(q^{\star \ell})  = 
    S^\star(q^{\star r(\xi)} - q^{\star \ell(\xi)})\,,\quad \xi=2,3,4\,,\\    
    \label{rhEm}
  && f^{E}(q^{\star r}) - f^{E}(q^{\star \ell})  = 
    S^\star(E^{\star r} - E^{\star \ell})\,,
\end{eqnarray}
\end{subequations}
where $f^E = (E+\alpha_1 p_1 + \alpha_2 p_2)u$ is the flux function associated to the mixture 
total energy  $E$.
Let us remark that instead Rankine--Hugoniot  conditions 
for the conservative portion
of the equations of the non-conserved quantities $\alpha_k \rho_k u$ and $\alpha_k E_k$ hold by construction across the external waves, but do not hold
in general across the middle wave.
This was inexactly reported in \cite{pelanti-shyue}, where we wrote incorrectly 
Rankine--Hugoniot conditions for the phasic energies
$\alpha_k E_k$ across the middle wave. 
As a final summarizing remark, we note that the simple HLLC-type solver illustrated here is obtained by neglecting the
non-conservative term $\Xi$ appearing in the phasic energy equations
and in the phasic momentum equations in the jump conditions for the external 1-wave and 3-wave of the Riemann solution, but not for the 2-wave. The solver construction implies indeed an approximation of the non-conservative
terms in the jump relations for the  middle wave which can be deduced by observing:
\begin{subequations}
\begin{eqnarray}
\label{rhekm}
  && f^{(4+k)}(q^{\star r}) - f^{(4+k)}(q^{\star \ell})  =
  ((\alpha_k E_k+\alpha_k p_k)^{\star r}
  - (\alpha_k E_k+\alpha_k p_k)^{\star \ell})S^\star\\
    &&=  S^\star(q^{\star r (4+k)} - q^{\star \ell (4+k)})+
  S^\star((\alpha_k p_k)^{\star r} - (\alpha_k p_k)^{\star \ell})\,, \quad k=1,2\,.
\end{eqnarray}
\end{subequations}
We find that the contribution to the jump across the 2-wave
representing the non-conservative term is approximated by this HLLC solver as
\begin{equation}
\label{eqnonc2w}
- S^\star((\alpha_k p_k)^{\star r}
 - (\alpha_k p_k)^{\star \ell}).
\end{equation}
This is a a reasonable approximation since across the 2-wave
$p_{\rm m}$ = constant, hence the non-conservative term
$\noncep$ in (\ref{eq:nonce}) reduces to $\noncep = -\dxs(\alpha_1 p_1)$
(and $-\noncep = \dxs(\alpha_1 p_1)=-\dxs(\alpha_2 p_2$)),
and (\ref{eqnonc2w}) can be then considered as a jump across the middle wave associated to 
the non-conservative terms   $-u\dxs(\alpha_k p_k)$, $k=1,2$.
Let us  finally remark that the simple HLLC-type that we have illustrated above
belongs to a more general class of HLLC-type Riemann solvers for the
six-equation two-phase flow model (\ref{eq:sysvect}), which we have 
introduced and assessed in \cite{delorenzo-hllc} by defining a Suliciu-type 
Riemann solver.

\section{Relaxation processes}

\label{sec:relax-step}

As indicated in Section~\ref{sec:numet}, after solving the homogeneous system (\ref{eq:homsys}), 
we solve a sequence of systems of ordinary differential equations
accounting for the relaxation source terms, namely the systems (\ref{eq:oderelp}), (\ref{eq:oderelT}), 
and~(\ref{eq:oderelg}).
 First of all, we observe that for any relaxation process we have
\begin{subequations}
\begin{eqnarray}
&&\dts \,\rho = 0\,,\\
&& \dts \,(\rho \vec{u}) = 0\,,\\
&&\dts\, E =0\,.
\end{eqnarray}  
\end{subequations}
Therefore, the mixture density, velocity, total energy and internal energy remain 
constant during the transfer processes:
\begin{equation}
\label{eq:inv_relax1}
\rho = {\rm const.}\,, \quad \vec{u} = {\rm const.}\,, \quad
E = {\rm const.}\,, \quad \Ei = {\rm const.}\,.
\end{equation}
Moreover, if chemical relaxation is not activated, also the partial densities
remain constant, since $\dts (\alpha_k \rho_k) =0$, $k=1,2$:
\begin{equation}
\label{eq:inv_relax2}
\alpha_k \rho_k = {\rm const.}\,.
\end{equation}
To completely  determine the relaxed states in the  mechanical and thermal relaxation steps 
we need to determine two independent variables (here we choose as unknowns the volume
fraction $\alpha_1$ and the equilibrium pressure $p$).  In the chemical relaxation step
we have to determine instead three variables, since the partial densities vary.
It is important to note that for consistency with the mixture equation
of state for flows in mechanical equilibrium (\ref{eq:mixpresimpl}) the equilibrium pressure $p$ 
determined in the all the relaxation steps should satisfy the energy relation (\ref{eq:mixencons2}).

\subsection{Instantaneous relaxation processes}

\label{sec:inst_rel}

Before illustrating our new relaxation procedures, let us recall briefly 
the methods presented in our previous work \cite{pelanti-shyue}
for instantaneous transfer processes.
The idea is to use for each process the   invariance relations
(\ref{eq:inv_relax1}),  with also (\ref{eq:inv_relax2}) for mechanical and thermal relaxation,
and  the corresponding equilibrium conditions
to obtain an algebraic system for the unknown relaxed variables.
Similar relaxation procedures for instantaneous processes can be also found for instance 
in \cite{met-mas-saur-proc,lemart-boil,saurelCF16}. We summarize here the equations to be used for each step
of the algorithm in Section~\ref{sec:numet}:

\begin{itemize}
\item[2(a)] Instantaneous mechanical relaxation. We use the invariance relations (\ref{eq:inv_relax1}) and 
(\ref{eq:inv_relax2})
plus the mechanical equilibrium condition $p_1=p_2=p$. In this step we also need to integrate 
the phasic energy equations $\dts \mathcal{E}_k=(-1)^k \pint \dts \alpha_1$, $k=1,2$ between the states
$0$ and $*$. To simplify the integration  we  make an assumption
on the interface pressure $\pint$, which we define as a convex combination
of the initial value $p^0$ and the equilibrium value $p^*$,
$\pint = \beta_p p^0+(1-\beta_p)p^*$, $\beta_p\in [0,1]$. 
In our previous work we set $\beta_p=\frac{1}{2}$ \cite{pelanti-shyue} or $\beta_p = 0$ \cite{pelanti-amc}.
We obtain an algebraic system of equations to be solved for two variables, for instance
$\alpha_1^*$ and $p^*$. In the case of the stiffened gas EOS, the system can be reduced to the
solution of a quadratic equation for the equilibrium pressure $p^{*}$.
\item[2(b)] Instantaneous mechanical and thermal relaxation. We use the invariance relations (\ref{eq:inv_relax1}) 
and (\ref{eq:inv_relax2}) plus the mechanical equilibrium condition $p_1=p_2=p$ and the thermal equilibrium
condition $T_1=T_2=T$. We obtain an algebraic system of equations to be solved for two variables, for instance
$\alpha_1^{**}$ and $p^{**}$. For the stiffened gas EOS, the system can be reduced to the
solution of a quadratic equation for $p^{**}$.
\item[2(c)] Instantaneous mechanical, thermal and chemical relaxation. We use the invariance relations 
(\ref{eq:inv_relax1}) plus the mechanical equilibrium condition $p_1=p_2=p$, the thermal equilibrium
condition $T_1=T_2=T$, and the chemical equilibrium condition $g_1=g_2$.
We obtain an algebraic system to be solved for three variables, for instance
$\alpha_1^{\otimes}$, $p^{\otimes}$, and $T^{\otimes}$. In general,
an iterative method is necessary for the solution.
\end{itemize} 
Let us remark that in all these relaxation procedures the energy relation (\ref{eq:mixencons2}) is ensured by construction,
 and the resulting algorithm is mixture-energy-consistent (we recall that (\ref{eq:mixencons1}) is
 guaranteed by the HLLC method).

\subsection{Arbitrary-rate  relaxation processes}

\label{sec:arbraterel}

As we have explained, we will always consider instantaneous mechanical 
relaxation processes, so we could adopt the mechanical relaxation procedure 
used in \cite{pelanti-shyue} and described above. Nonetheless, we will present 
below a new procedure for pressure relaxation based on an analytical semi-exact exponential solution, 
which is particularly advantageous for complex equations of state.
It  could be also potentially used for finite-rate pressure relaxation processes (e.g.\ \cite{pandare-et-al}),
which however are not of interest here.

Concerning heat and mass transfer, we wish to model here 
processes with arbitrary relaxation times, hence we wish to design  algorithms capable
of handling  both instantaneous (stiff) processes and slow finite-rate ones.
Let us consider the systems of ODEs in (\ref{eq:oderelT}) and (\ref{eq:oderelg}).
We see that during thermal and chemical relaxation we need to
guarantee pressure equilibrium, as represented by the presence in these systems of the 
pressure relaxation term $\psi_{\mu}$ with $\mu\rightarrow+\infty$.
In addition, during chemical relaxation, we need to account for 
the thermal relaxation effect, and in the limit of instantaneous thermal
relaxation ($\vartheta \rightarrow +\infty$), we need to guarantee temperature equilibrium,
 as represented by the presence of
the thermal source term $\psi_{\vartheta}$ in (\ref{eq:oderelg}).
Due to these constraints, for thermal and chemical relaxation it is not possible to 
use a simple fractional
step method where each source term $\psi_{\vartheta}$ and $\psi_{\nu}$
in the six-equation model is integrated individually. On the other hand it appears very complicated to try
to solve the ODEs with all the relaxation terms, which  have very different characteristic
time scales. Our idea consists in  modifying the thermal and chemical relaxation 
terms $\psi_{\vartheta}$ and $\psi_{\nu}$ of the six-equation model to translate on 
them
the effect of the instantaneous
pressure relaxation term $\psi_{\mu}\,$, and also to translate  the effect 
of  the thermal relaxation term $\psi_{\vartheta}$ on $\psi_{\nu}\,$. Hence, we replace  the system in (\ref{eq:oderelT}) with a new system
$\dts q  = \tilde{\psi}_{\vartheta}\,$, where $\tilde{\psi}_{\vartheta}$ models thermal relaxation
under pressure equilibrium, and we replace the system in (\ref{eq:oderelg})  with a new system
$\dts q  = \tilde{\psi}_{\nu}\,$, where $\tilde{\psi}_{\nu}$
 models chemical relaxation under pressure equilibrium and a thermal constraint, in particular thermal
 equilibrium. Our technique consists in employing in this approach the equations of the
 $p$-relaxed (\ref{eq:sisp})  and $pT$-relaxed (\ref{eq:sysT}) two-phase models. 
By using then some simplifying assumptions, analytical semi-exact exponential solutions are obtained
to describe the relaxation processes.

\subsubsection{Mechanical relaxation}

We propose here a new numerical procedure to model instantaneous mechanical relaxation.
We start by writing the ordinary differential equations governing the relaxation process in terms
of the volume fraction $\alpha_1$  and the phasic pressures $p_1$ and $p_2$. 
Based on 
(\ref{eq:2phase_alpha}) and (\ref{eq:2phase_pres}) we have the equations:
\begin{subequations}
\label{eq:prels}
\begin{eqnarray}
\label{eq:prelalapha}
&&\dts \alpha_1  =\mu (p_1-p_2),\\
[2mm]
&&\dts p_1 = \mu \frac{\Gamma_1}{\alpha_1}\left[\pint -\rho_1^2
\left(\frac{\partial \varepsilon_1}{\partial \rho_1}\right)_{p_1}  \right](p_2-p_1),\\
[2mm]
&&\dts p_2 = -\mu \frac{\Gamma_2}{\alpha_2}\left[\pint -\rho_2^2
\left(\frac{\partial \varepsilon_2}{\partial \rho_2}\right)_{p_2}  \right](p_2-p_1),
\end{eqnarray}
\end{subequations}
where $\mu$ is considered a constant.
The initial condition for the above system corresponds to the solution  of the homogeneous system,
denoted with superscript $0$.  
Let us now introduce the quantities $\xi_k^p$, whose inverse values correspond to
the terms multiplying
$(p_2-p_1)$ in the last two equations of the system (\ref{eq:prels}):
\begin{equation}
\frac{1}{\xi_k^p} = \frac{\Gamma_k}{\alpha_k}\left[\pint -\rho_k^2
\left(\frac{\partial \varepsilon_k}{\partial \rho_k}\right)_{p_k}  \right]=
\frac{\Gamma_k}{\alpha_k}\left[\pint -p_k+ \frac{\rho_k c_k^2}{\Gamma_k}  \right]=
\frac{\rho_k}{\alpha_k} \cint{pk}^2\,,\quad, k=1,2\,.
\end{equation}
Here we have used the relation 
$\rho_k^2 \left(\frac{\partial \varepsilon_k}{\partial \rho_k}\right)_{p_k}
=p_k-\frac{c_k^2 \rho_k}{\Gamma_k}$, and we have defined
\begin{equation}
\cint{pk}^2 = \Gamma_k \frac{\Ei_k +\pint}{\rho_k}+ \chi_k\,,\quad k=1,2\,.
\end{equation}
We now introduce an approximation by assuming $\xi_k^p$  constant in time,
 $\xi_k^p = \xi_k^{p0}$. Note that this means that we consider constant the
 impedance $\rho_k \cint{pk}$ (since $\alpha_k\rho_k$ is constant). Hence we consider the solution of 
\begin{subequations}
\label{eq:prelax}
\begin{eqnarray}
&&\dts p_1 = \mu \frac{1}{\xi_1^{p0}}(p_2-p_1),\\
[2mm]
&&\dts p_2 = -\mu \frac{1}{\xi_2^{p0}}(p_2-p_1).
\end{eqnarray}
\end{subequations}
From (\ref{eq:prelax}) we obtain the following ordinary equation for the
pressure difference $\Delta p=p_2-p_1$:
\begin{equation}
\label{eq:eqrelaxp}
\dts \Delta p = -\mu\left(\frac{1}{\xi_1^{p0}}+\frac{1}{\xi_2^{p0}}\right) \Delta p,
\end{equation}
which has the exact solution after a time interval $\Delta t$:
\begin{equation}
\label{eq:solrelaxp}
 \Delta p^* = \Delta p^0 {\rm e}^{-K_p \,\Delta t},
\end{equation}
where 
\begin{equation}
K_p = \mu \left(\frac{1}{\xi_1^{p0}}+\frac{1}{\xi_2^{p0}}\right).
\end{equation}
We can now solve the partial differential equation for the volume fraction in (\ref{eq:prelalapha})
by using  the solution for $\Delta p^*$ in (\ref{eq:solrelaxp}). By integrating we find
\begin{equation}
\alpha_1^* = \alpha_{1}^0  -\frac{\Delta p^0}{\left(\frac{1}{\xi_1^{p0}}+\frac{1}{\xi_2^{p0}}\right)}
(1-{\rm e} ^{-K_p\,\Delta t}).
\end{equation} 
In the limit of instantaneous pressure relaxation 
$\mu \rightarrow +\infty$ the above expression for $\alpha_1$ 
gives the equilibrium value
\begin{equation}
\label{eq:alpha1eqp}
\alpha_{1,\,\mu\rightarrow \infty}^* = \alpha_1^0 - \frac{\Delta p^0}{\left(\frac{1}{\xi_1^{p0}}+\frac{1}{\xi_2^{p0}}\right)},
\end{equation}
and  the limit equilibrium pressure is:
\begin{equation}
\label{eq:peqp}
p^{*}_{\mu\rightarrow \infty} =  \frac{\xi_1^{p0} p_1^0 + \xi_2^{p0} p_{2}^0}{\xi_1^{p0}+\xi_2^{p0}}.
\end{equation}
We might then use the two equilibrium quantities (\ref{eq:alpha1eqp}) and (\ref{eq:peqp}) 
to define the $p$-relaxed solution in the step (2a)
of the algorithm. However, in contrast to the techniques for instantaneous relaxation
described in Section~\ref{sec:inst_rel}, in general the pair $\alpha^{*}_{1,\,\mu\rightarrow \infty}$  and $p^{*}_{\mu\rightarrow \infty}$  
does not satisfy by construction the relation (\ref{eq:mixencons2}) (with $\#=*$), due to the approximations
in the ODEs solution. Hence here we
define the updated  volume fraction $\alpha_1^*$ by using
the exponential solution (\ref{eq:alpha1eqp}), but we update the equilibrium pressure by using the
value $p^*$ determined by the energy relation
\begin{equation}
\label{eq:mixencons2p}
\Ei^{0} = \alpha_1^{*}\Ei_1\left(p^{*},\frac{(\alpha_1\rho_1)^0}{\alpha_1^*}\right)+
\alpha_2^{*}\Ei_2\left(p^*,\frac{(\alpha_2\rho_2)^0}{\alpha_2^*}\right)\,.
\end{equation}
The pressure relaxation procedure presented here is particularly convenient when complex equations
of state are used. In fact the procedure used in \cite{pelanti-shyue} and 
recalled in Section~\ref{sec:inst_rel} might lead to a complex
algebraic system needing an iterative method for its solution. 
In  Appendix~\ref{sec:pressinv}  we also show that the relaxation procedure proposed here allows us to ensure velocity 
and pressure invariance at material interfaces at least when the stiffened gas equation of state
is used.   
Let us finally observe that a similar mechanical relaxation procedure was proposed in \cite{delor-laf-pelanti-JCP19}.
However, in \cite{delor-laf-pelanti-JCP19} the exponential solution for $\alpha_1$ is assumed \textit{a priori},
whereas here it is deduced from the model equations by assuming 
the quantities $\xi_k^p$  constant in time during the relaxation process.
Moreover in \cite{delor-laf-pelanti-JCP19} the instantaneous equilibrium case was 
modeled by an exponential solution with very small relaxation time, whereas here
we use the analytical limit (\ref{eq:alpha1eqp}).

\subsubsection{Thermal relaxation}

To describe the thermal relaxation process under the constraint of mechanical equilibrium
$p_1=p_2$ we consider here the ordinary differential equations
with heat transfer source term corresponding to the reduced five-equation  pressure equilibrium model (\ref{eq:sisp}).
Specifically, we write the ODEs governing the thermal relaxation process for this model
in terms of  the volume fraction $\alpha_1$, 
and the phasic temperatures $T_1$ and $T_2$. Based on (\ref{eq:sisp_alpha}) and 
(\ref{eq:sisp_temp}) we have the equations:
\begin{subequations}
\label{eq:Trelsys}
\begin{eqnarray}
\label{eq:Trelalapha}
&&\dts \alpha_1  =\vartheta \frac{Z}{D}(T_2-T_1),\\
[2mm]
&&\dts T_1 = \vartheta \frac{1}{\phider_1 D}
\left[-\frac{\rho_1}{\alpha_1}Z 
-\zetader_1 (\Gamma_1\rho_2c_2^2-\Gamma_2\rho_1c_1^2)\right](T_2-T_1),\\
[2mm]
&&\dts T_2 = \vartheta\frac{1}{\phider_2 D}
\left[\frac{\rho_2}{\alpha_2} Z
-\zetader_2 (\Gamma_1\rho_2c_2^2-\Gamma_2\rho_1c_1^2)\right] (T_2-T_1).
\end{eqnarray}
\end{subequations}
where $D$ is given in (\ref{eq:Dpar}) and we have defined:
\begin{equation}
Z = \alpha_2\Gamma_1+\alpha_1\Gamma_2\,.
\end{equation}
The initial condition here corresponds to the pressure equilibrium solution coming from the mechanical relaxation step,
denoted with the  superscript $*$.  
Let us introduce the quantities $\xi_k^T$, whose inverse values correspond to
the terms multiplying
$(T_2-T_1)$ in the last two equations of the above system:
\begin{equation}
\xi_k^T = -\frac{\phider_k D}{\frac{\rho_k}{\alpha_k}Z
 +(-1)^{k-1}\zetader_k(\Gamma_1\rho_2c_2^2-\Gamma_2\rho_1c_1^2)} \,,\quad k=1,2\,.
\end{equation}
Analogously to the pressure relaxation step we assume that the quantities $\xi_k^T$ are constant in time,
 $\xi_k^T = \xi_k^{T*}$. Hence we consider the solution of 
\begin{subequations}
\label{eq:Trelax}
\begin{eqnarray}
&&\dts T_1 = \vartheta \frac{1}{\xi_1^{T*}}(T_2-T_1),\\
[2mm]
&&\dts T_2 = -\vartheta \frac{1}{\xi_2^{T*}}(T_2-T_1).
\end{eqnarray}
\end{subequations}
We also consider that $\vartheta$ is a constant. If $\vartheta$ is a given function of some variables
instead of a constant parameter, then we fix $\vartheta$ to its value at the state $*$.
From (\ref{eq:Trelax}) we obtain the following ordinary equation for the
temperature difference $\Delta T=T_2-T_1$:
\begin{equation}
\label{eq:eqrelaxp}
\dts \Delta T = -\vartheta\left(\frac{1}{\xi_1^{T*}}+\frac{1}{\xi_2^{T*}}\right) \Delta T,
\end{equation}
which has the following exact solution after a time step $\Delta t$:
\begin{equation}
\label{eq:solrelaxT}
 \Delta T^{**} = \Delta T^* {\rm e}^{-K_T \,\Delta t},
\end{equation}
where 
\begin{equation}
K_T = \vartheta \left(\frac{1}{\xi_1^{T*}}+\frac{1}{\xi_2^{T*}}\right).
\end{equation}
We now need to solve the partial differential equation for the volume fraction in (\ref{eq:Trelalapha}).
Assuming that $Z/D$ is constant and 
by using  the solution for $\Delta T^{**}$ in (\ref{eq:solrelaxT})  we obtain
\begin{equation}
\label{eq:alpha1teqT}
\alpha_1^{**} = \alpha_{1}^*  +{\textstyle \left(\frac{Z}{D}\right)^*}
\frac{\Delta T^*}{\left(\frac{1}{\xi_1^{T*}}+\frac{1}{\xi_2^{T*}}\right)}
(1-{\rm e} ^{-K_T\,\Delta t}).
\end{equation} 
In the limit of instantaneous temperature relaxation 
$\vartheta \rightarrow +\infty$ the above expression for $\alpha_1$ 
gives the equilibrium value
\begin{equation}
\label{eq:alpha1eqT}
\alpha_{1,\, \vartheta \rightarrow \infty} = \alpha_1^* + {\textstyle\left(\frac{Z}{D}\right)^*}
\frac{\Delta T^*}{\left(\frac{1}{\xi_1^{T*}}+\frac{1}{\xi_2^{T*}}\right)},
\end{equation}
and  the equilibrium temperature is:
\begin{equation}
T_{\vartheta \rightarrow \infty} =  \frac{\xi_1^{T*} T_1^* + \xi_2^{T*} T_{2}^*}{\xi_1^{T*}+\xi_2^{T*}}.
\end{equation}
To update the solution at the step (2b) of the algorithm we use the relaxed value
$\alpha_1^{**}$ in (\ref{eq:alpha1teqT}) (or in (\ref{eq:alpha1eqT}) in case of instantaneous relaxation), and the
pressure value $p^{**}$ determined by the energy relation
\begin{equation}
\label{eq:mixencons2t}
\Ei^{0} = \alpha_1^{**}\Ei_1\left(p^{**},\frac{(\alpha_1\rho_1)^0}{\alpha_1^{**}}\right)+
\alpha_2^{**}\Ei_2\left(p^{**},\frac{(\alpha_2\rho_2)^0}{\alpha_2^{**}}\right)
\end{equation}
in order to ensure 
mixture-energy-consistency.

\subsubsection{Chemical relaxation}

\label{sec:chemrelax}

The chemical relaxation process occurs under mechanical equilibrium and it is coupled to
a thermal relaxation process which is assumed much faster than the chemical relaxation one.
Hence, it is often reasonable to consider a chemical relaxation process occurring under
both mechanical and thermal equilibrium, and here we  consider this case. 
To describe this process we use then the  ordinary differential equations with mass transfer source term corresponding to the reduced four-equation pressure and temperature equilibrium 
model~(\ref{eq:sysT}).
Specifically, we wish to write the ODEs governing the chemical relaxation process for this model in terms of the  volume fraction $\alpha_1$, 
the partial density $\alpha_1 \rho_1$, and  the phasic chemical potentials $g_1$ and $g_2$.  First, let us note that the ordinary differential equations
for  $\alpha_1$ and $\alpha_1\rho_1$ are given by $\dts \alpha_1  = \mathcal{S}_{\alpha} \mM$ 
and $\dts (\alpha_1 \rho_1)= \mM$, respectively, based on (\ref{eq:sourcep2}) and (\ref{eq:sysT}).
We can then write the ordinary differential equations governing the phasic 
chemical potentials $g_k$, by writing $\dts g_k = \frac{1}{\rho_k} \dts p_k -s_k \dts T_k$
and by using the equations $\dts T = \mathcal{S}_T\mM$ and $\dts p =\mathcal{S}_p$ obtained from (\ref{eq:sourcep2}).
We obtain  the system:
\begin{subequations}
\label{eq:grelsysf}
\begin{eqnarray}
\label{eq:grelalapha}
&&\dts \alpha_1  =\nu \mathcal{S}_{\alpha}(g_2-g_1),\\
[2mm]
&&\dts g_1 = \nu\left(\frac{\mathcal{S}_p}{\rho_1}  -s_1 \mathcal{S}_T\right) (g_2-g_1),\\
[2mm]
&&\dts g_2 = \nu\left(\frac{\mathcal{S}_p}{\rho_2}  -s_2 \mathcal{S}_T\right) (g_2-g_1),\\
[2mm]
\label{eq:grelalprho}
&& \dts (\alpha_1 \rho_1) = \nu(g_2-g_1). 
\end{eqnarray}
\end{subequations}
 Note that the
interface chemical potential $\gint$ does not appear anymore in the equations 
(cf.\ also the $pT$-relaxed model), hence a definition for this quantity is not needed.
The initial condition for (\ref{eq:grelsysf}) corresponds to the solution coming 
from the thermal relaxation step, denoted  with superscript~$**$.  
Let us introduce the quantities $\xi_k^g$, whose inverse values correspond to
the terms multiplying
$(g_2-g_1)$ in the two equations for $g_k$ of the above system:
\begin{equation}
\frac{1}{\xi_k^g} = \pm\frac{1}{\rho_k} \mathcal{S}_p -s_k \mathcal{S}_T\, ,\quad k=1,2\,.
\end{equation}
Analogously to the pressure and temperature relaxation step we assume that the 
quantities $\xi_k^g$ are constant in time,
 $\xi_k^g = \xi_k^{g**}$. Hence we consider the solution of 
\begin{subequations}
\label{eq:grelax3}
\begin{eqnarray}
&&\dts g_1 = \nu \frac{1}{\xi_1^{g**}}(g_2-g_1),\\
[2mm]
&&\dts g_2 = -\nu \frac{1}{\xi_2^{g**}}(g_2-g_1).
\end{eqnarray}
\end{subequations}
We also consider that $\nu$ is a constant. If $\nu$ is a given function of some variables
instead of a constant parameter, then we fix $\nu$ to its value at the state $**$.
From (\ref{eq:grelax3}) we obtain the following ordinary equation for the
chemical potential  difference $\Delta g=g_2-g_1$:
\begin{equation}
\label{eq:eqrelaxp}
\dts \Delta g = -\nu\left(\frac{1}{\xi_1^{g**}}+\frac{1}{\xi_2^{g**}}\right) \Delta g,
\end{equation}
which has the exact solution after a time interval $\Delta t$:
\begin{equation}
\label{eq:solrelaxg}
 \Delta g^{\otimes} = \Delta g^{**} {\rm e}^{-K_g \,\Delta t},
\end{equation}
where 
\begin{equation}
K_g = \nu \left(\frac{1}{\xi_1^{g**}}+\frac{1}{\xi_2^{g**}}\right).
\end{equation}
We now need to solve the partial differential equation for the volume fraction in (\ref{eq:grelalapha})
and the partial density $\alpha_1\rho_1$ in (\ref{eq:grelalprho}).
By using  the solution for $\Delta g^\otimes$ in (\ref{eq:solrelaxg})  we immediately obtain
\begin{equation}
\label{eq:alpha1rho1relg}
(\alpha_1\rho_1)^{\otimes} = (\alpha_1\rho_1)^{**} +  
\frac{\Delta g^{**}}{\left(\frac{1}{\xi_1^{g**}}+\frac{1}{\xi_2^{g**}}\right)}
(1-{\rm e} ^{-K_g\,\Delta t}).
\end{equation}
Assuming now that $\mathcal{S}_{\alpha}$ is constant, for the volume fraction we get
\begin{equation}
\label{eq:alpha1relg}
\alpha_1^{\otimes} = \alpha_{1}^{**}  +\mathcal{S}_{\alpha}^{**}
\frac{\Delta g^{**}}{\left(\frac{1}{\xi_1^{g**}}+\frac{1}{\xi_2^{g**}}\right)}
(1-{\rm e} ^{-K_g\,\Delta t}).
\end{equation}  
In the limit of instantaneous chemical relaxation 
$\nu \rightarrow +\infty$ the above expressions for $\alpha_1 \rho_1$
and $\alpha_1$  
give the equilibrium values
\begin{eqnarray}
&&
\label{eqalpharhorelg1}
(\alpha_{1}\rho_1)_{\rm \nu \rightarrow \infty} = (\alpha_1\rho_1)^{**} + 
\frac{\Delta g^*}{\left(\frac{1}{\xi_1^{g**}}+\frac{1}{\xi_2^{g**}}\right)},\\
[2mm]
&&
\label{eqalpharhorelg2}
\alpha_{1,\,\nu \rightarrow \infty} = \alpha_1^* + \mathcal{S}_{\alpha}^{**}
\frac{\Delta g^*}{\left(\frac{1}{\xi_1^{g**}}+\frac{1}{\xi_2^{g**}}\right)},
\end{eqnarray}
and the equilibrium chemical potential is:
\begin{equation}
g_{\nu \rightarrow \infty} =  \frac{\xi_1^{g**} g_1^{**} + \xi_2^{g**} g_{2}^{**}}{\xi_1^{g**}+\xi_2^{g**}}.
\end{equation}
To update the solution at the step (2c) of the algorithm we use the relaxed values 
 $(\alpha_1 \rho_1)^\otimes$ and
$\alpha_1^\otimes$ in (\ref{eq:alpha1rho1relg}) and (\ref{eq:alpha1relg}) (or in (\ref{eqalpharhorelg1})
and (\ref{eqalpharhorelg2}) in
case of instantaneous relaxation), and the pressure value $p^\otimes$  is determined by the
energy relation
\begin{equation}
\label{eq:mixencons2t}
\Ei^{0} = \alpha_1^{\otimes}\Ei_1\left(p^{\otimes},\frac{(\alpha_1\rho_1)^\otimes}{\alpha_1^{\otimes}}\right)+
\alpha_2^{\otimes}\Ei_2\left(p^{\otimes},\frac{(\alpha_2\rho_2)^\otimes}{\alpha_2^{\otimes}}\right)
\end{equation}
in order to ensure 
mixture-energy-consistency.
Let us remark that although the equations (\ref{eq:grelsysf}) above have been obtained by
assuming  temperature equilibrium $T_2=T_1=T$ (in addition to pressure equilibrium), 
they hold  more generally in the hypothesis of a chemical relaxation process
occurring  at  constant temperature difference $T_2-T_1=\Delta T$,
since in this case $\partial T_1=\partial T_2$ and  the derivation 
of (\ref{eq:sourcep2}) in Appendix~\ref{sec:source4eq} can be easily extended to the
case in which the  variables associated to the phase $k$ are functions of $p$ and
$T_k$ with the constraint $T_2=T_1+\Delta T$, $\Delta T$ = constant.
Hence the  procedure described in this Section could be also employed 
to model chemical relaxation processes
with thermal disequilibrium, provided consistent relaxation times are chosen, since thermal relaxation 
is physically faster than chemical relaxation.

\subsubsection*{Complete evaporation or condensation}

It is physically possible that the mass transfer process leads to complete evaporation 
or complete condensation, thus to the disappearance of one phase. In such a case
the computation via (\ref{eq:alpha1relg}) (or (\ref{eqalpharhorelg2})) produces a value 
$\alpha_1^\otimes \notin (0,1)$. To handle this case numerically 
we proceed as follows. First, let us remark that in our numerical model
both phases must always be present, therefore we model a pure phase $k$ as
a mixture with a volume fraction $\alpha_k=1-\epsilon_{\alpha}$, where
$0<\epsilon_{\alpha}\ll 1$ (for instance $\epsilon_{\alpha}=10^{-8}$).
If in the numerical computation we find  $\alpha_1^\otimes \notin (0,1)$ we consider that one phase
vanishes hence we set $\alpha_1^\otimes = \bar{\alpha}_1$ where
$\bar{\alpha}_1=\epsilon_{\alpha}$ if $s_2^{**}>s_1^{**}$ (vanishing of phase 1),
or  $\bar{\alpha}_1=1-\epsilon_{\alpha}$ if $s_1^{**}>s_2^{**}$ (vanishing of phase 2).
Then we determine the value of the relaxation parameter $K_g$
that gives the value $\bar{\alpha}_1$ through the formula (\ref{eq:alpha1relg}).
We find:
\begin{equation}
\label{eq:chemrelaxpure}
\bar{K}g = -\frac{1}{\Delta t}
\log \left( 1-\frac{\bar{\alpha}_1 - \alpha_1^{**}}{\mathcal{S}_{\alpha}^{**}
\frac{\Delta g^{**}}{\left(\frac{1}{\xi_1^{g**}}+\frac{1}{\xi_2^{g**}}\right)}} \right)\,.
\end{equation}
Finally we set $(\alpha_1\rho_1)^\otimes$ by using the formula (\ref{eq:alpha1rho1relg})
 with $K_g = \bar{K}_g$.
Let us remark that if an unphysical value of $\alpha_1$ is computed in the 
pressure and temperature relaxation procedures previously presented, it  suffices
to set $\alpha_1 = \bar{\alpha}_1$.

\section{Numerical experiments}

\label{sec:numexp}

We present in this section several numerical experiments in one and two dimensions. 
The algorithm has been implemented by using the basic routines of the \textsc{clawpack} software
\cite{rjl:clawpack12}. All the computations have been performed with the second-order wave propagation 
scheme  with the minmod limiter. In some one-dimensional Riemann problems we plot the exact solution
for the $p$-relaxed model (\ref{eq:sisp}) (see \cite{petit-fslm}) and the $pT$-relaxed model (\ref{eq:sysT}).
These exact solutions have been computed by extending to the models (\ref{eq:sisp}) and (\ref{eq:sysT})
the methodology detailed in \cite{kamm,EPkamm}
for the compressible single-phase Euler equations with general equation of state.

\subsection{Numerical  tests with only mechanical relaxation}

We begin by presenting some numerical experiments where we activate only
instantaneous mechanical relaxation. The aim in particular is to show the good performance  of the numerical model when
complex equations of state are employed in problems with strong shocks and interfaces.

\subsubsection{Detonation gas-water shock tube}

\begin{figure}[h!]
\centering 
\includegraphics[height=5.4cm]{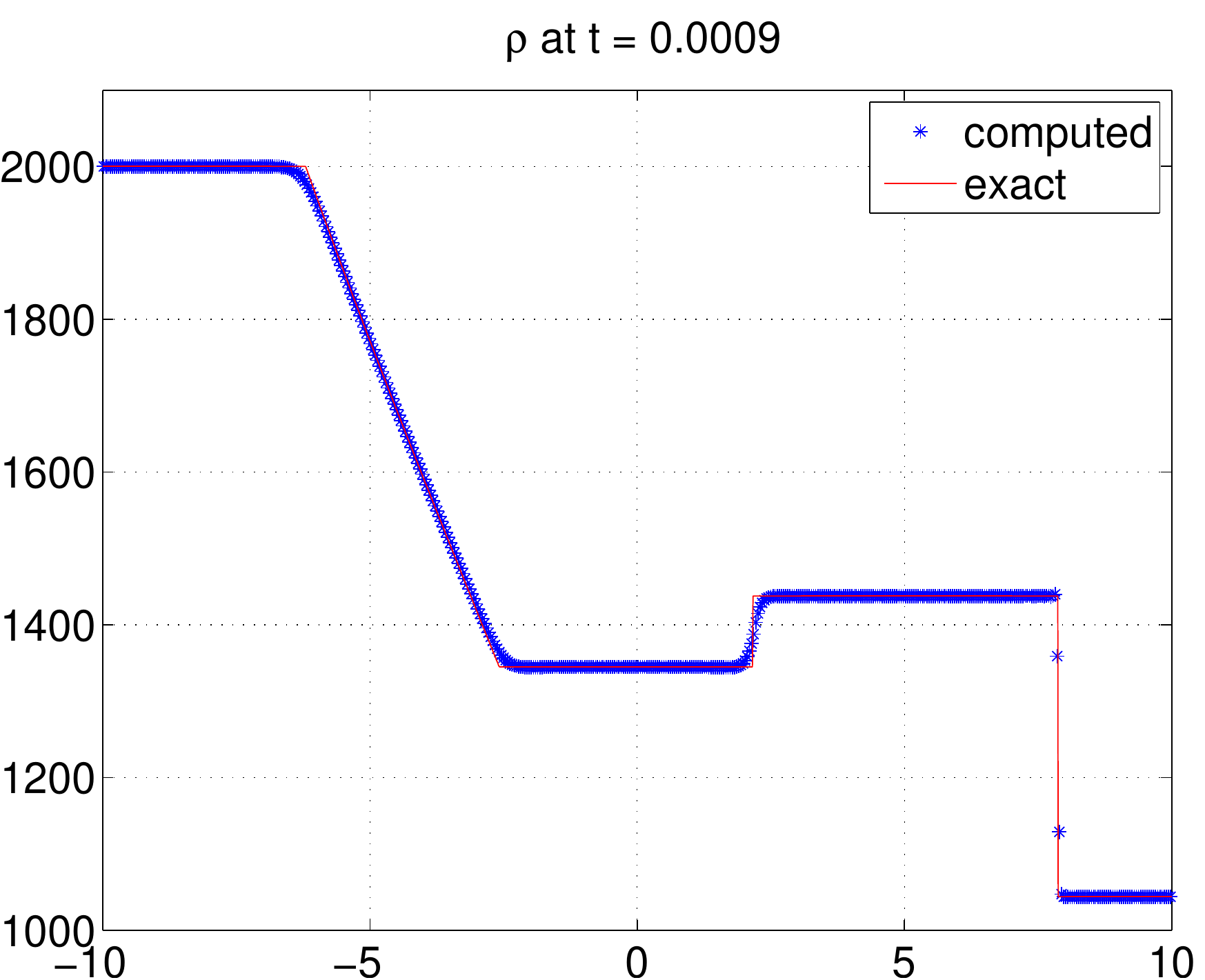}
\includegraphics[height=5.4cm]{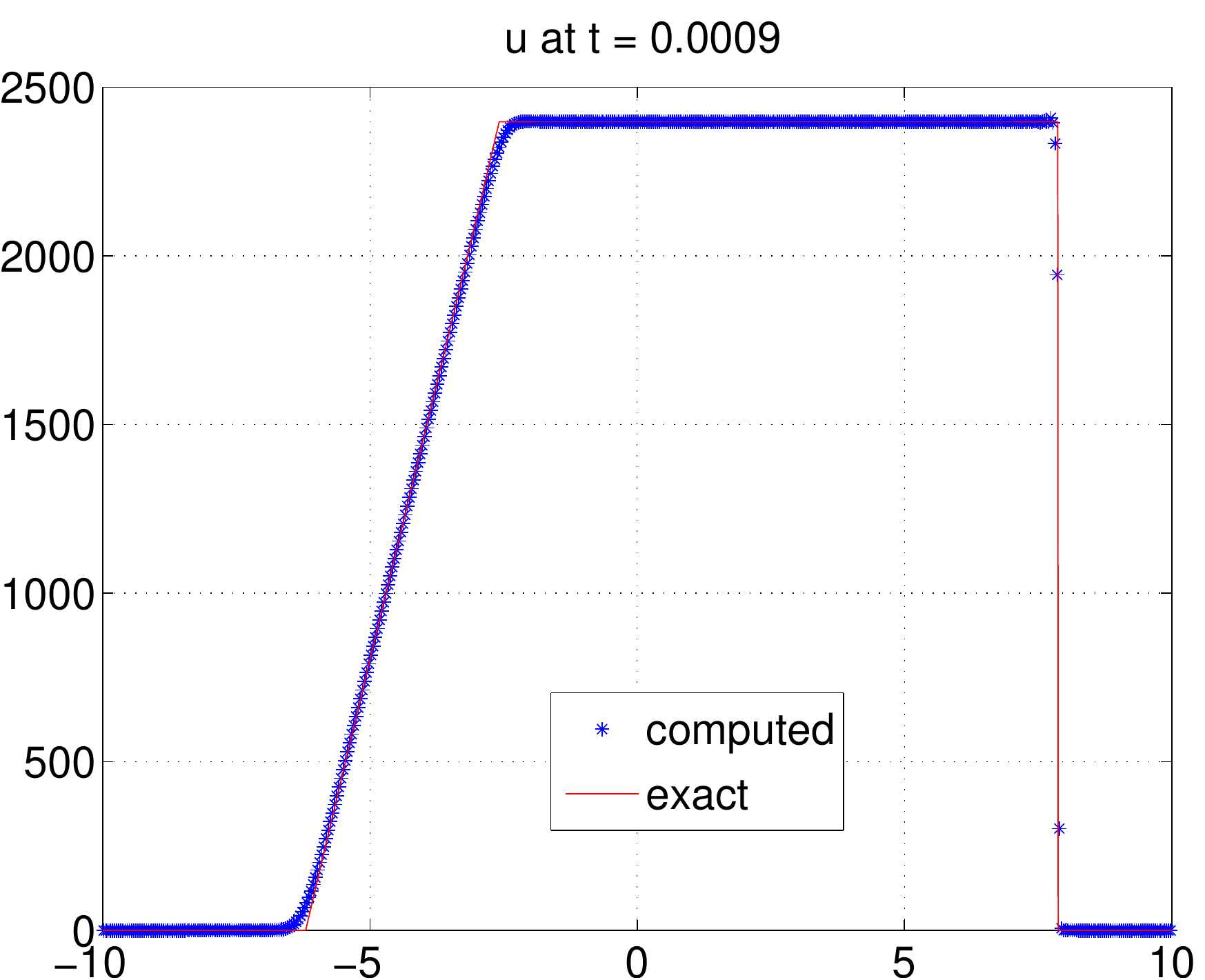}

\includegraphics[height=5.4cm]{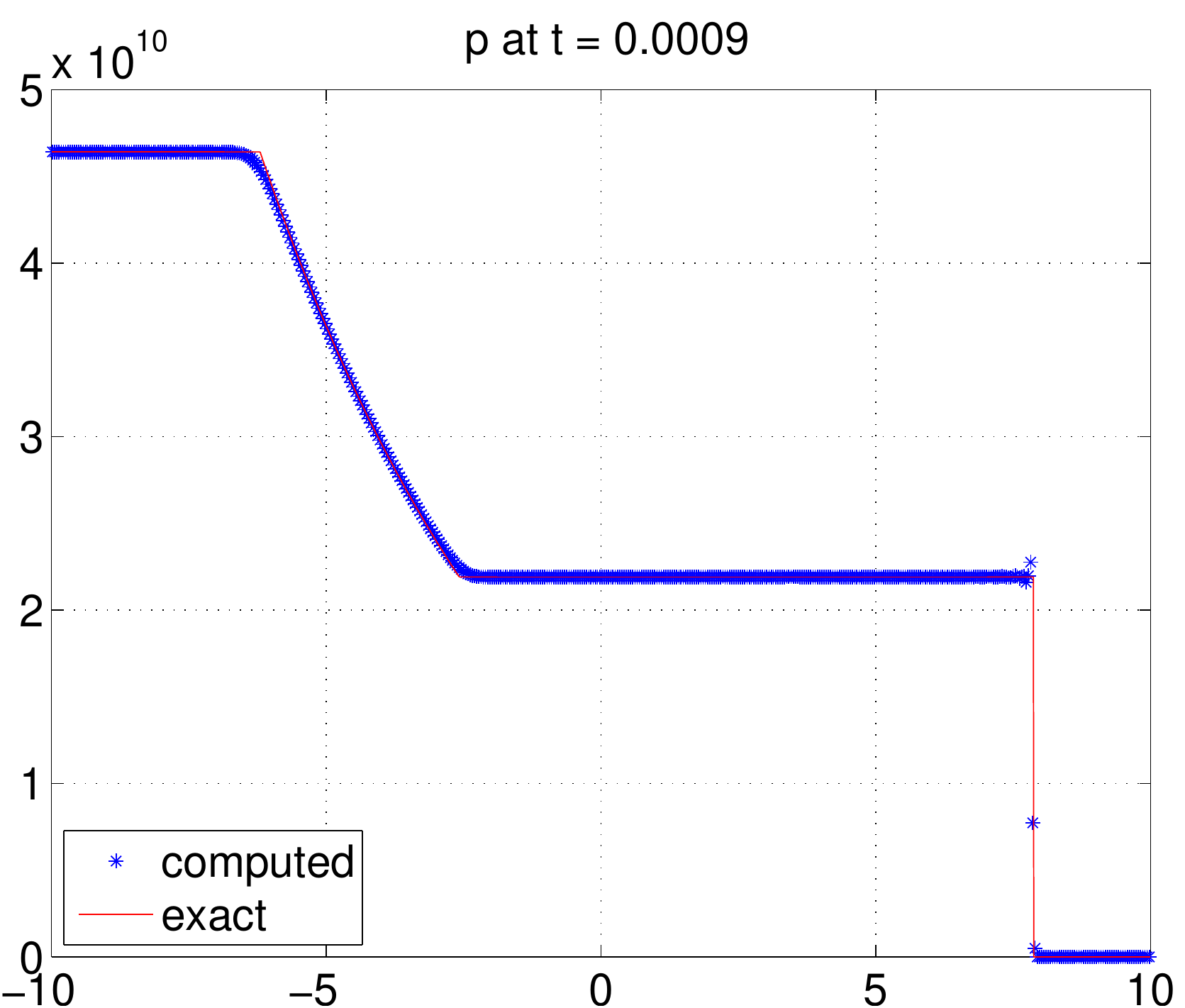}
\includegraphics[height=5.4cm]{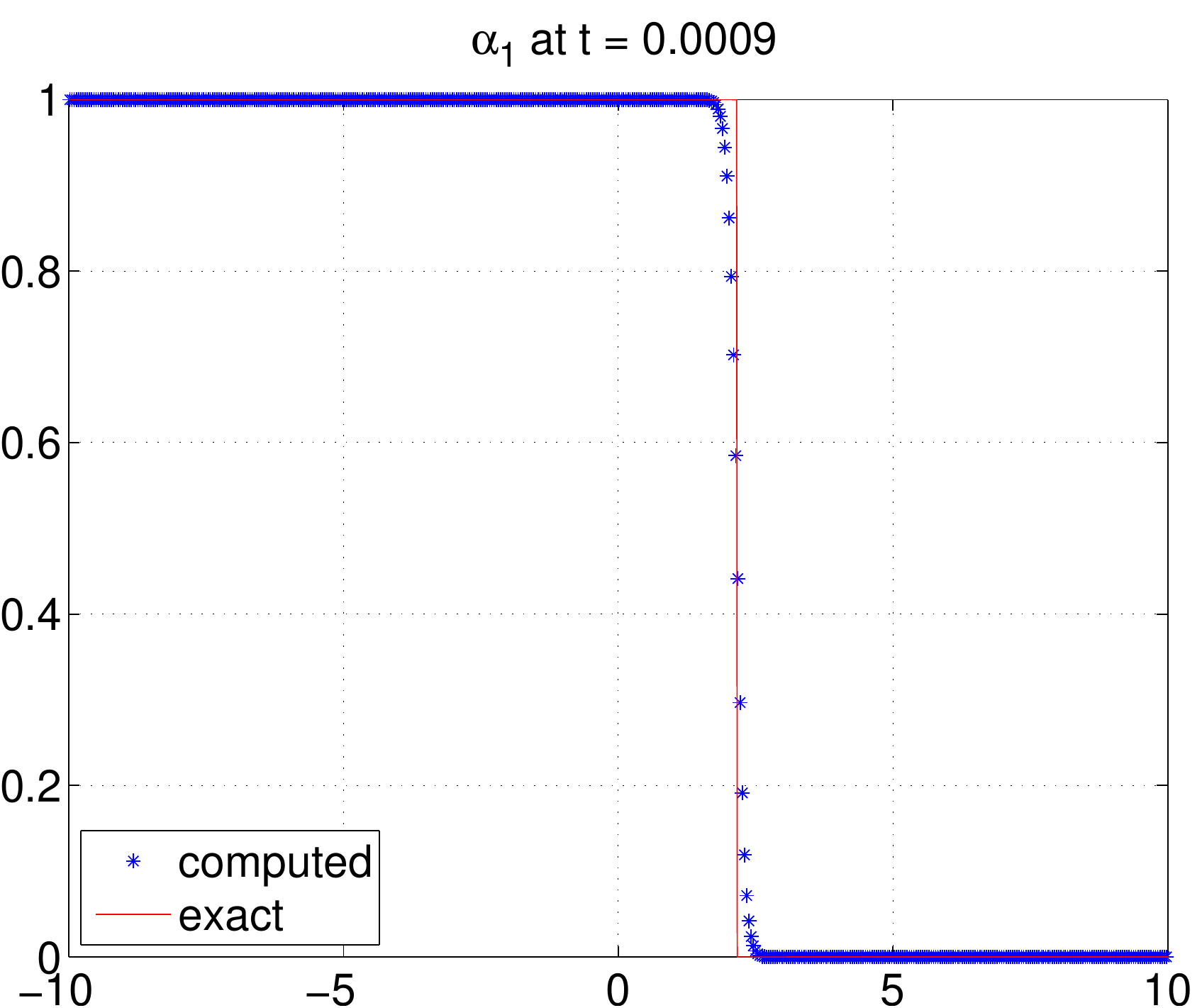}
\caption{Detonation gas-water shock tube problem. Computed results for the density $\rho$, velocity
$u$, pressure $p$, and gas volume fraction $\alpha_1$
at time $t= 0.9$~ms (blue marks $*$), compared with the exact solution (solid red line).}
\label{fig:st_gas_liq}
\end{figure}

We consider a two-fluid one-dimensional shock tube problem over the domain $[-10,10]$~m.
There is an initial discontinuity at $x=0$ that separates a left region
filled with detonation gases with density  $\rho_{\rm gas}= 2000\,{\rm kg\cdot m^{-3}}$
and pressure $p_{\rm gas}= 4.6406 \cdot 10^{10}$~Pa, and a right region filled with liquid water
with density  $\rho_{\rm liq}= 1044\,{\rm kg\cdot m^{-3}}$ and pressure $p_{\rm liq}=  10^{5}$ Pa.
Detonation gases are modeled by the JWL equation of state with the parameters in Table~\ref{tab_jwl}
(taken from \cite{sa-ab:multi}),
and the liquid water is modeled by the NASG equation of state with the parameters in
Table~\ref{tab_water}. In each region we consider a nearly pure fluid with volume
fraction $1-10^{-8}$.  The initial velocity is $u=0$. 
We compute the solution with $500$ grid cells and CFL number = $0.46$.
In Figure~\ref{fig:st_gas_liq} we show results at $t= 0.9$~ms for the density, velocity,
pressure, and gas volume fraction, together with the exact solution for this problem.
We observe the agreement with the exact solution and in particular the ability of the method
to preserve velocity and pressure invariance across and around the material interface.

\begin{table}[h!]
\centering
\caption{Material properties for the JWL equation of state  
modeling explosive.} 
\label{tab_jwl}
\begin{tabular}{|c|c|c|}
\hline
 Parameter  & Value (Explosive) & Units\\
 \hline
 $\rho_0$ & $1840$& $\rm [kg/m^3]$\\
 $\Gamma_0$ & $0.25$& \phantom{a}\\
 $a$ &$854.5 \times 10^9$ & $\rm [Pa]$\\
 $b$ & $20.5 \times 10^9$ & $\rm [Pa]$\\
 $r_1$ &$4.6$ &\phantom{a}\\
 $r_2$ &$1.35$ & \phantom{a}\\
 $\eint_0$ & 0 & $[\rm J/kg]$\\
 \hline
\end{tabular}
\end{table}

 \begin{table}[h!]
\caption{Parameters for the NASG EOS for liquid and vapor water in the temperature range 300-500 K}
\label{tab_water}
\centering
\begin{tabular}{|c |c |c |c |c|c|c|}
\hline
    phase  & $\gamma$ & $\varpi$ [Pa] & $\eta$ [J/kg]  & $\etapp \,\, {\rm[J/(Kg \cdot K)]}$ & 
    $\kappa_v\,\,{\rm [J/(Kg \cdot K )]}$ & $b\,\,{\rm [m^3/kg]}$\\ \hline 
   liquid & $1.187$ & $7028 \times 10^{5}$ 
      & $-1177788$ & 0 & $3610$ & $6.61 \times 10^{-4}$ \\ 
   vapor &  $1.467$ & $0$ & $2077616$ & $14317$   
      &  $955$ & 0 \\ \hline
\end{tabular}
\end{table}

\subsubsection{Underwater explosion close to a rigid  wall}

\begin{figure}
\centering
\includegraphics[height=3.7cm]{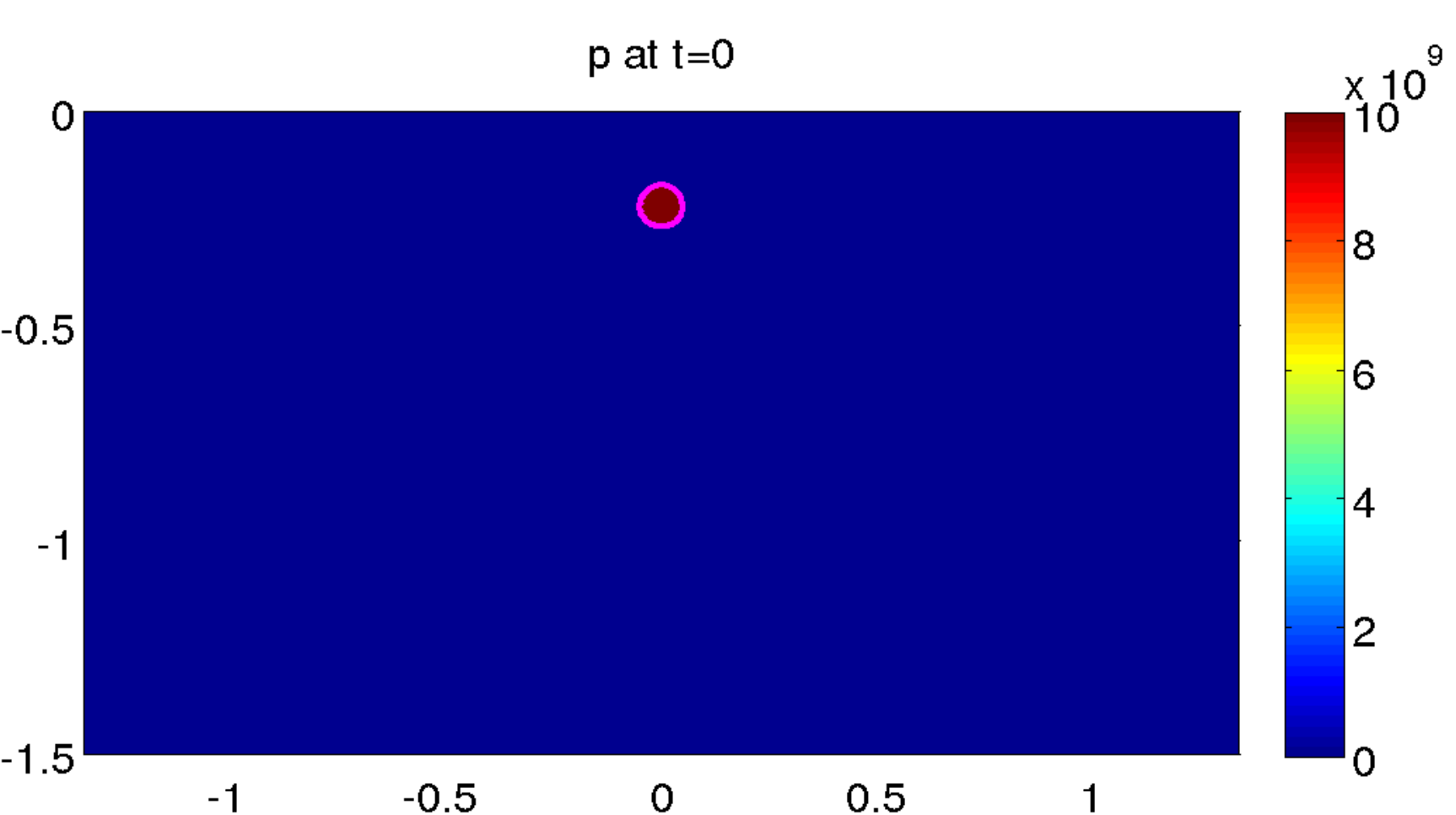}
\includegraphics[height=3.7cm]{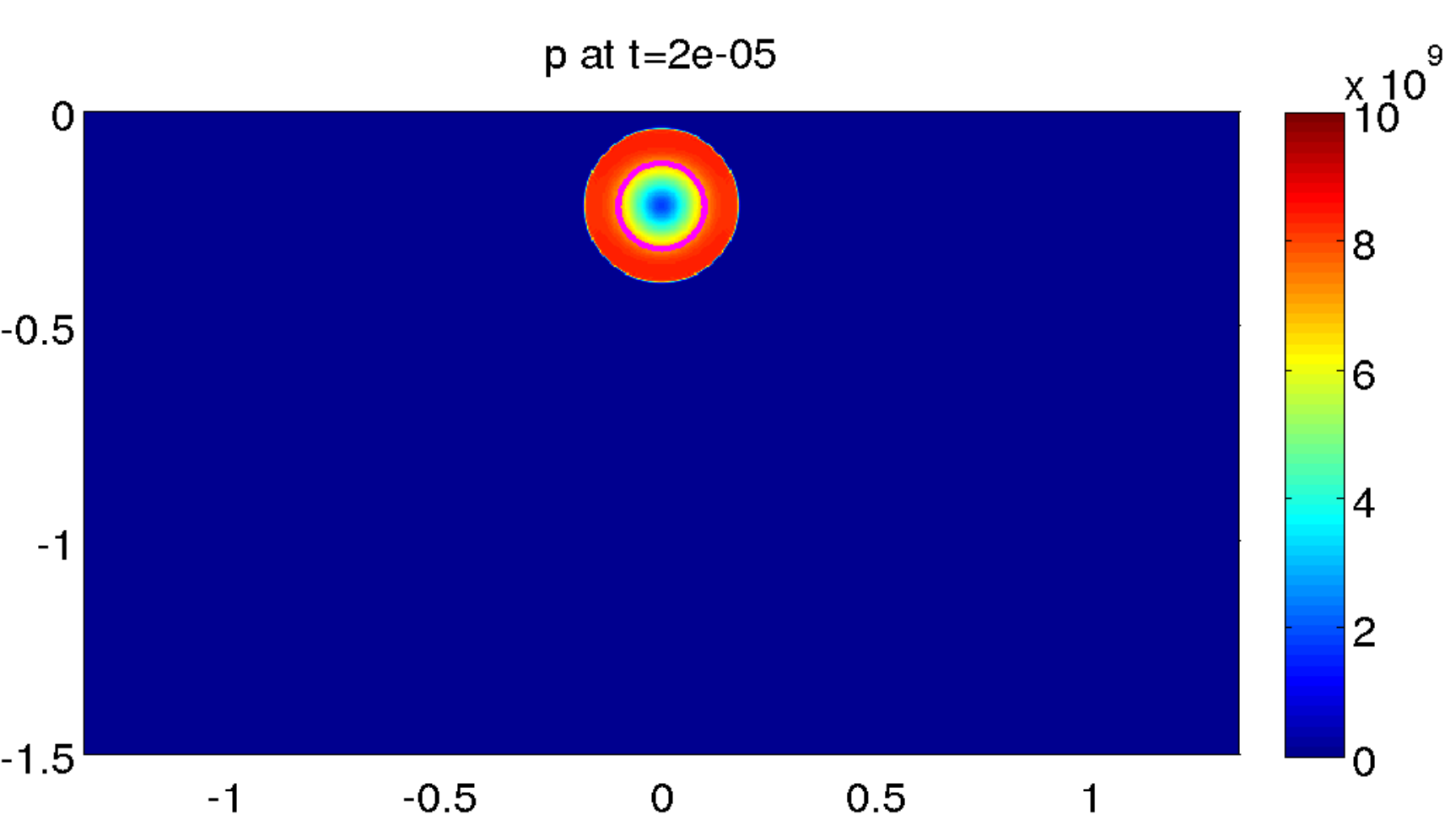}

\includegraphics[height=3.7cm]{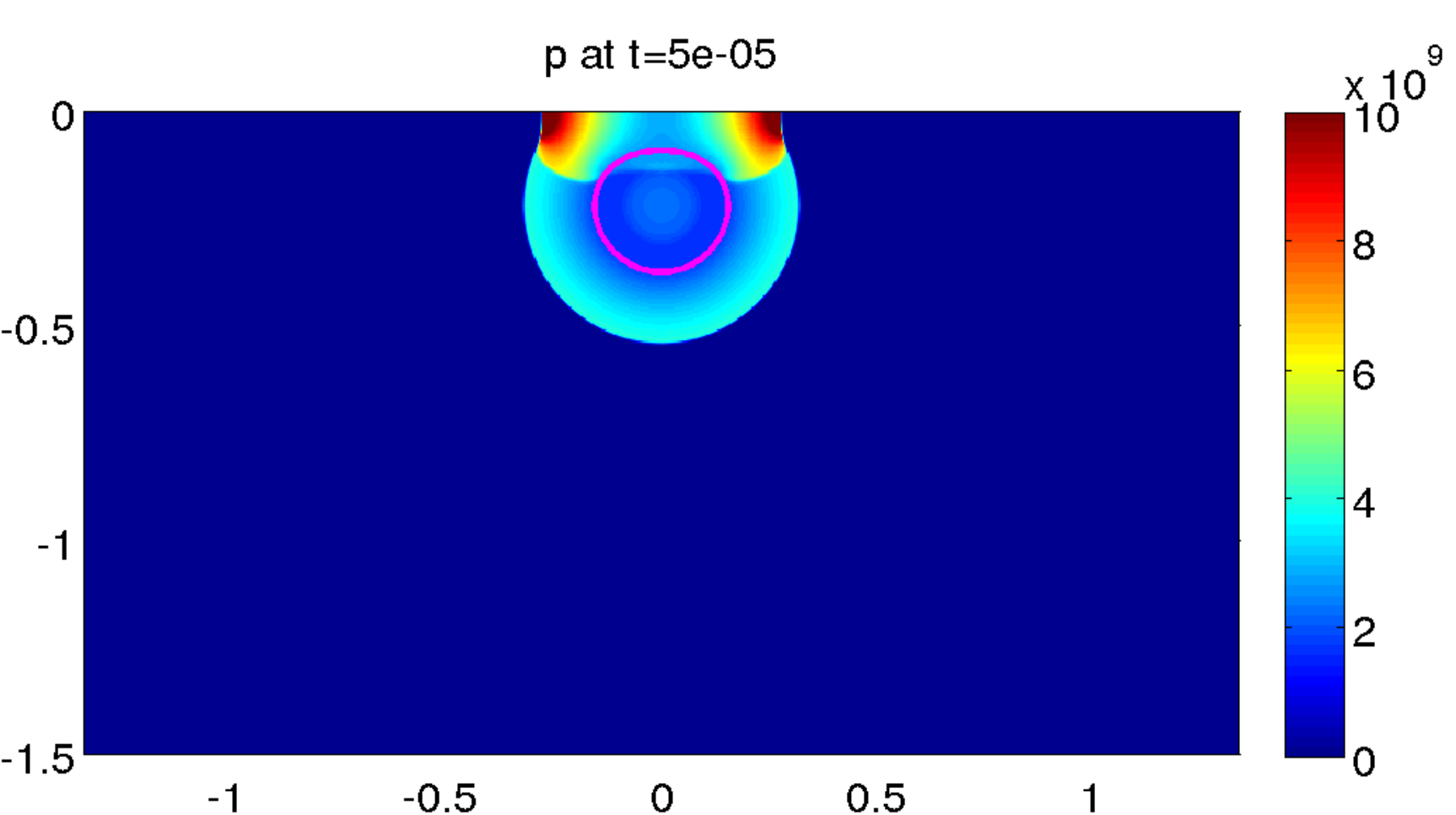}
\includegraphics[height=3.7cm]{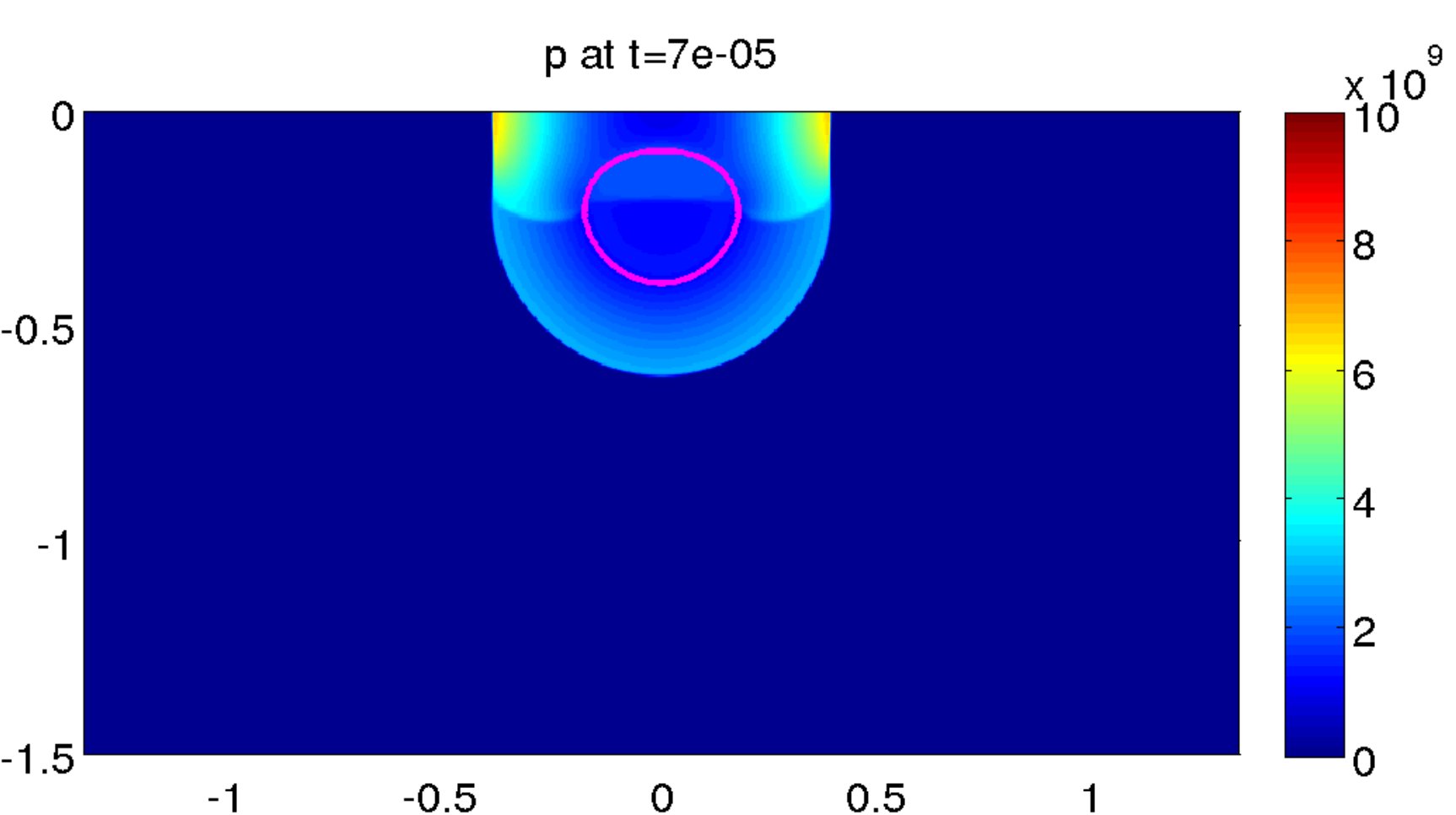}

\includegraphics[height=3.7cm]{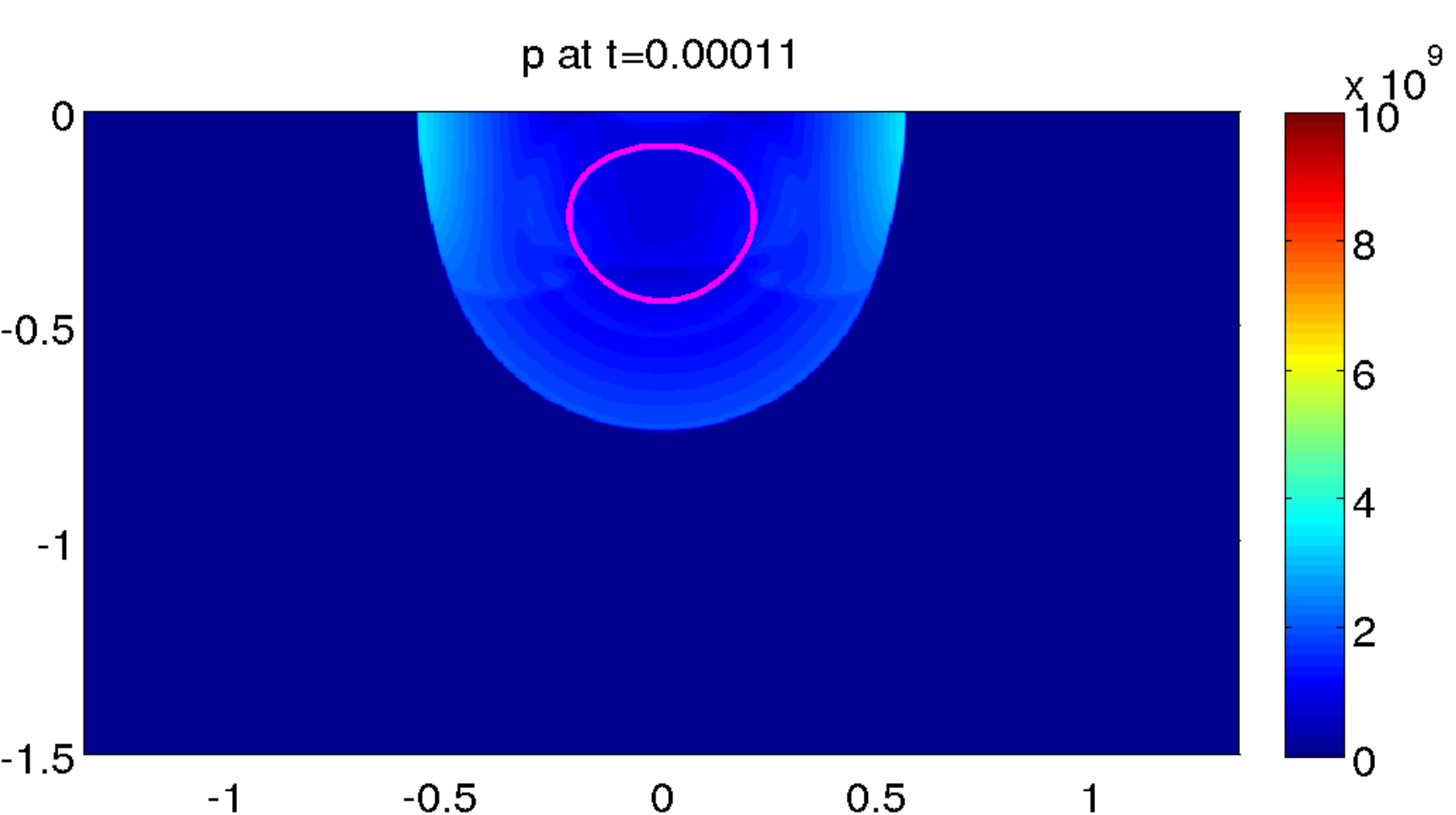}
\includegraphics[height=3.7cm]{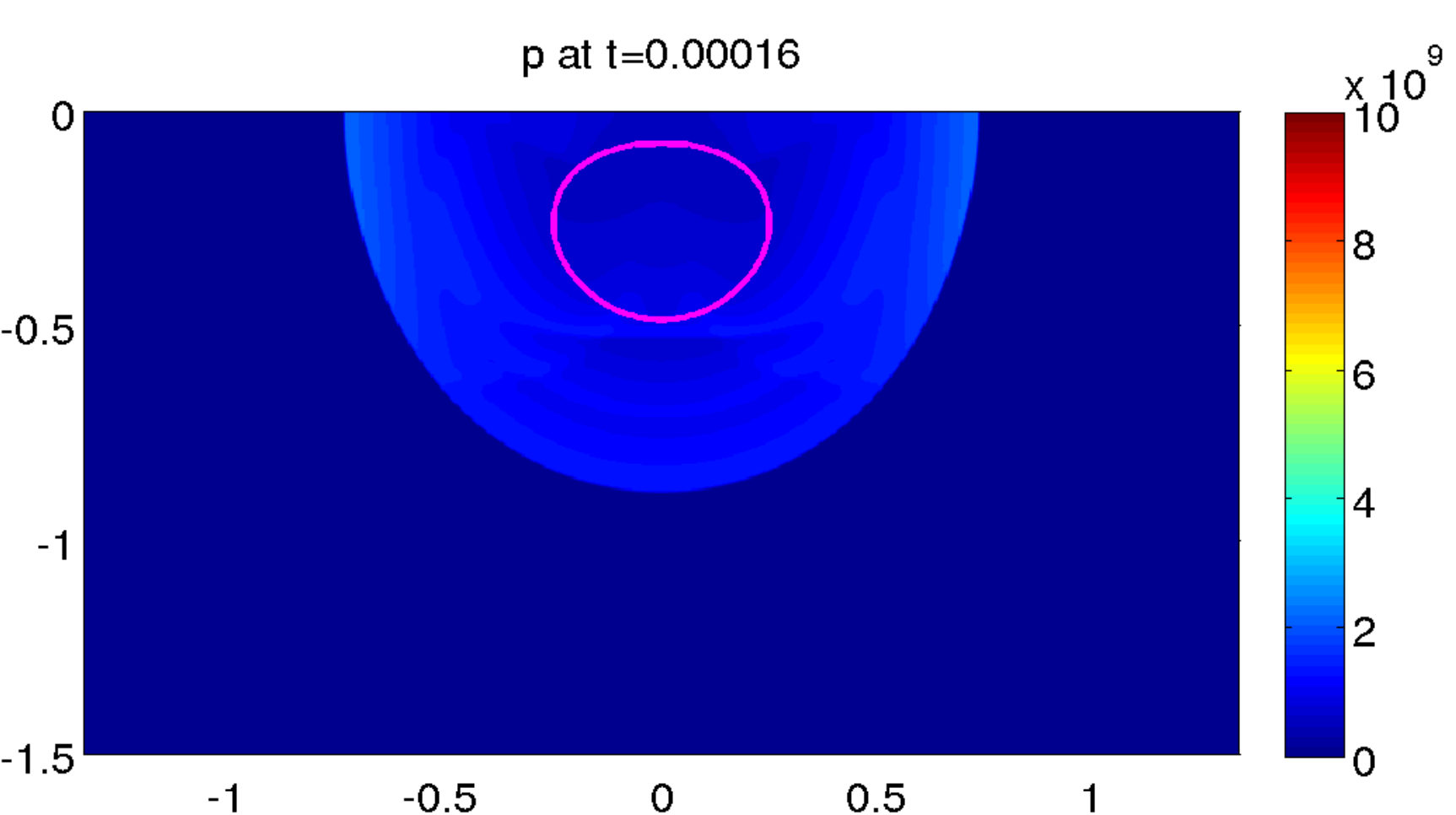}

\caption{Underwater explosion  near a rigid upper flat surface. Computed  pressure field at times 
$t = 0,\,0.02,\,0.05, \,0.07,\,\,0.11,\,0.16$~ms.
The thick solid line (magenta color) indicates the water/bubble interface.}
\label{fig:undex}
\end{figure}

We now perform a two-dimensional experiment. In this test we simulate a cylindrical underwater explosion (UNDEX) 
close to a rigid upper surface. Following \cite{xie-liu-khoo:cf2006}, we consider an initial bubble of highly pressurized gas (combustion products)
 surrounded by liquid water and located near an upper flat wall. 
 Combustion gases are modeled by the JWL equation of state with the parameters in Table~\ref{tab_jwl},
 while liquid water is modeled by the NASG EOS with the parameters in Table~\ref{tab_water}. The considered domain is $[-1.35,1.35]\times[-1.5,0]$~${\rm m}^2$,
 with the wall at $y=0$~m. The bubble initially is located at 
 $(x_{\rm b},y_{\rm b})= (0,-0.22)$~m, and it has radius $r_{\rm b} = 0.05$~m. 
 Inside the bubble we set initially a pressure $p = 4.6406 \cdot 10^{10}$~Pa, a gas density 
 $\rho_{\rm gas} = 2000\,{\rm kg/ m^3}$, and a gas volume fraction $\alpha_{\rm gas} = 1-10^{-6}$.
 Outside the bubble we set a pressure $p = 10^5$~Pa, a liquid density $\rho_{\rm liq} = 1044\,{\rm kg/ m^3}$
 and a liquid volume fraction $\alpha_{\rm liq} = 1-10^{-6}$. This explosion problem is characterized by a complex pattern of shocks and rarefaction waves \cite{xie-liu-khoo:cf2006}. We use $721 \times 400$ grid cells
  with CFL number = $0.4$. We show in Figure~\ref{fig:undex} plots of the pressure at six different times.
 At $t = 0.02$ ms (upper right plot) we can observe the circular shock created by the explosion. 
 At $t = 0.05$~ms (middle left plot) this shock has reflected from the wall. The shock
 reflection gives rise to strong rarefactions that cause the formation of 
 a low pressure  region close to the wall, which is later re-compressed.
 
\subsection{Numerical tests with thermo-chemical relaxation} 
 
We now present several numerical experiments  where we activate thermal
and chemical relaxation, simulating both instantaneous and finite-rate transfer processes.
Concerning non instantaneous transfers we report  primarily tests with finite-rate mass transfer
under thermal equilibrium, 
which is our principal interest.

\subsubsection{Water two-phase cavitation tube}

\begin{figure}[h!]

\centering

\includegraphics[height=5.3cm]{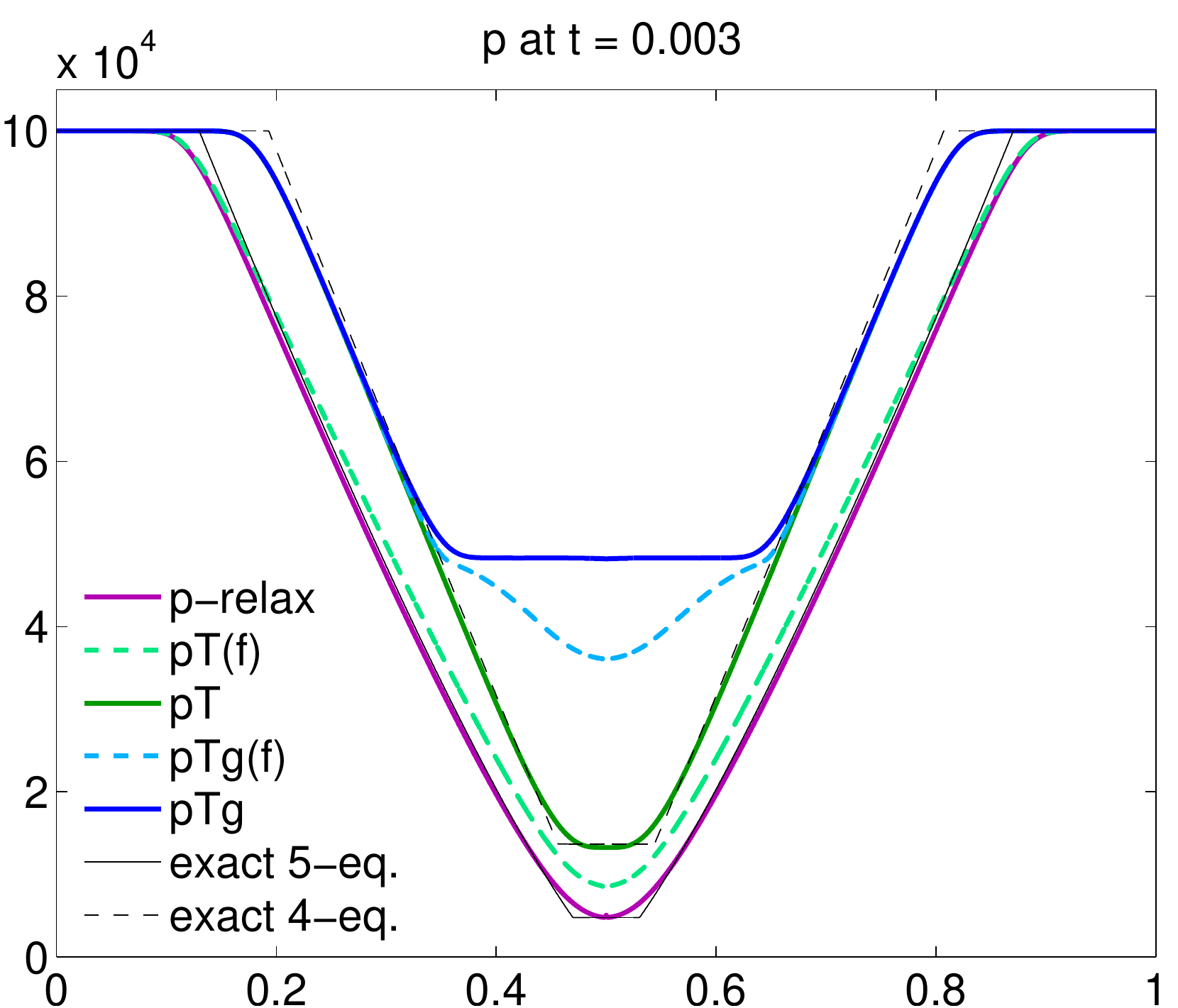}
\includegraphics[height=5.3cm]{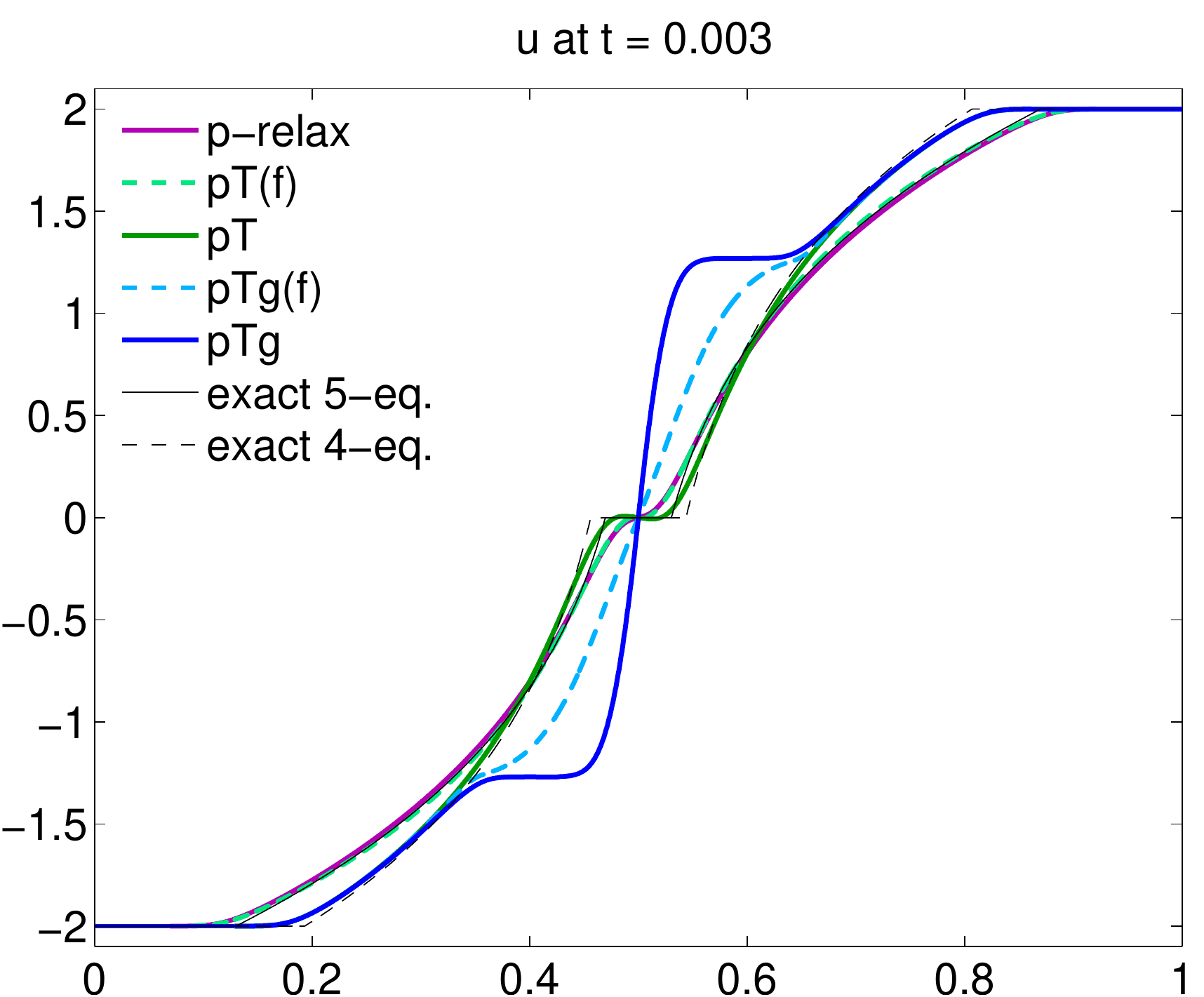}

\includegraphics[height=5.3cm]{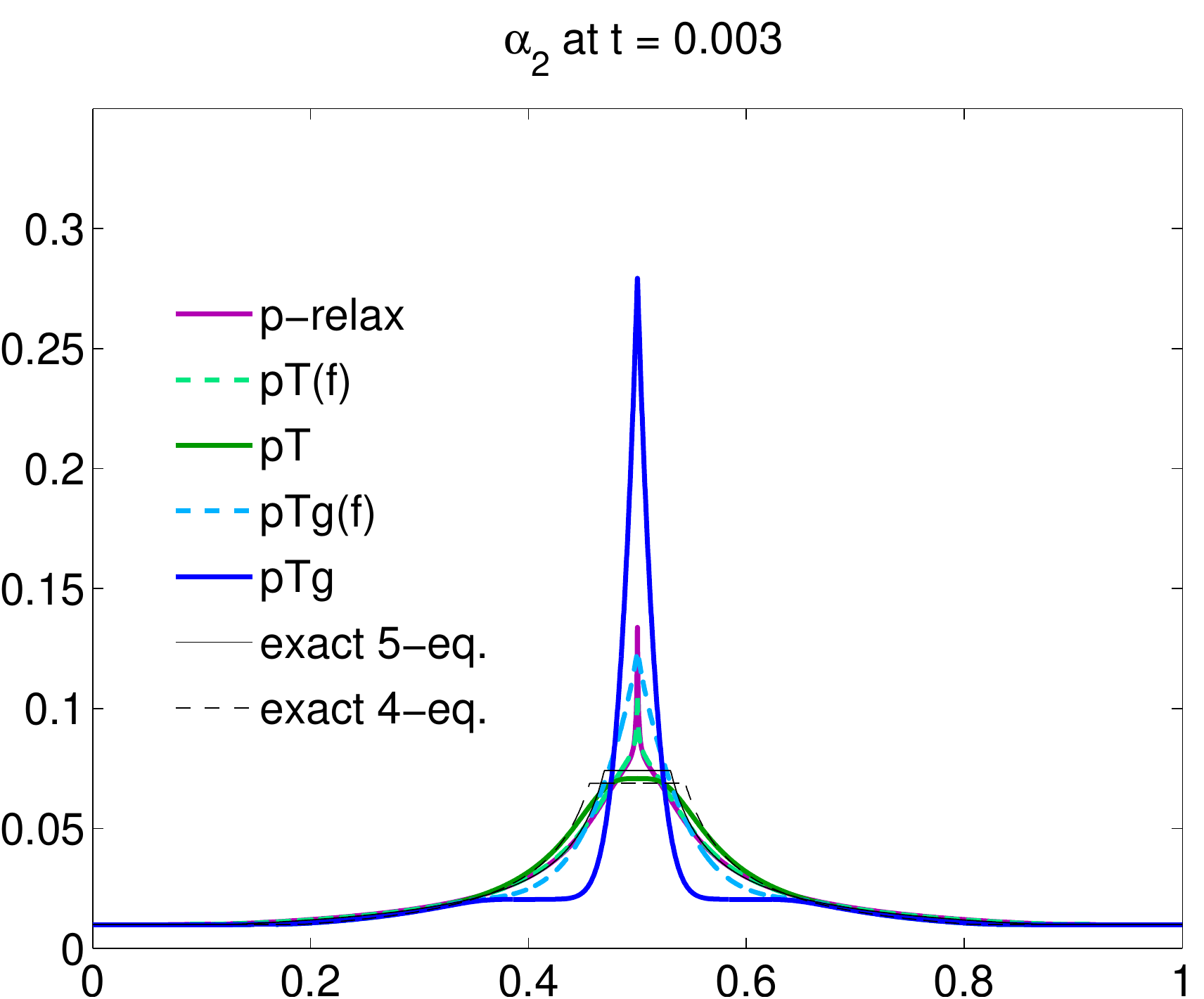}
\includegraphics[height=5.3cm]{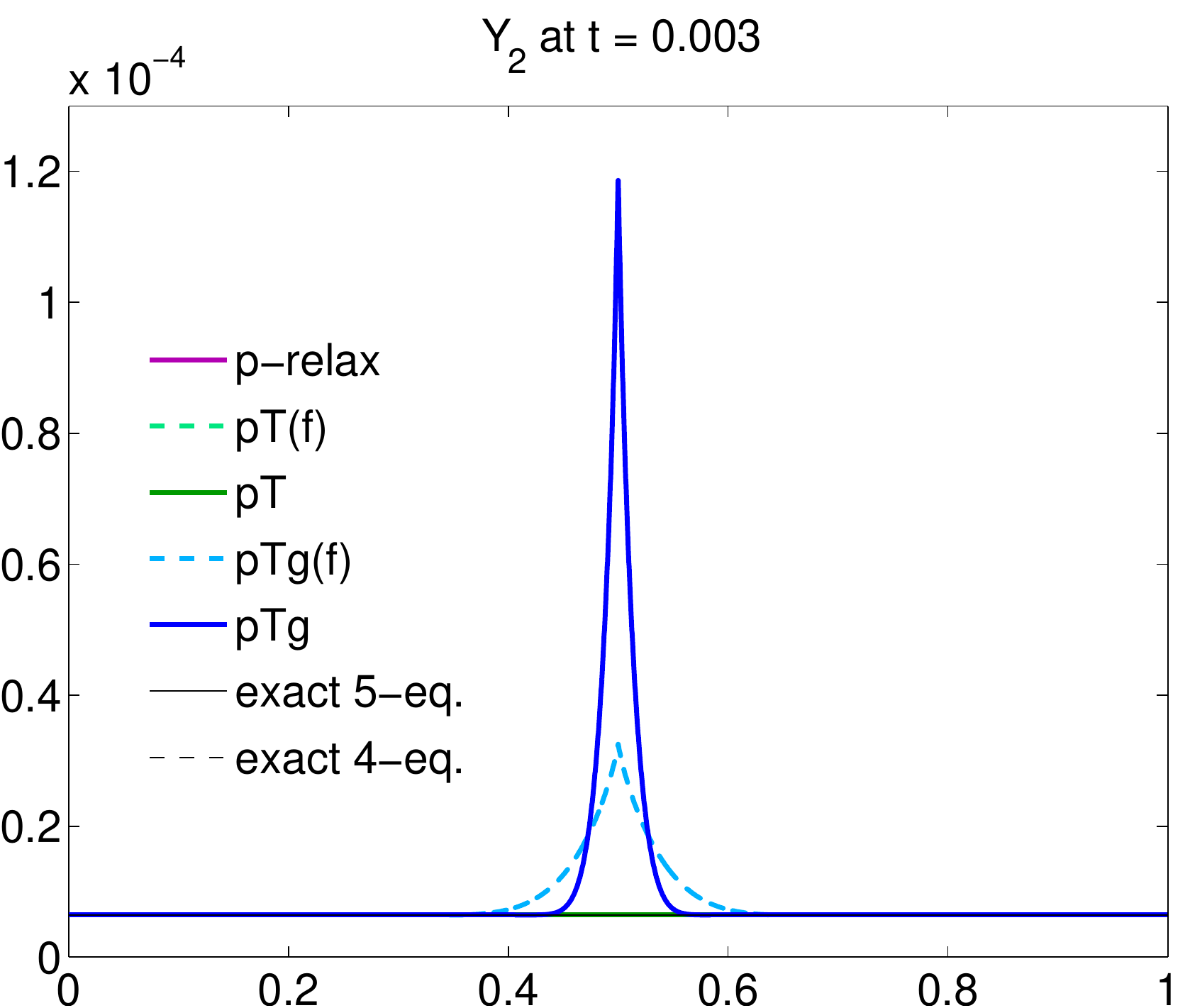}

\includegraphics[height=5.2cm]{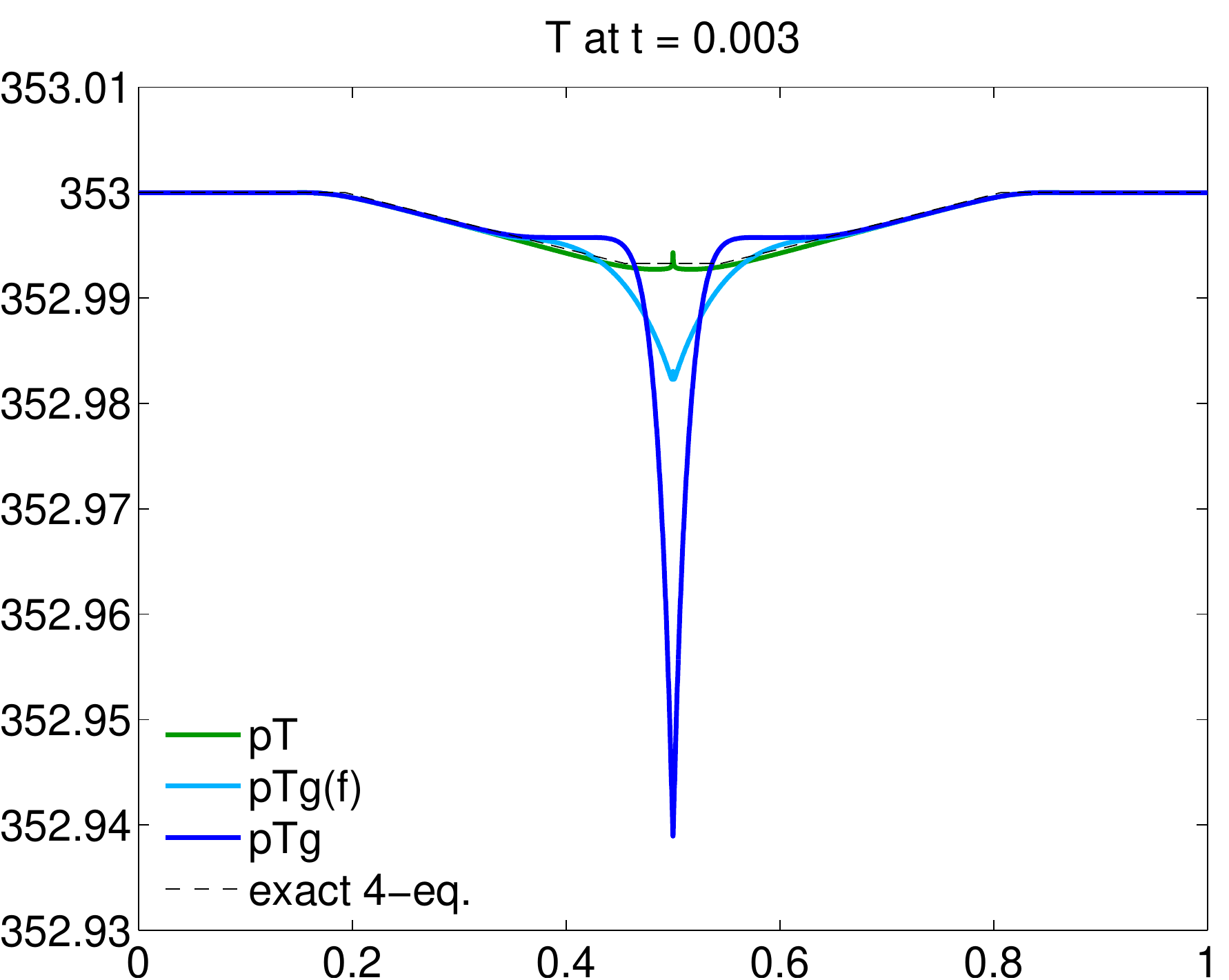}
\includegraphics[height=5.2cm]{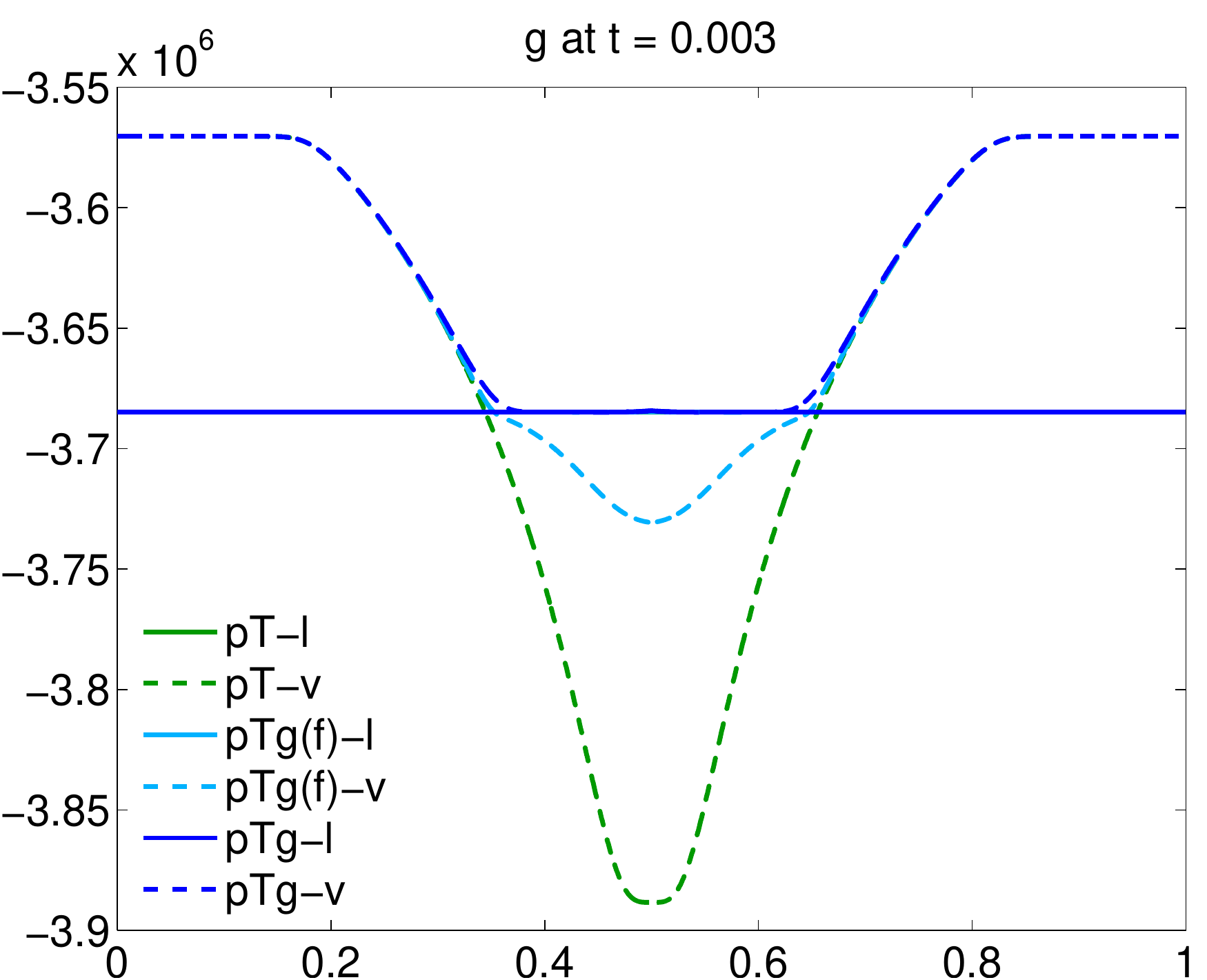}

\caption{Water cavitation tube test,  results at $t=0.003$~s. 1st and 2nd rows: pressure $p$,
velocity $u$, vapor volume fraction $\alpha_2$, vapor mass fraction $Y_2$ for five  levels of relaxation.
3rd row: temperature $T$ and chemical potentials $g_k$ for the cases with
temperature equilibrium. $p$-relax (violet solid line): instantaneous mechanical relaxation;
$pT{\rm (f)}$ (dashed light green line): instantaneous mechanical relaxation and finite-rate thermal relaxation;
$pT$ (solid dark green line): instantaneous mechanical and thermal relaxation; 
$pTg{\rm (f)}$ (dashed light blue line): instantaneous mechanical and thermal relaxation and finite-rate 
chemical relaxation; $pTg$ (solid dark blue line): instantaneous mechanical, thermal, and chemical relaxation.
Only in the plot of the chemical potentials $g_k$ the solid line indicates
the liquid ($l$) and the dashed line  the vapor 
($v$).  
 The exact solution
of the $p$-relaxed model (solid black line) and of the $pT$-relaxed model (dashed black line) is
also plotted.}
\label{fig:cavinasg}
\end{figure}

We consider a one-dimensional water cavitation tube problem, which is a variant of a numerical test taken
from  \cite{saurel-PA} that we performed in \cite{pelanti-shyue} with the stiffened gas 
equation of state and instantaneous relaxation processes. 
Initially we have a tube over the interval $[0,1]$~m filled 
with liquid water at pressure $p = 10^5$~Pa and temperature $T=T_1=T_2 = 353$~K.
The liquid  contains a uniformly distributed small amount of vapor,  with a volume
fraction $\alpha_{\rm vap}=10^{-2}$ in the whole domain.  A velocity discontinuity is set at $x = 0.5$~m at initial time,
with $u=-2\,\,{\rm m/s}$ on the left and $u=2\,\,{\rm m/s}$ on the right.
We use here the NASG equation of state for the
liquid water and water vapor phases, with the parameters in Table~\ref{tab_water}. 
We perform computations with five different levels of relaxation by using the  techniques
detailed in Section~\ref{sec:arbraterel}: only instantaneous  mechanical  relaxation 
(case denoted as $p$-relax in the plots), instantaneous mechanical relaxation and finite-rate thermal relaxation with
$\vartheta=2000\,\,{\rm Pa/(s\cdot K)}$ ($pT{\rm(f)}$),
instantaneous mechanical and thermal relaxation ($pT$),   instantaneous mechanical and thermal relaxation 
and finite-rate chemical relaxation with $\nu = 10^{-4}\,\,{\rm Pa\cdot kg^2/(s \cdot J^2)}$ ($pTg{\rm (f)}$),
and full thermodynamic relaxation ($pTg$). 
Phase transition is hence activated only in  the two last cases.
Let us remark also that in this test (following \cite{saurel-PA}) chemical relaxation is activated when the equilibrium temperature is greater than the saturation temperature, $T>T_{\rm sat}(p)$ (this happens in the middle region
of the tube).
We use $5000$ grid cells and we set the CFL number  = 0.5. Results for the pressure,
velocity, vapor volume fraction and vapor mass fraction at time $t=0.003$~s are displayed in the  top and middle  rows of Figure~\ref{fig:cavinasg}.
In all the cases the solution involves two rarefactions going in opposite directions that cause
a decrease of the pressure in the middle of the tube, and correspondingly an increase of the vapor volume
fraction. Let us note that if mass transfer is not activated then the vapor mass fraction remains constant, and the cavitation process is only mechanical. In contrast, 
if mass transfer is activated then the solution involves also two evaporation waves.
In this case the vapor mass fraction increases in the middle of the tube, 
and moreover here the pressure is driven to its saturation value, whereas without mass transfer
the pressure continues to decrease in the center of the tube.
In the bottom row of Figure~\ref{fig:cavinasg} we show the equilibrium temperature and the chemical potentials for
the three test cases with instantaneous pressure and temperature equilibrium ($pT$, $pTg{\rm (f)}$, $pTg$), 
with or without mass transfer.
Computed liquid and vapor temperatures are found to be overlapped, this proving the ability
to impose numerically thermal equilibrium. By observing the plot  of the 
liquid and vapor chemical potentials 
we notice the region of chemical potential equilibrium in the middle of the tube for the 
$pTg$-relaxation case,
corresponding to the region of activation of chemical relaxation under the 
evaporation condition  $T>T_{\rm sat}\,$.
For the case with finite rate mass transfer ($pT{\rm (f)}$-relaxation) we notice that liquid and vapor chemical potentials
are being driven to equilibrium and their difference is correctly reduced with respect to the case with
no mass transfer ($pT$-relaxation).    
In Figure~\ref{fig:cavinasg} we also plot the exact solution for this problem of the five-equation
pressure equilibrium model (\ref{eq:sisp}) and of the four-equation
pressure and temperature equilibrium model (\ref{eq:sysT}).
We observe good agreement of the results computed with activation of instantaneous
mechanical relaxation  with the exact solution of the five-equation $p$-relaxed model,
and of the results computed with activation of instantaneous mechanical and thermal
relaxation with the exact solution
of the four-equation $pT$-relaxed model. This shows the capability of the numerical model
to approximate solutions of the limit equilibrium models in the limit on instantaneous relaxation processes. 
Let us also observe that the different speeds of the leading edges of the rarefactions
for different levels of activation of relaxation processes is consistent
with the subcharacteristic condition in (\ref{eq:subchar}) $c_{pT}\leq c_p$.
Note that also in the tests with phase transition
 in correspondence of these leading edges chemical relaxation  is not activated since $T<T_{\rm sat}\,$, hence for  the tests with mass transfer the sound speed in these regions is the pressure and temperature equilibrium sound speed $c_{pT}$.

\subsubsection{Dodecane liquid-vapor shock tube}

\begin{figure}[h!]

\centering

\includegraphics[height=5.5cm]{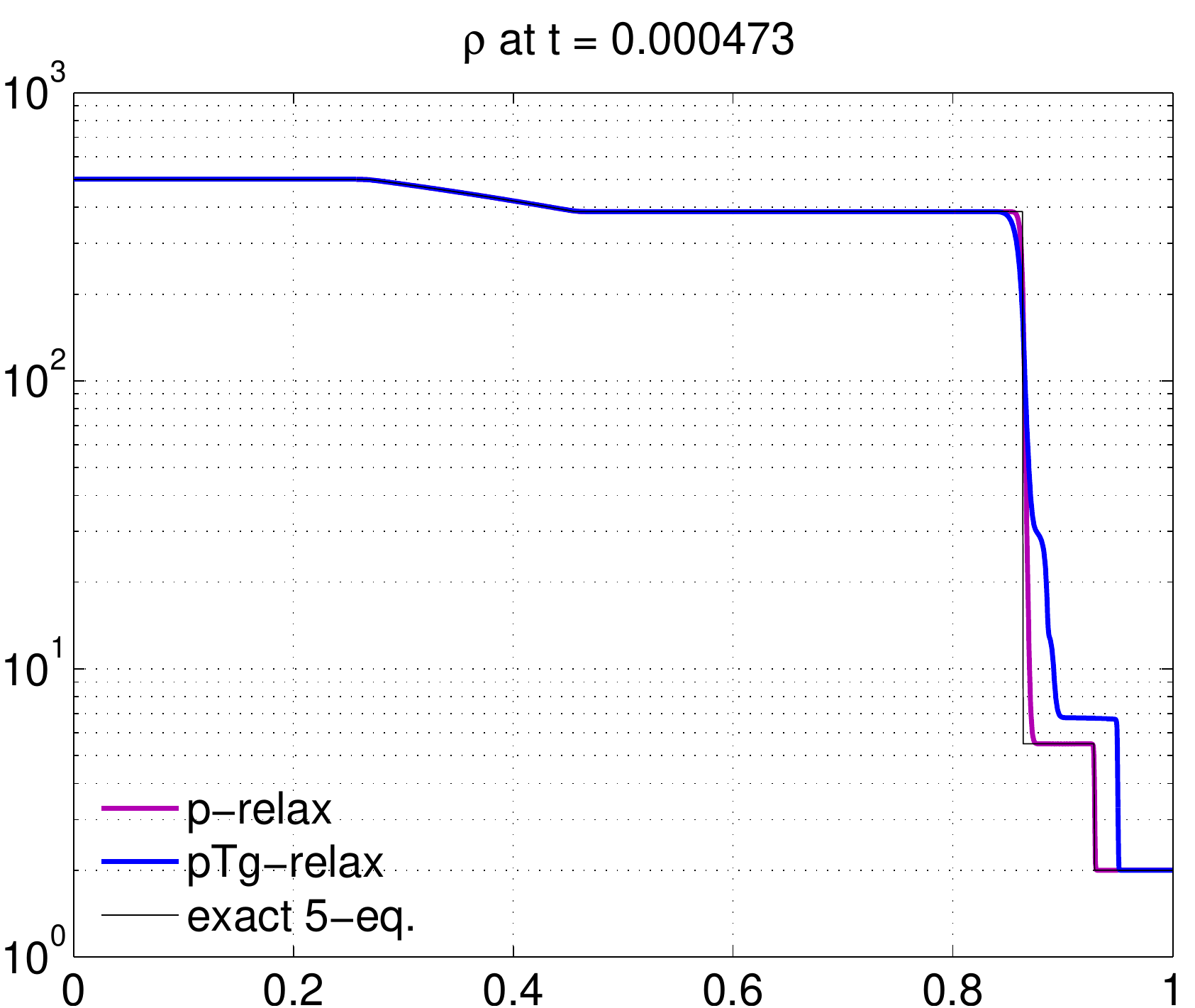}
\includegraphics[height=5.5cm]{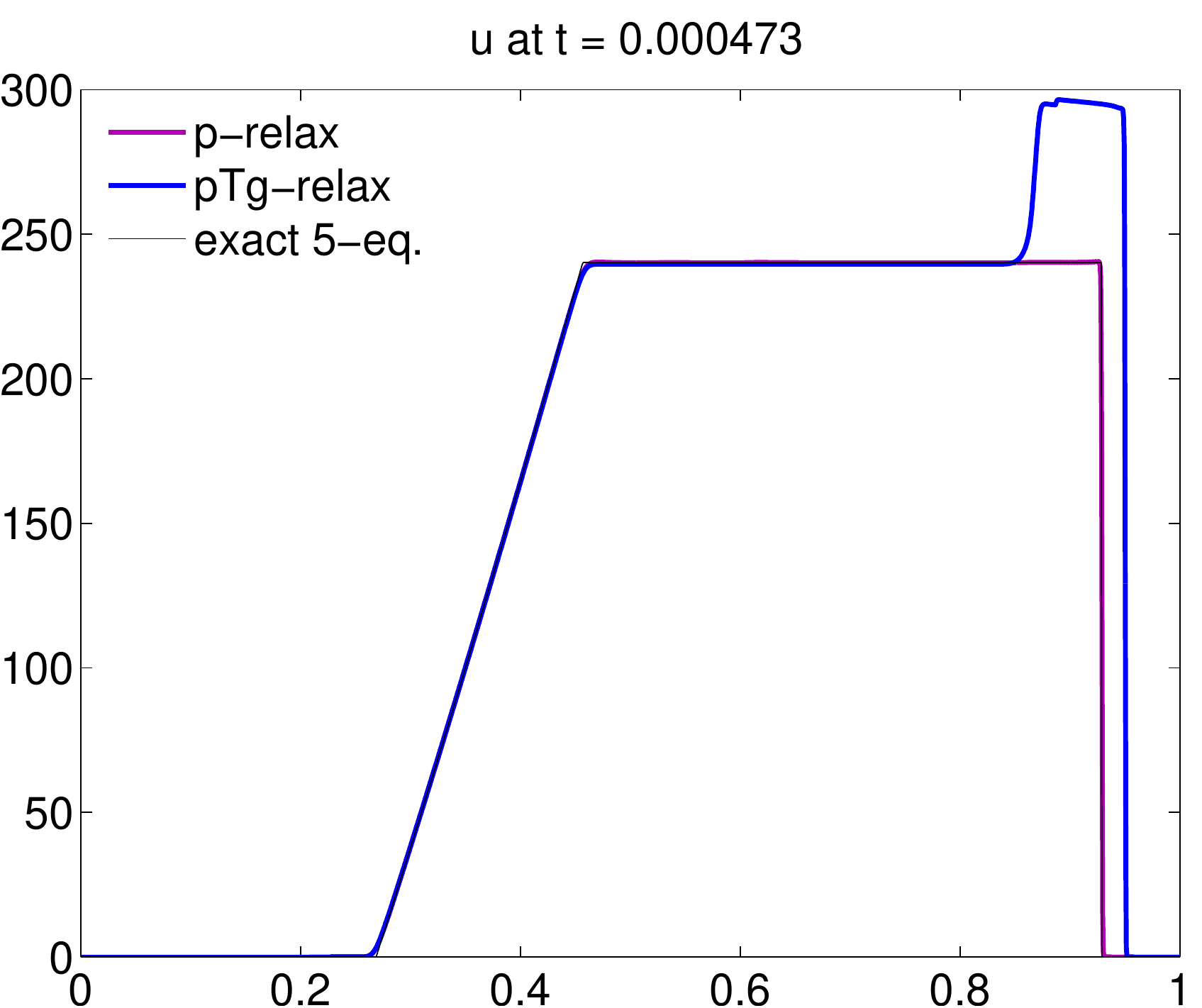}
\includegraphics[height=5.5cm]{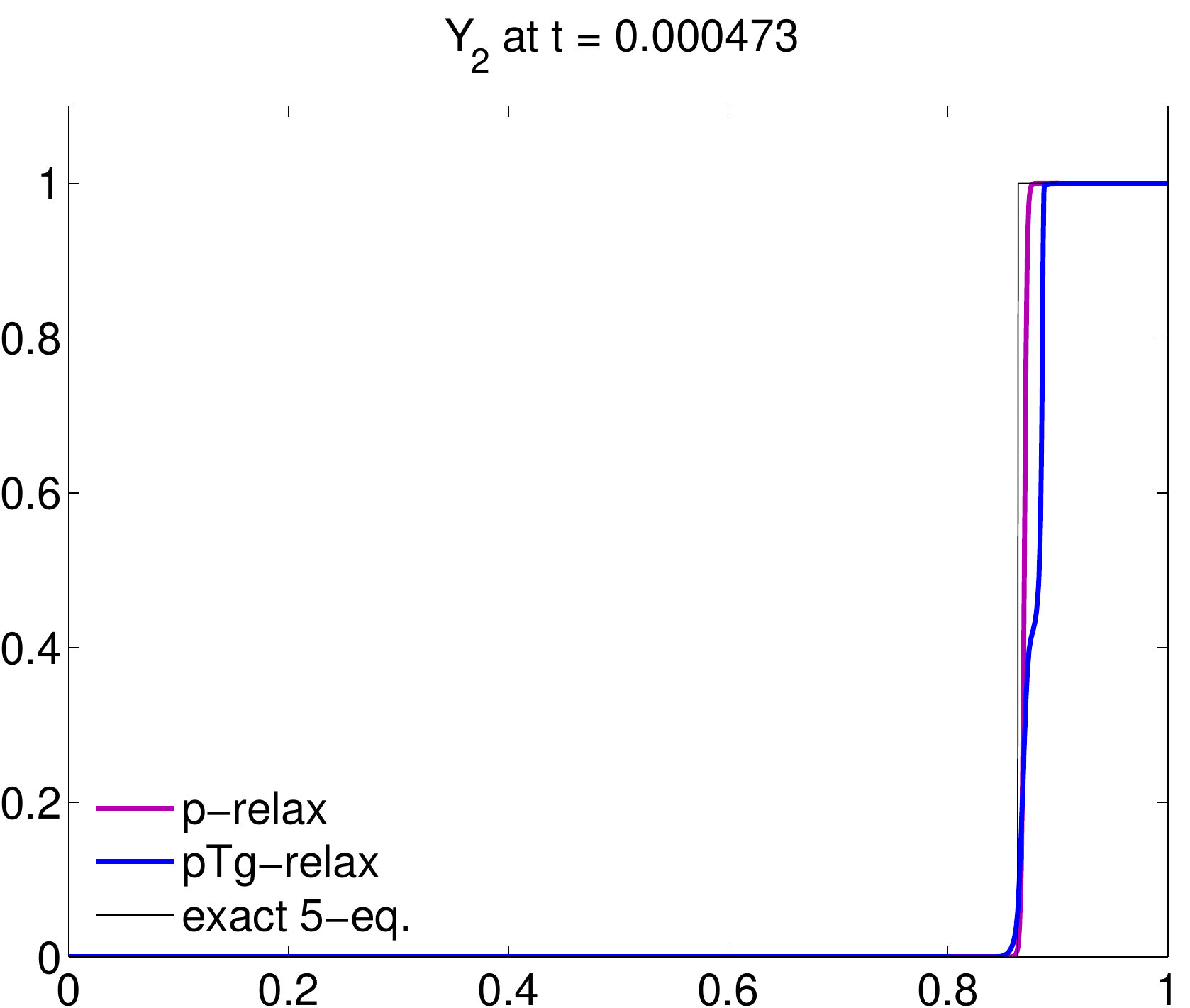}
\includegraphics[height=5.5cm]{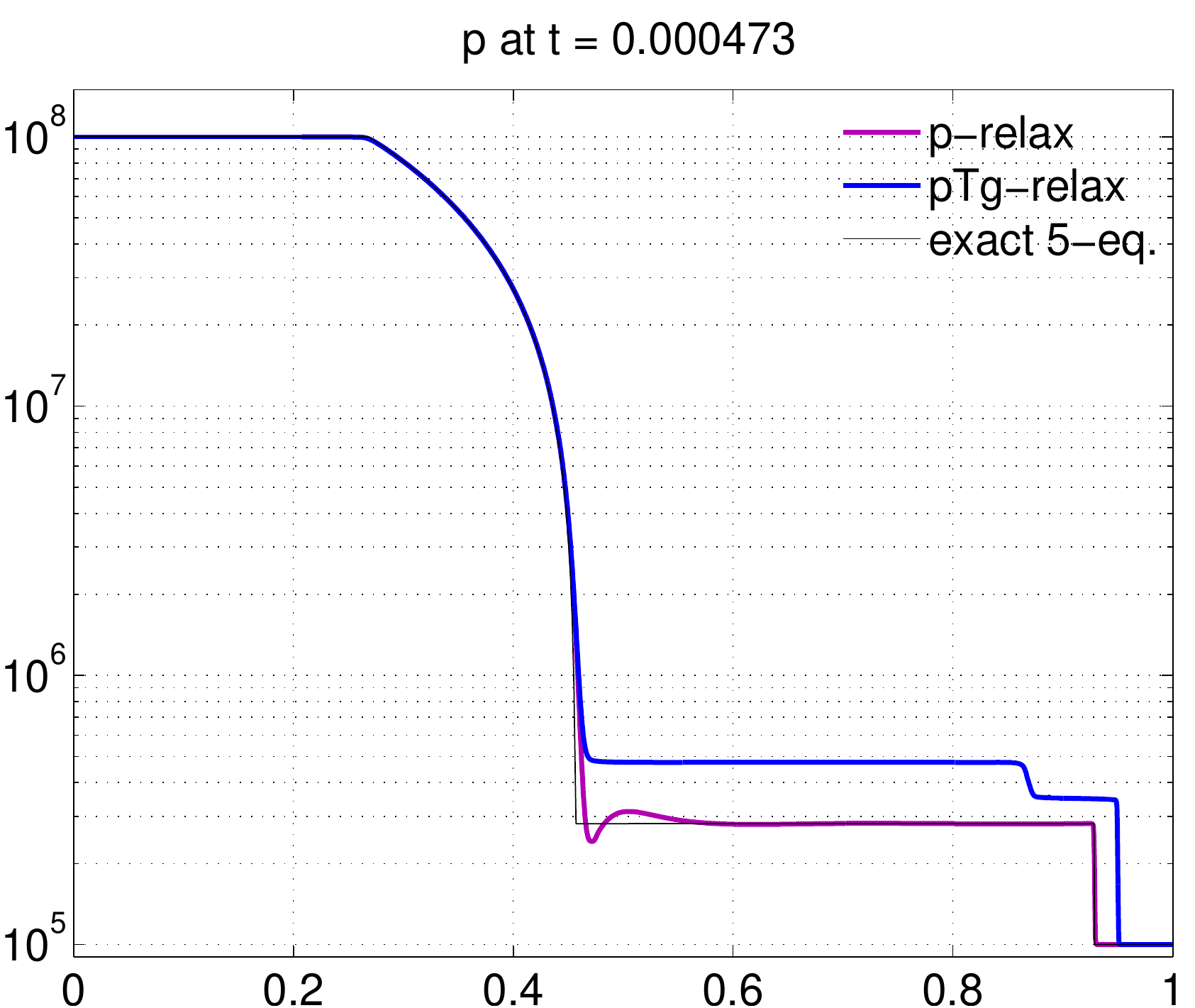}

\includegraphics[height=5.5cm]{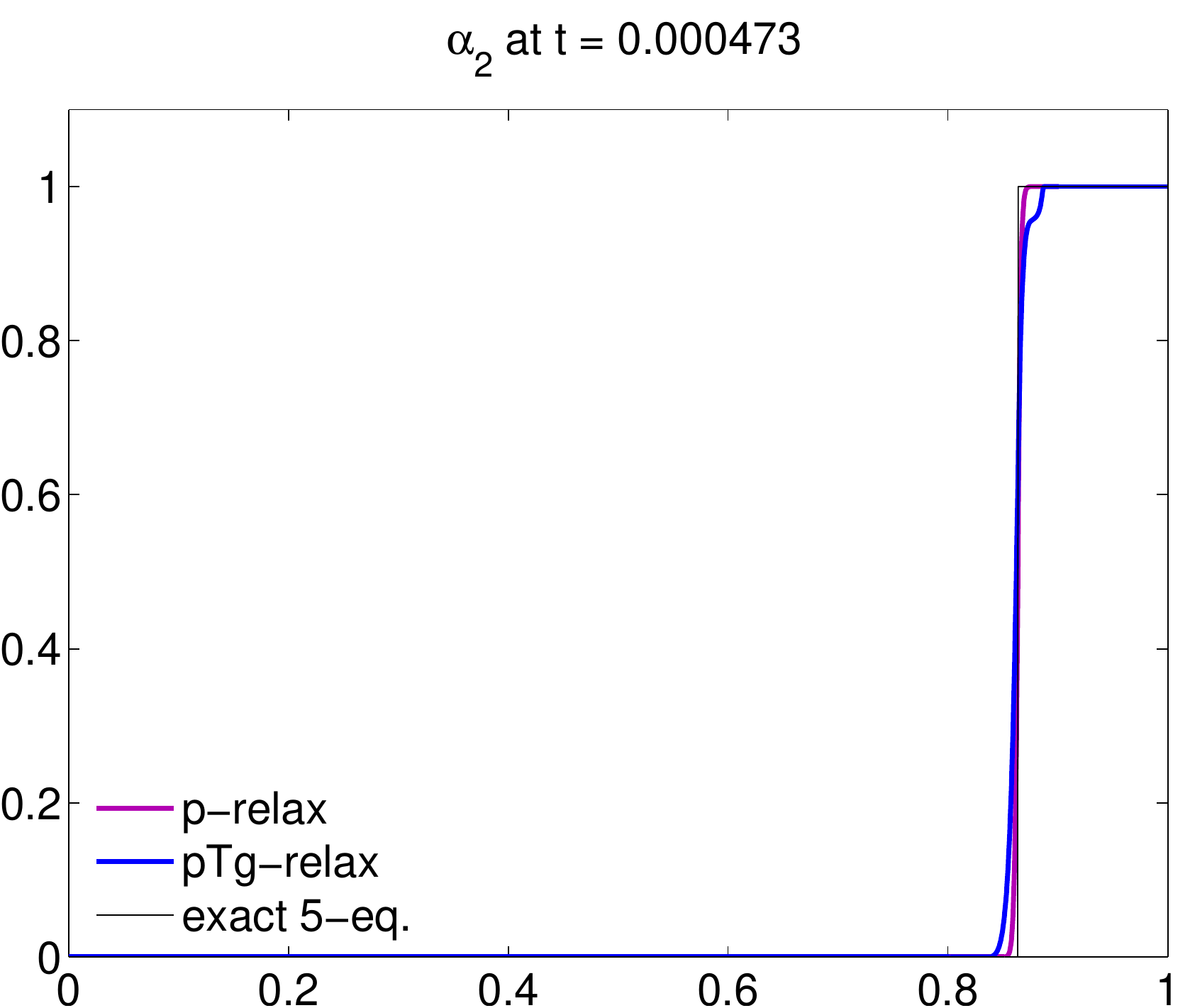}
\includegraphics[height=5.5cm]{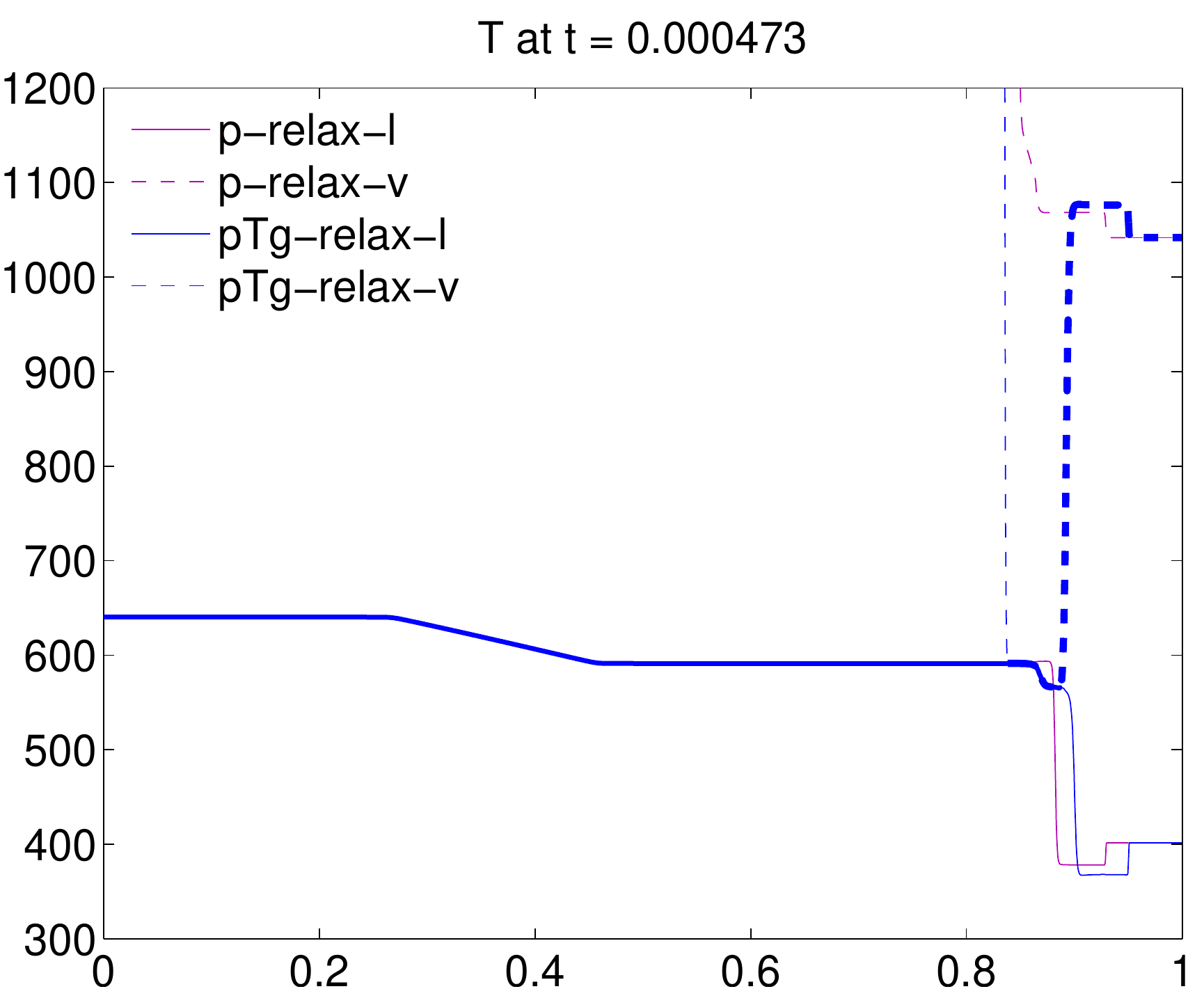}
\caption{Dodecane shock  tube test. Results at $t=473\,\,\mu {\rm s}$. Density $\rho$, velocity $u$, 
vapor mass fraction $Y_2$, pressure $p$,
vapor volume fraction $\alpha_2$, vapor and liquid temperatures  $T_k$.
The density and pressure plots are 
 in  semi-logarithmic scale. $p$-relax (solid violet line): instantaneous mechanical relaxation;
 $pTg$-relax (solid blue line): instantaneous mechanical, thermal, and chemical relaxation.
 Only in the plot of the temperatures $T_k$ the solid line indicates the liquid ($l$) and the dashed line the vapor ($v$). Moreover, for the $pTg$-relaxed case (blue) thinner lines indicates regions where one phase is
 almost absent. 
 The exact solution
of the 5-equation $p$-relaxed model (solid black line)  is
also plotted.}
\label{fig:dodest}
\end{figure}

We consider here a dodecane liquid-vapor shock tube problem also proposed in \cite{saurel-PA},
which we solved in \cite{pelanti-shyue} with the stiffened gas equation of state. This test involves a unit length
shock tube  with an initial discontinuity located at $x = 0.75$~m that separates a left region filled with liquid dodecane and a
right region filled with vapor dodecane. As in \cite{saurel-PA}, 
for numerical reasons each fluid region contains a small amount of the
phase that fills the region on the other side of the discontinuity ($\alpha = 10^{-8}$).
The initial condition consists of two constant states on the two sides of the discontinuity with pressure $p=10^8$~Pa on the left and $p=10^5$~Pa on the right.
The initial velocity is $u=0$, and the initial values of the vapor and liquid densities
are $\rho_{\rm vap} = 2 \,\,{\rm kg\cdot m^ 3}\,$ and 
 $\rho_{\rm liq} = 500 \,\,{\rm kg\cdot m^ 3}\,$, respectively.
The liquid and vapor phases of
dodecane are modeled by the NASG equation of state with 
the parameters given in Table~\ref{tab_dode}.
Figure~\ref{fig:dodest} shows numerical results at time $t = 473\,\,\mu{\rm s}$
obtained by employing our numerical model with and without 
heat and mass transfer effects. More precisely, we plot
 results for two different 
levels of relaxation: instantaneous mechanical relaxation ($p$-relax in the plots),
 and instantaneous full thermodynamical relaxation ($pTg$-relax). 
 Let us note  that for the latter case  
 thermo-chemical relaxation 
is activated only at interfaces, defined by  $\min(\alpha_{\rm vap}, \alpha_{\rm liq})>\epsilon$,
$\epsilon=10^{-4}$, and chemical relaxation is activated under the condition $T>T_{\rm sat}$.
We can observe for both cases with and without phase
transition that the solution consists of
a leftward going rarefaction wave, a rightward going
contact discontinuity, and a shock wave. 
Moreover, when thermal 
and chemical effects are activated liquid-vapor phase
change occurs, generating an additional evaporation wave between the rarefaction wave and the contact discontinuity. This
evaporation front produces a liquid-vapor saturation region at higher speed.  
Note that the left-going rarefaction occurs in a region of almost
pure liquid ($\alpha_{\rm vap}$ nearly zero) and,  as noted above,
 only mechanical relaxation is activated in this zone for any case ($p$- and $pTg$-relaxation), 
 hence the phases have different temperatures. Let us also remark that in this left zone the vapor temperature has no physical
meaning, and since the value that it reaches is
unphysically high, of the order of $10^6$~K, we have plotted the temperatures in Figure~\ref{fig:dodest}
only over the  physical range  with a maximum temperature value of 1200~K, to be able
to observe the relevant liquid and vapor temperature curves.  
For the $pTg$-relaxation case the liquid-vapor transition leads  to  complete evaporation
(handled with the technique illustrated in Section~\ref{sec:chemrelax}, see (\ref{eq:chemrelaxpure})),
and on the right end of the interval there is a region of nearly pure vapor.
Again, in this region of nearly pure vapor ($\alpha_{\rm liq}$ nearly zero) only mechanical relaxation is activated.
We can notice from the temperatures plot in Figure~\ref{fig:dodest} the small liquid-vapor
 mixture zone where thermo-chemical relaxation is activated and where thermal
 and chemical equilibrium is imposed. This corresponds to the zone where the vapor and liquid
 temperature curves of the $pTg$-relaxed computation overlap
 around $\approx x=0.83$~m   (solid and dashed blue lines). 
In the plots we also display the exact solution of the five-equation pressure equilibrium 
model for this problem, and we observe good agreement with this solution of
our results with mechanical relaxation only.

 \begin{table}[h!]
\caption{Parameters for the NASG EOS for liquid and vapor dodecane in the temperature range 400-600 K}
\label{tab_dode}
\centering
\begin{tabular}{|c |c |c |c |c|c|c|}
\hline
    phase  & $\gamma$ & $\varpi$ [Pa] & $\eta$ [J/kg]  & $\etapp \,\, {\rm[J/(Kg \cdot K)]}$ & 
    $\kappa_v\,\,{\rm [J/(Kg \cdot K )]}$ & $b\,\,{\rm [m^3/kg]}$\\ \hline 
   liquid & $1.206$ & $1681 \times 10^{5}$ 
      & $-996054$ & 0 & $2532$ & $7.51 \times 10^{-4}$ \\ 
   vapor &  $1.021$ & $0$ & $-384592$ & $-4301$   
      &  $2274$ & 0 \\ \hline
\end{tabular}
\end{table}

\subsubsection{Bart\'ak's depressurization experiment}
 
We simulate here the Bart\'ak's blowdown experiment presented in \cite{bartak}.
This laboratory experiment consists in the rapid depressurization of a pressurized water
pipe initiated by a disk rupture. The study of this type of blowdown experiment
is relevant in particular in the context of the hazard assessment of
water-cooled reactors of nuclear power plants. 
One characteristic feature of the flow in this test is the rapid fall of the pressure 
  to a value
below the saturation pressure, so that for a certain time there is metastable superheated liquid,
before vaporization starts. As we observe numerically, to simulate this problem
it is important to be able to model non-instantaneous mass transfer processes.    
Initially in the tube there is liquid water at a pressure $p=12.5 \times 10^6$~Pa
and at temperature $T= T_1=T_2 =563.15$~K.    
There is an initial uniformly distributed small amount of vapor in the tube,
with volume fraction $\alpha_{\rm vap}=10^{-3}$. The tube has a length of $1700$~mm,
and it is permanently closed on one side, here the right side. On the other side,
here the left side, the tube is suddenly opened, hence we consider atmospheric
pressure conditions at the left boundary, with $p=10^5$~Pa.
We use the NASG equation of state for water with the parameters in Table~\ref{tab_water2}.

\begin{figure}[!ht]
\centering

\includegraphics[height=5.6cm]{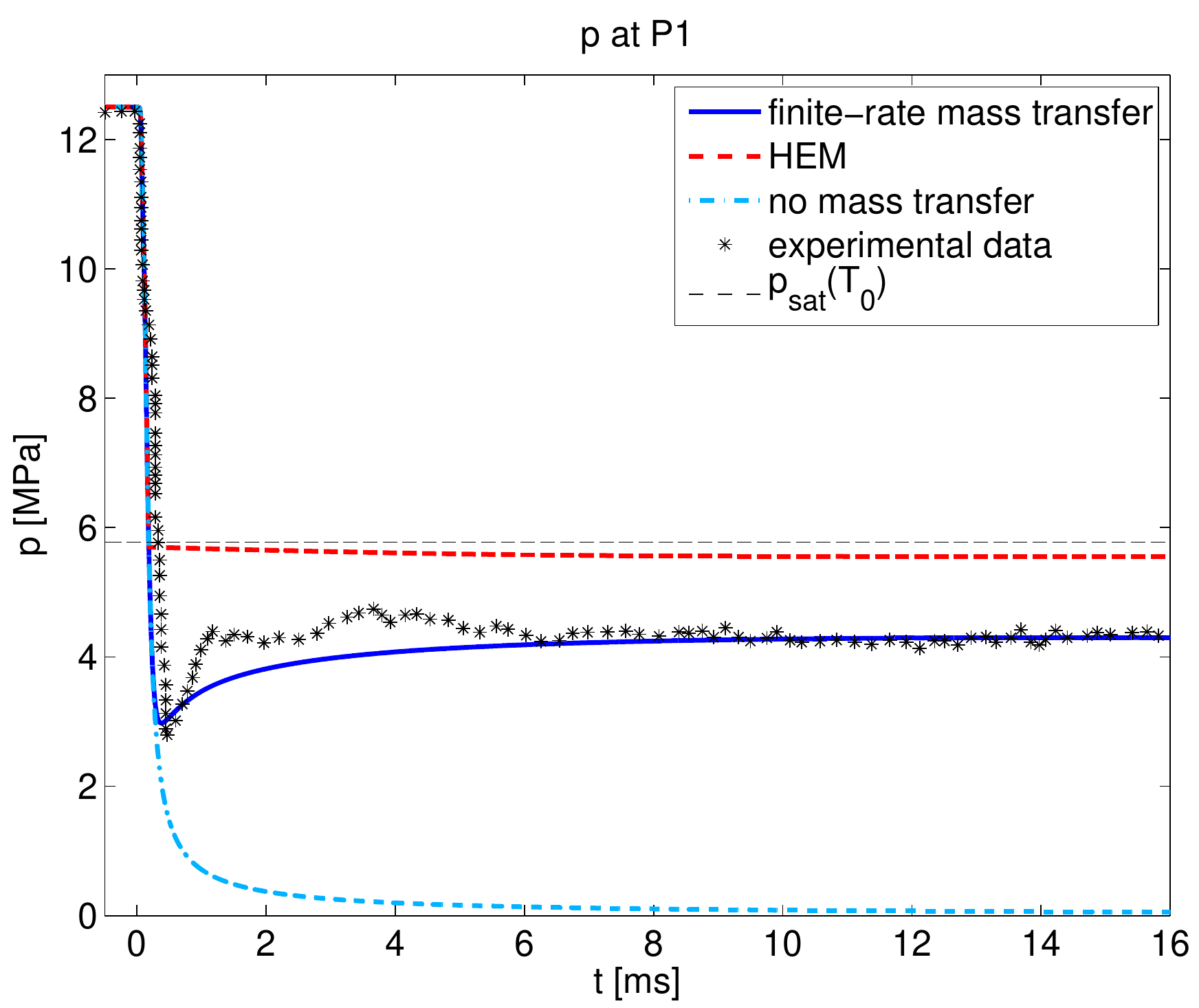}
\includegraphics[height=5.6cm]{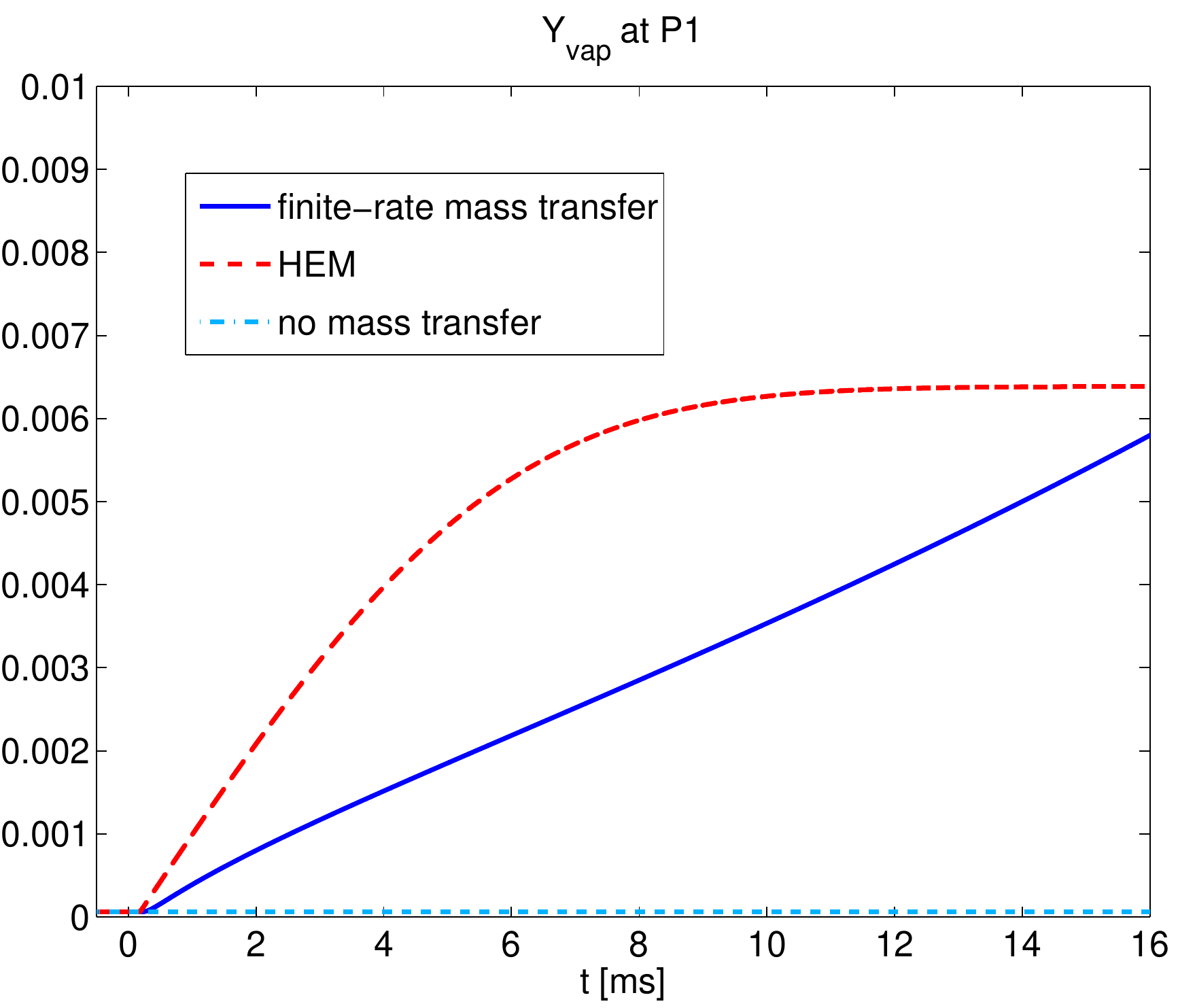}
\caption{Bart\'ak depressurization experiment. Computed results for the pressure history (left) 
and vapor mass fraction history (right) at $x=48$~mm (location of the first pressure gauge P1 in the
experimental apparatus), 
and comparison with the experimental
results of \cite{bartak}. Solid blue line: finite-rate mass transfer ($\nu$ defined in (\ref{eq:nubart}));
dash-dot light blue line: no mass transfer; dashed red line: instantaneous mass transfer ($\nu\rightarrow \infty$), correspondending to the solution of the HEM model. The value corresponding to the saturation pressure at the
initial temperature $T_0 =563.15$~K is also indicated in the left plot (dashed fine black line).} 
\label{fig:bartak}
\end{figure}

We compute solutions for this test with 1000 grid cells until a final time $t=17.5$~ms
(note that in this test we study the very first stage of this type of transient flow). 
In Figure~\ref{fig:bartak}
we  plot results for the pressure history (left) and the vapor mass fraction history (right)
at a fixed location corresponding to the first pressure gauge of the experimental apparatus,
at $x=48$~mm. The
solid dark blue line represents results obtained by activating instantaneous heat transfer
($\vartheta \rightarrow \infty$) 
and finite-rate mass transfer with the mass transfer relaxation function $\nu$
expressed by the following relation, which is a modified version of the relations
presented in \cite{HRMmodel,delor-laf-pelanti-IJMF}:
\begin{equation}
\label{eq:nubart}
\nu= C_r \alpha_2^{0.6}\!\left(
\frac{p_{\rm sat} -p}{p_{\rm crit}-p_{\rm sat}}\right)^{\!1.76}\!,
\quad C_r=1\,\,{\rm Pa\cdot kg^2/(s \cdot J^2)}.
\end{equation} 
We observe the qualitative agreement of these results with the experimental data (black marks~$*$),
and in particular the ability of the numerical model to predict the occurrence of
a metastable superheated state with $p<p_{\rm sat}(T)$ (region of the pressure undershoot), followed by vaporization.
In Figure~\ref{fig:bartak} we also plot results computed with no mass transfer 
($\vartheta \rightarrow \infty$, $\nu=0$, dashed light blue line)
and results computed with  instantaneous heat and mass transfer 
($\vartheta\rightarrow \infty$, $\nu \rightarrow \infty$,
solid red line), this corresponding to the solution of the homogeneous equilibrium model (HEM)
($pTg$-relaxed model (\ref{eq:pTgHEM})). 
We notice in particular that the activation of instantaneous mass transfer
does not allow the description of metastable states.

\begin{table}[h!]
\caption{Parameters for the NASG EOS for liquid and vapor water in the temperature range 350-550 K}
\label{tab_water2}
\centering
\begin{tabular}{|c |c |c |c |c|c|c|}
\hline
    phase  & $\gamma$ & $\varpi$ [Pa] & $\eta$ [J/kg]  & $\etapp \,\, {\rm[J/(Kg \cdot K)]}$ & 
    $\kappa_v\,\,{\rm [J/(Kg \cdot K )]}$ & $b\,\,{\rm [m^3/kg]}$\\ \hline 
   liquid & $1.387$ & $8899 \times 10^{5}$ 
      & -1244191$$ & 0 & $3202$ & $4.78 \times 10^{-4}$ \\ 
   vapor &  $1.954$ & $0$ & $2287484$ & $6417$   
      &  $462$ & 0 \\ \hline
\end{tabular}
\end{table}


\subsubsection{High-pressure fuel injector}

\label{sec:injector}

\begin{figure}
\centering

\includegraphics[height=3.6cm]{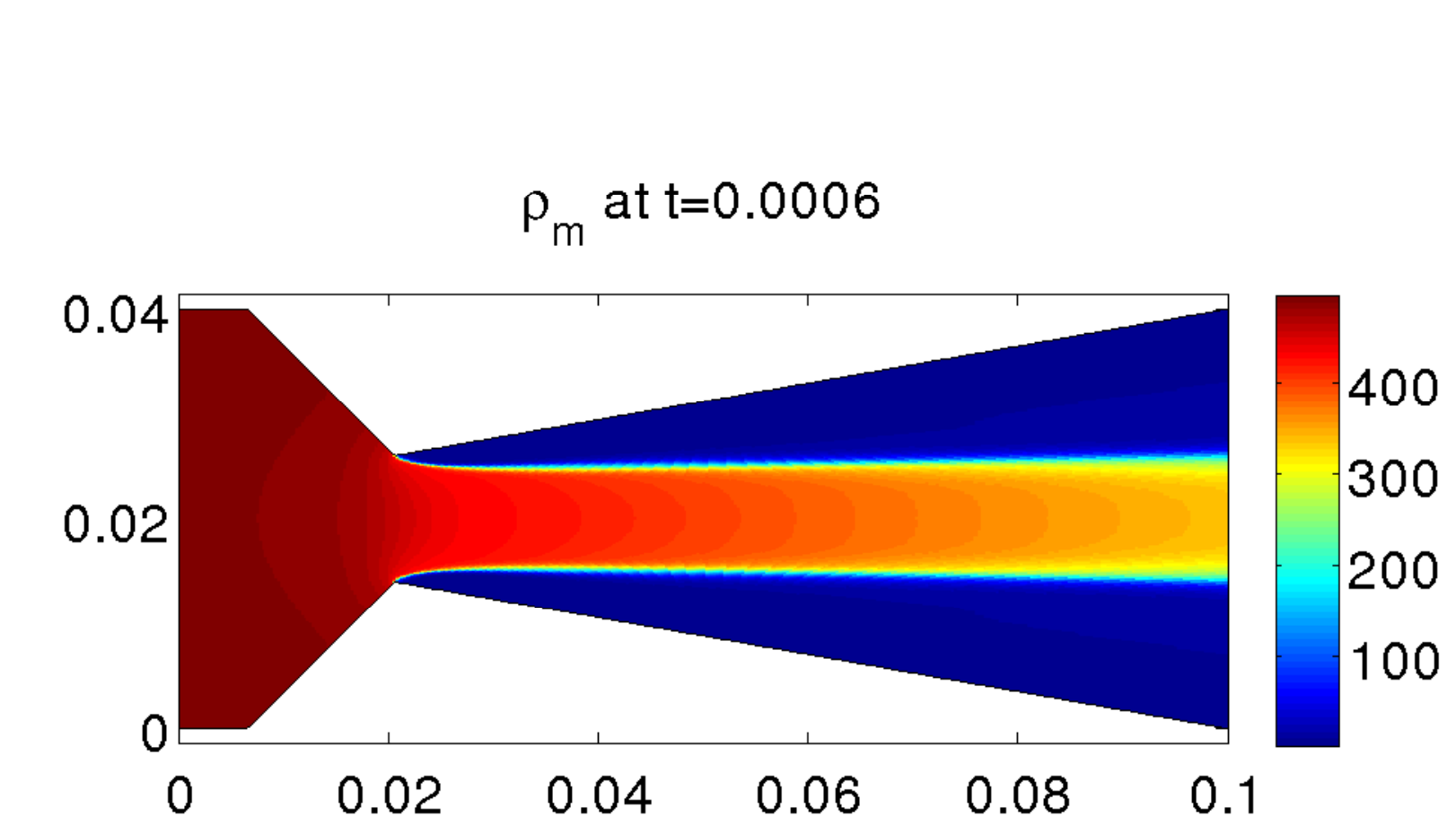}
\includegraphics[height=3.6cm]{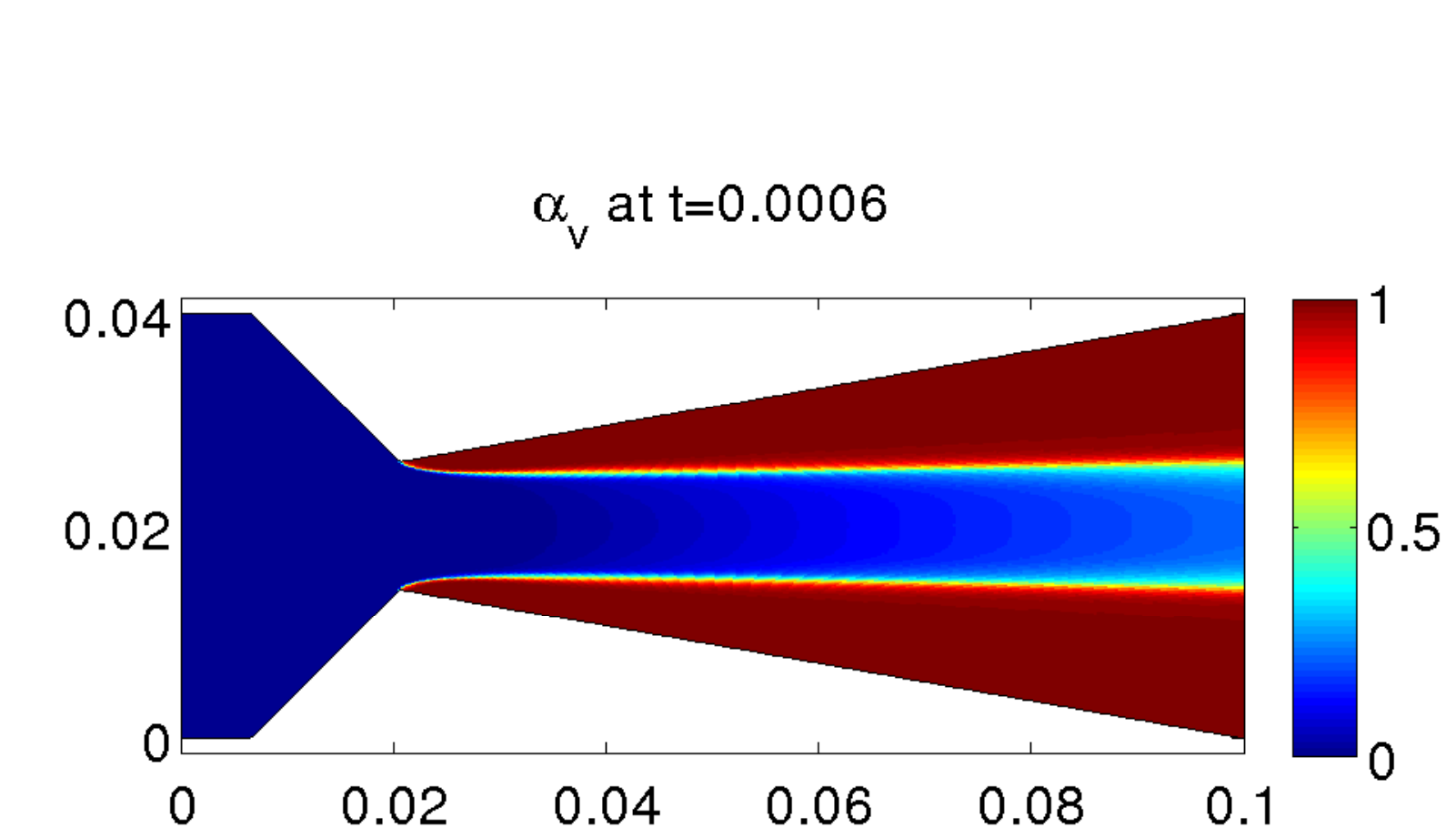}

\includegraphics[height=3.6cm]{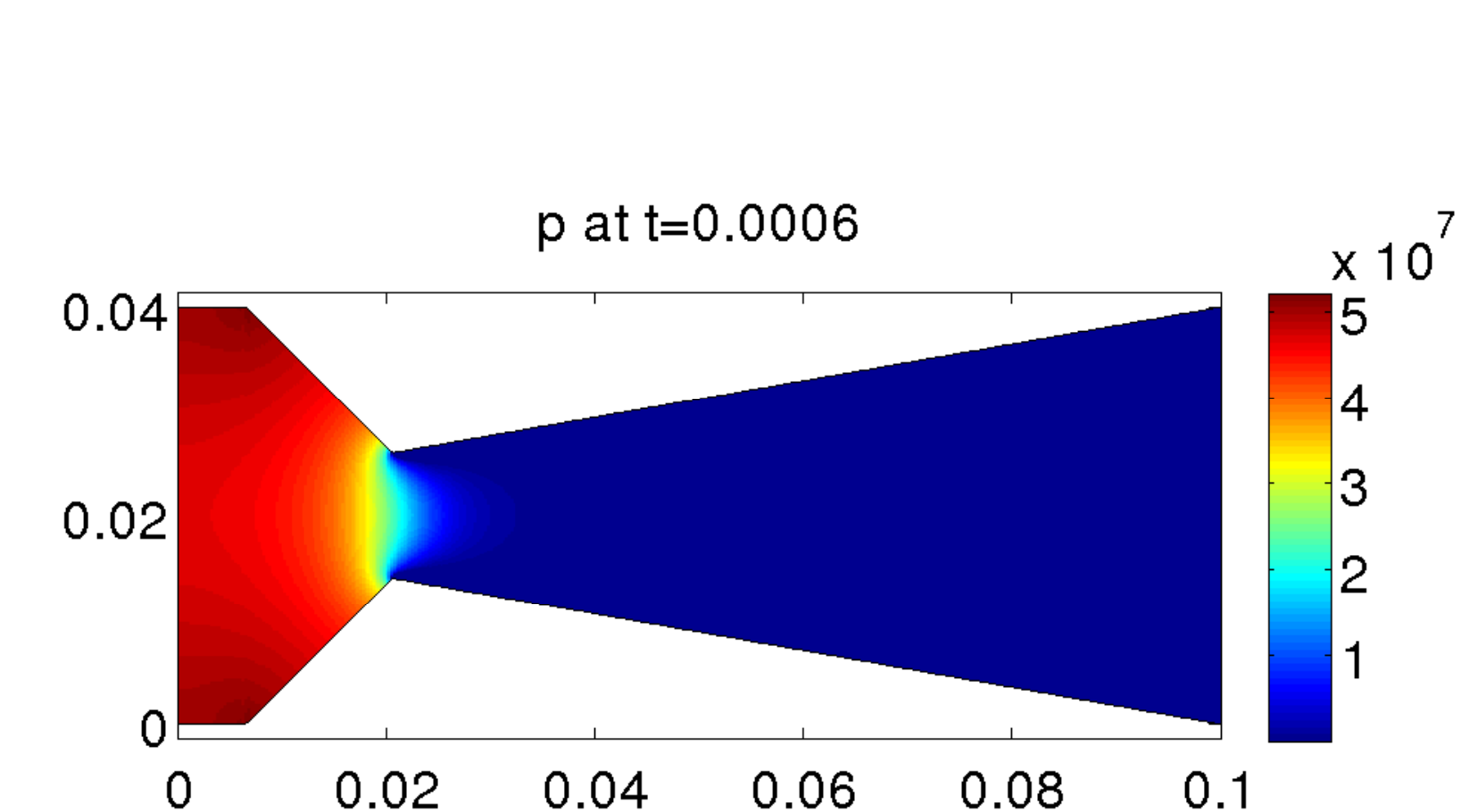}
\includegraphics[height=3.6cm]{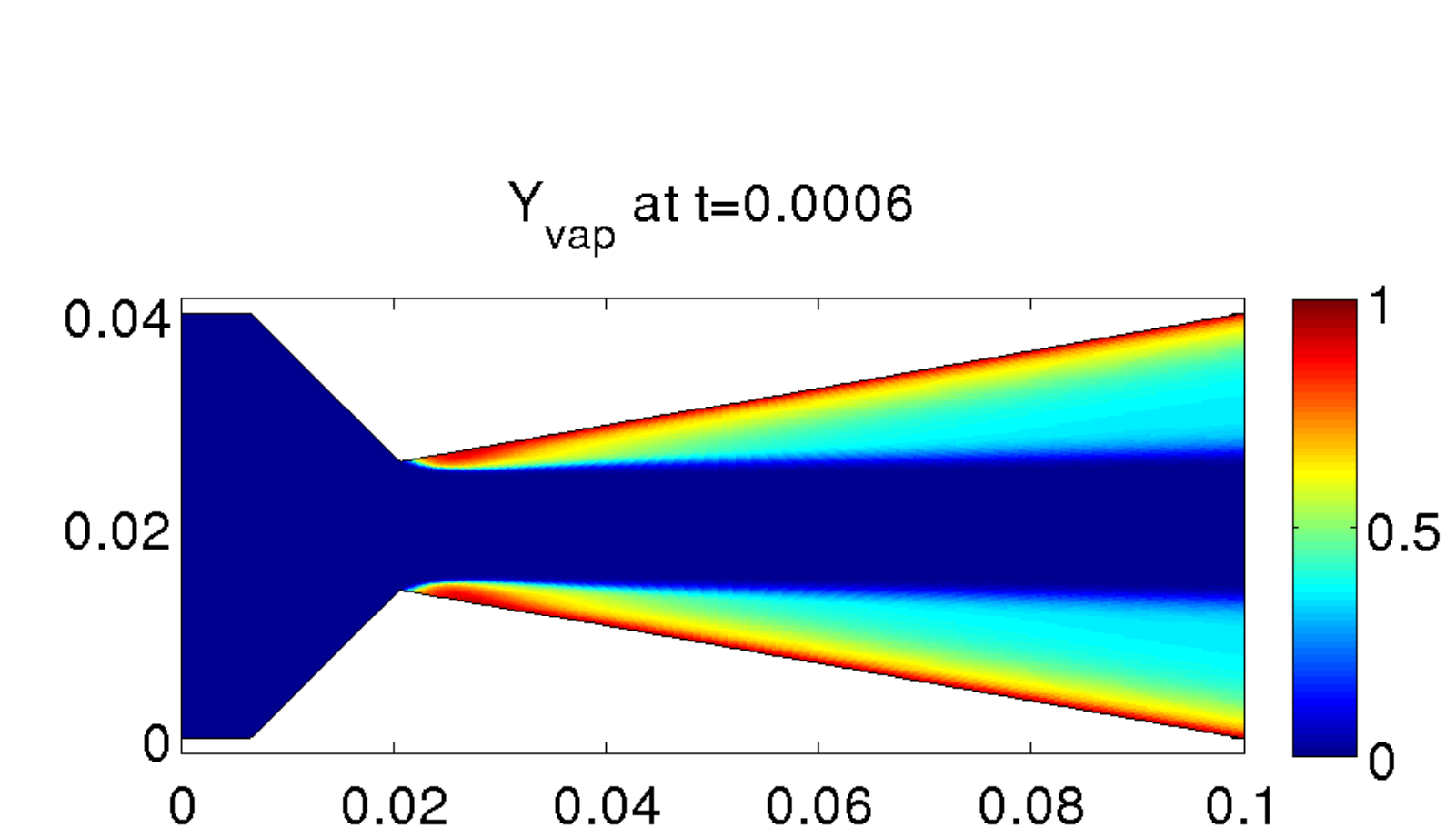}

\includegraphics[height=3.6cm]{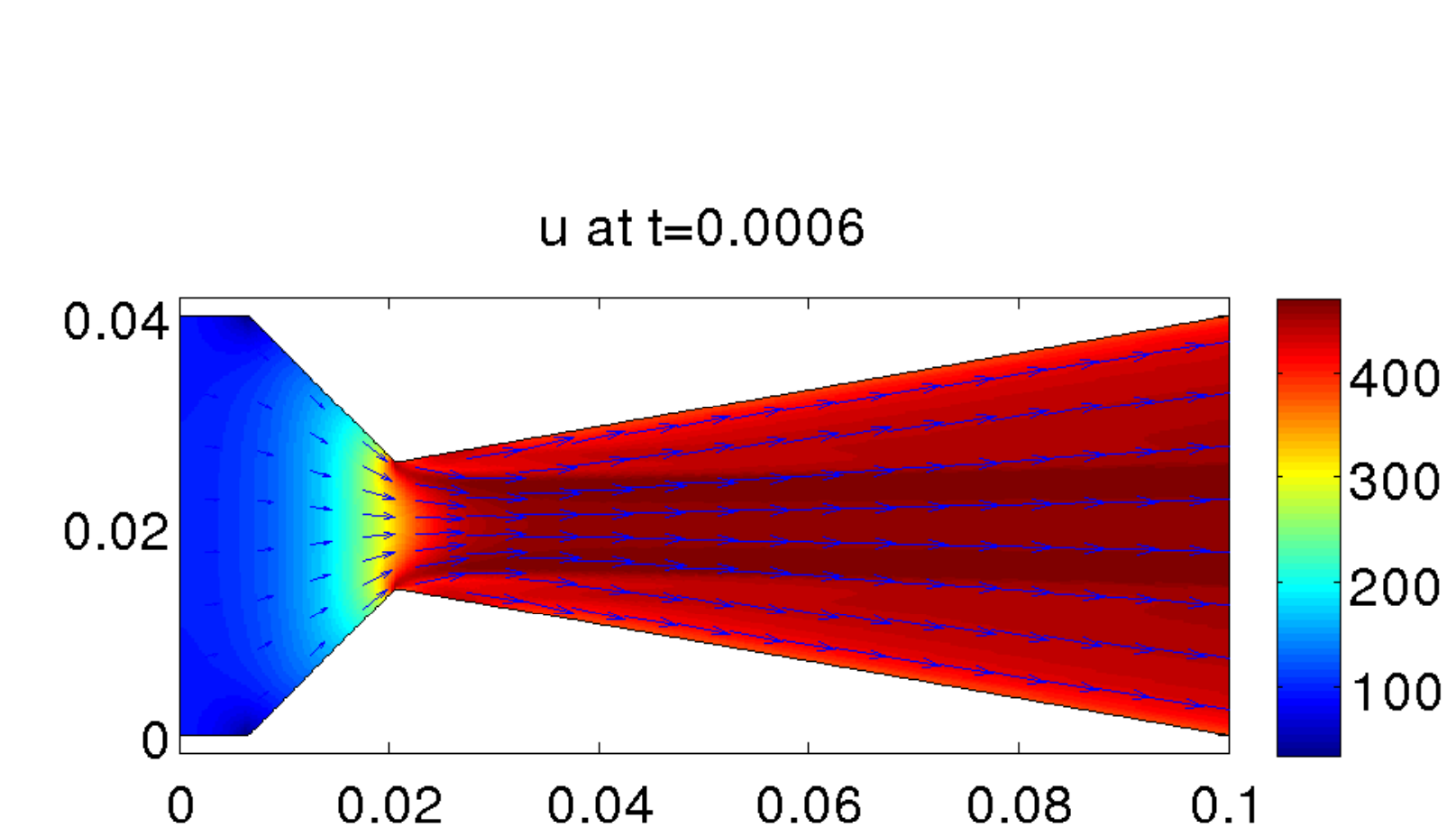}
\includegraphics[height=3.6cm]{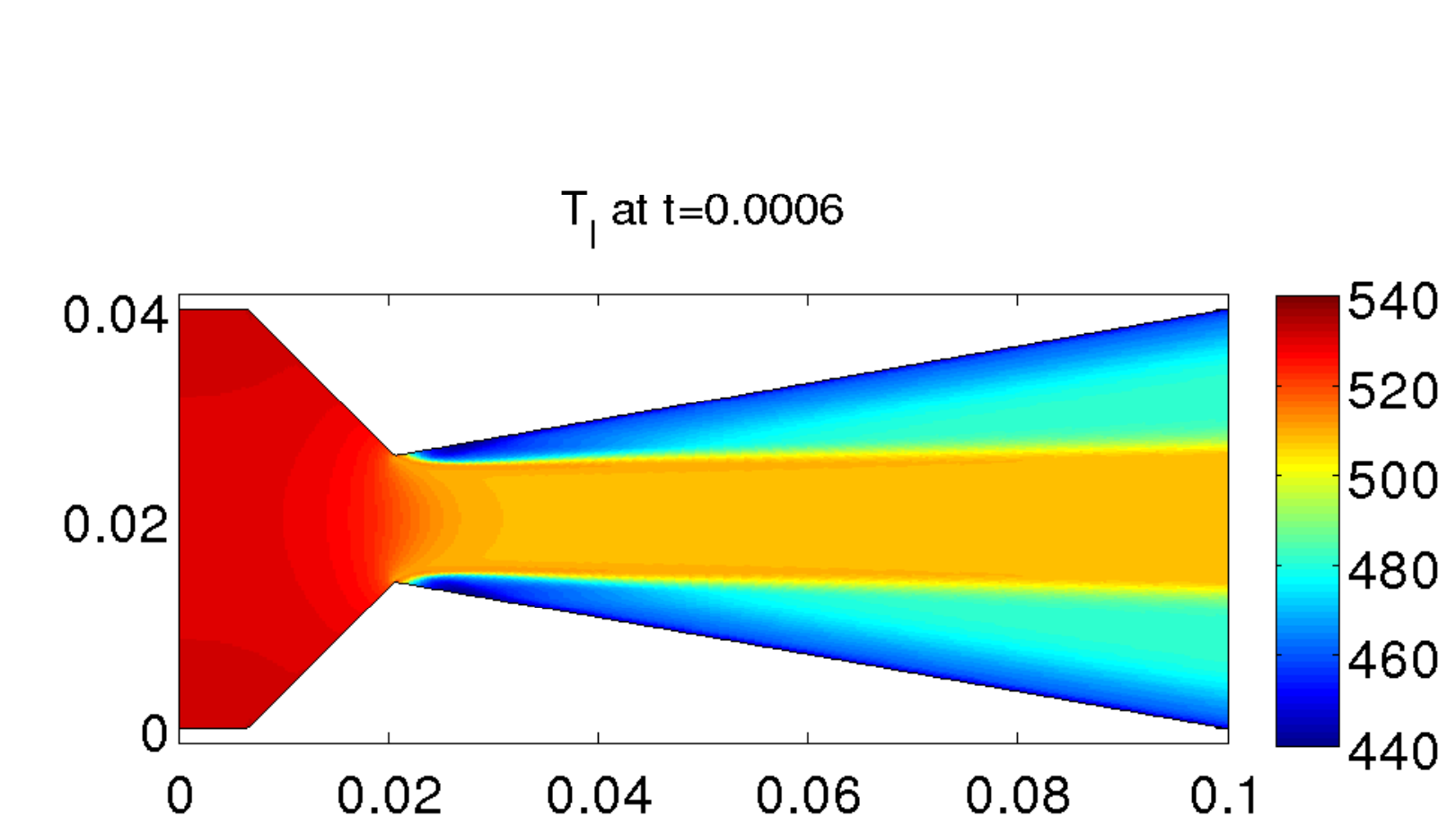}




\caption{High-pressure fuel injector experiment. Computed results with instantaneous mass transfer for the
density $\rho$,
the vapor volume fraction $\alpha_{\rm v}$,  the pressure $p$, the vapor mass fraction $Y_{\rm vap}$, the velocity field
$u$,
and the liquid temperature $T_{\rm l}$ at time $t=0.0006\,{\rm s}$ using a $400\times 160$ grid.}
\label{fig:fueldode1}
\end{figure}

\begin{figure}

\centering

{\small $\nu=0$} 

\includegraphics[height=2.4cm,viewport=10 0 460 250,clip=]{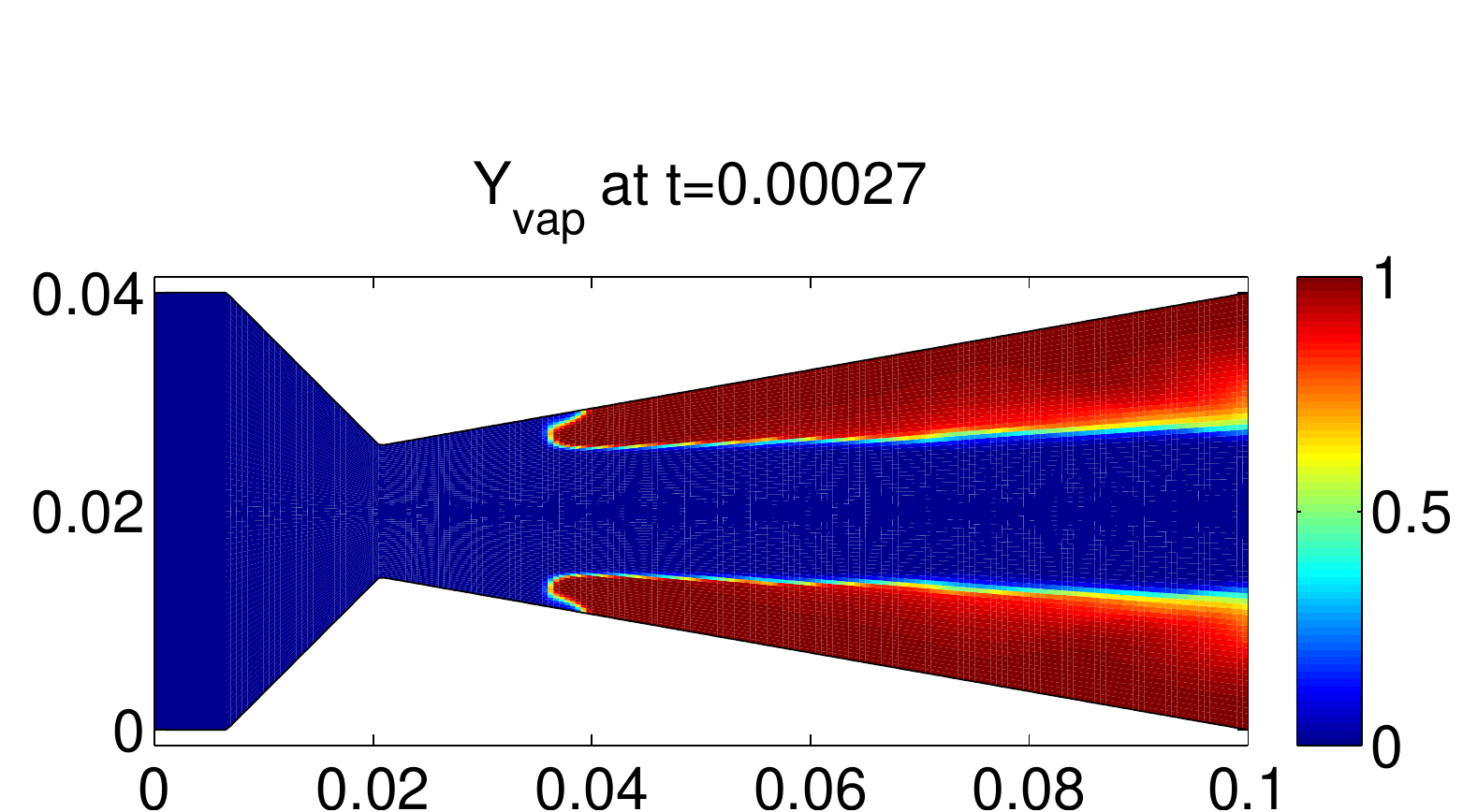}
\includegraphics[height=2.4cm,viewport=10 0 460 250,clip=]{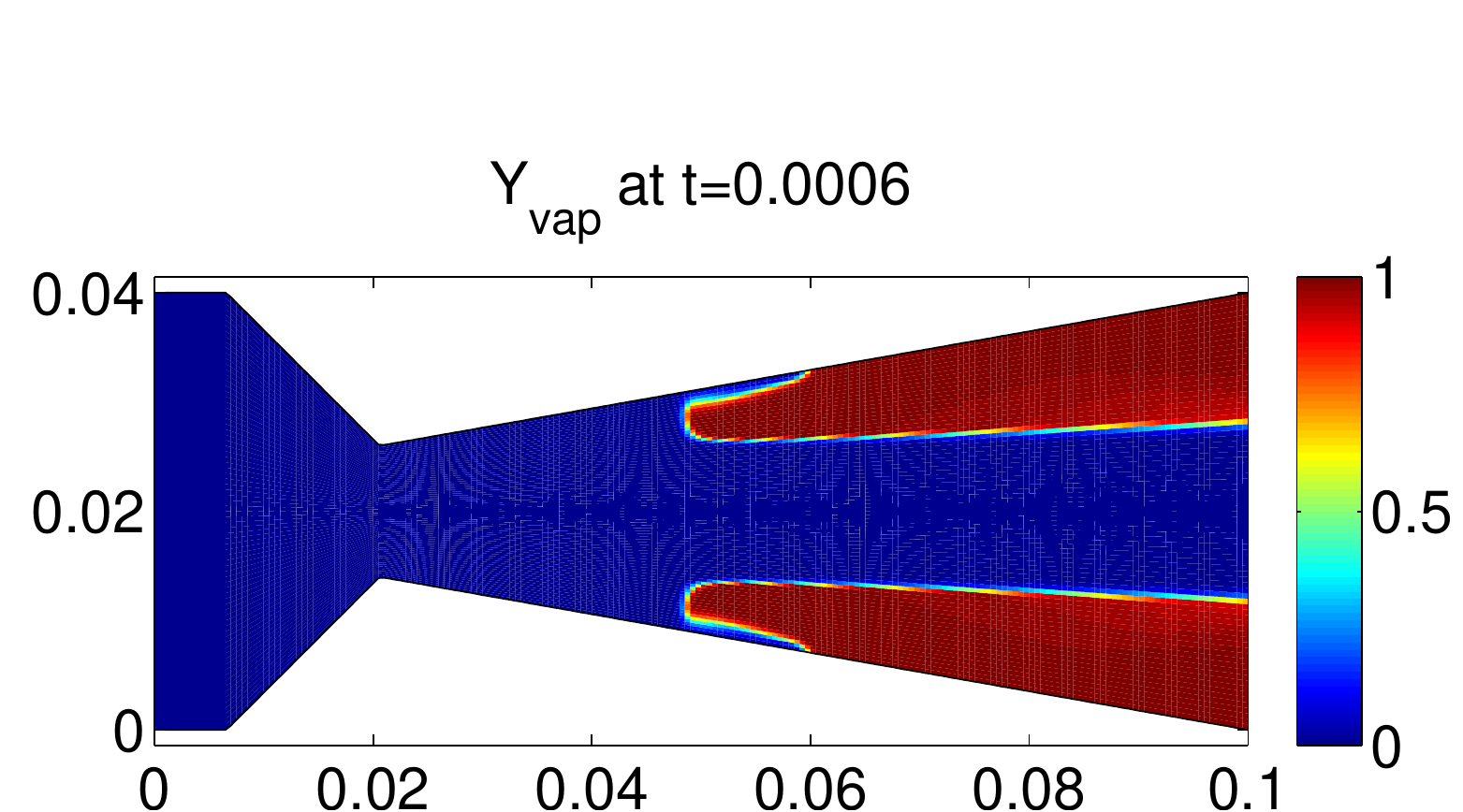}
\includegraphics[height=2.4cm,viewport=10 0 460 250,clip=]{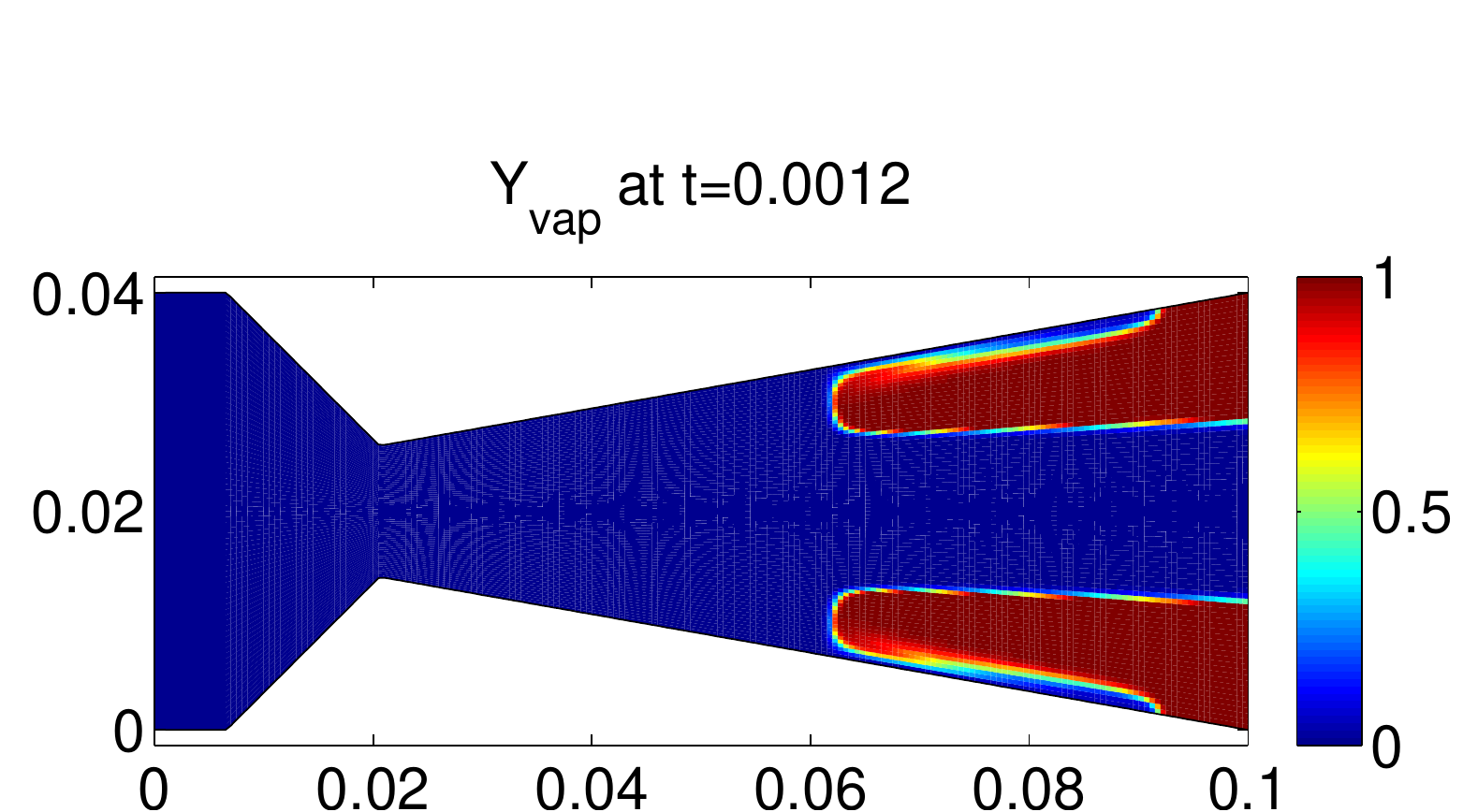}

\vspace{5mm}

{\small $\nu=0.1$}

\includegraphics[height=2.4cm,viewport=10 0 460 250,clip=]{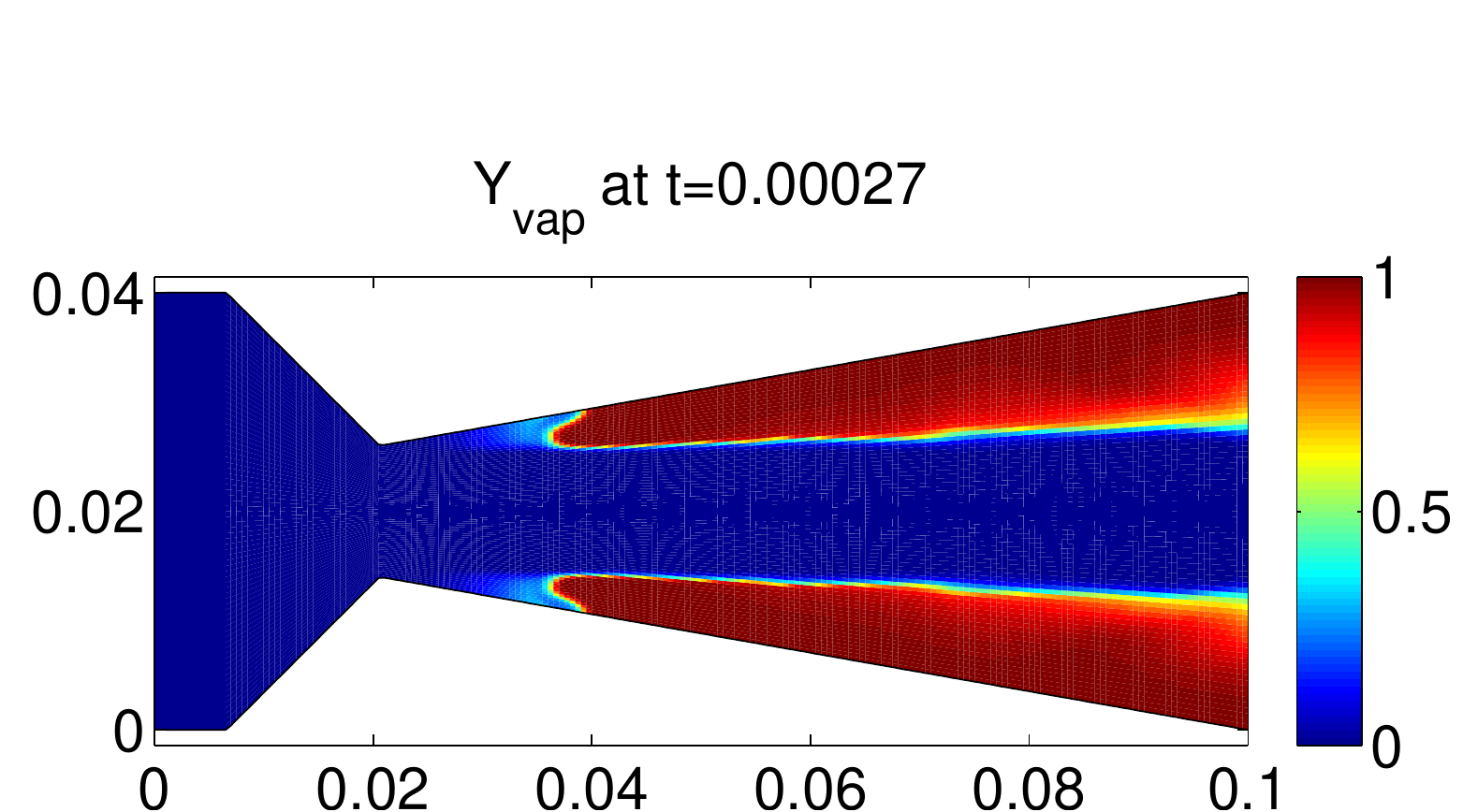}
\includegraphics[height=2.4cm,viewport=10 0 460 250,clip=]{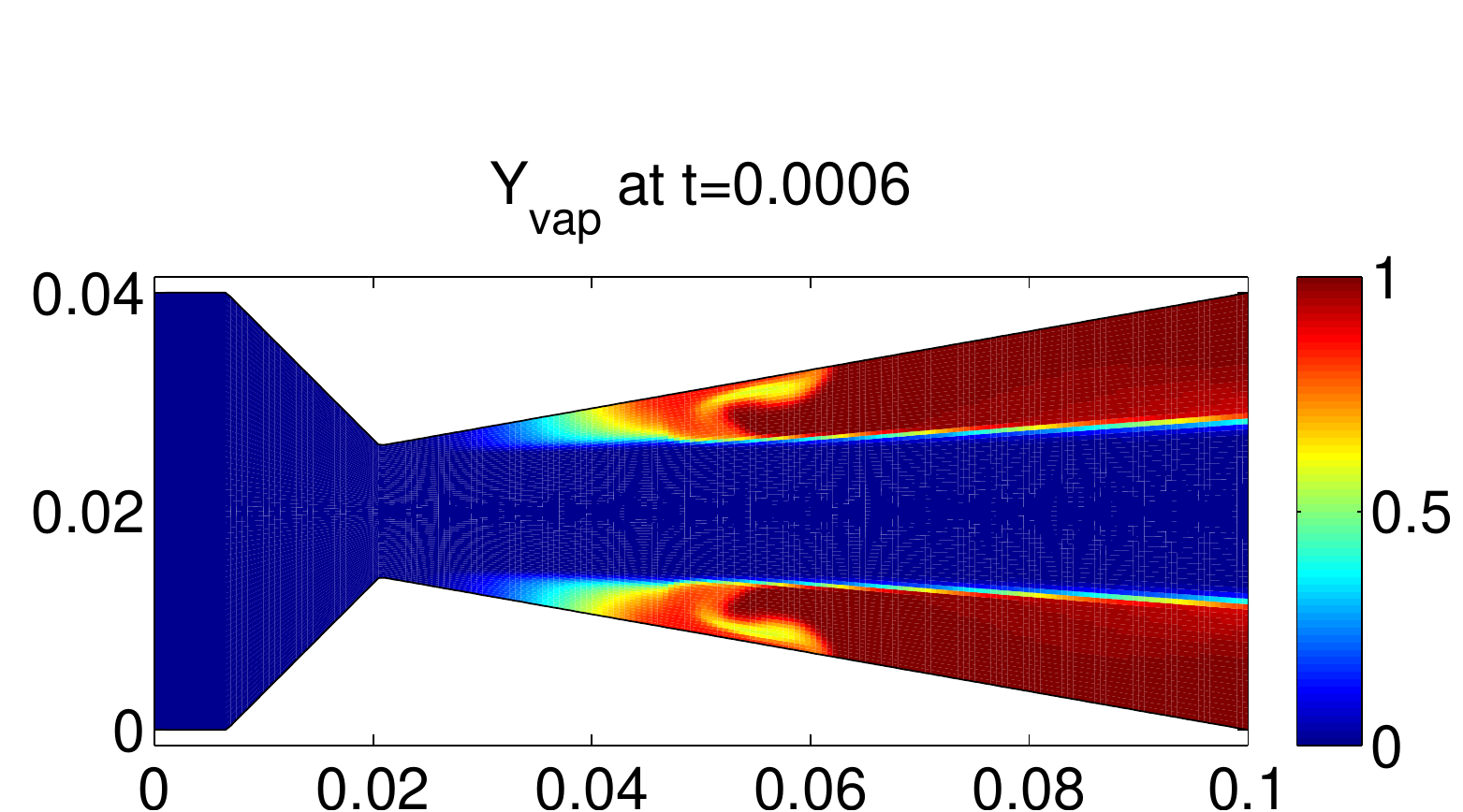}
\includegraphics[height=2.4cm,viewport=10 0 460 250,clip=]{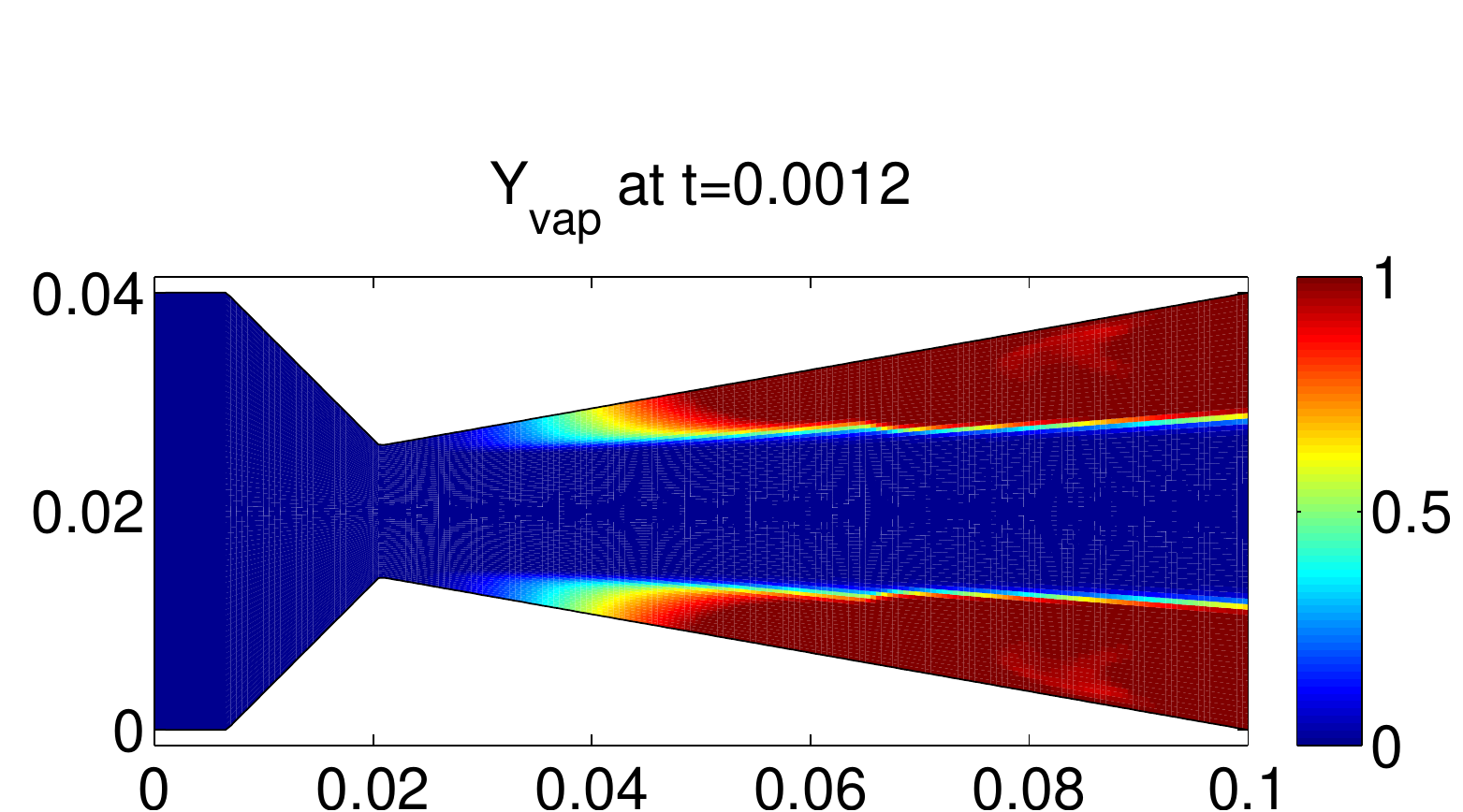}

\vspace{5mm}

{\small $\nu=25$}

\includegraphics[height=2.4cm,viewport=10 0 460 250,clip=]{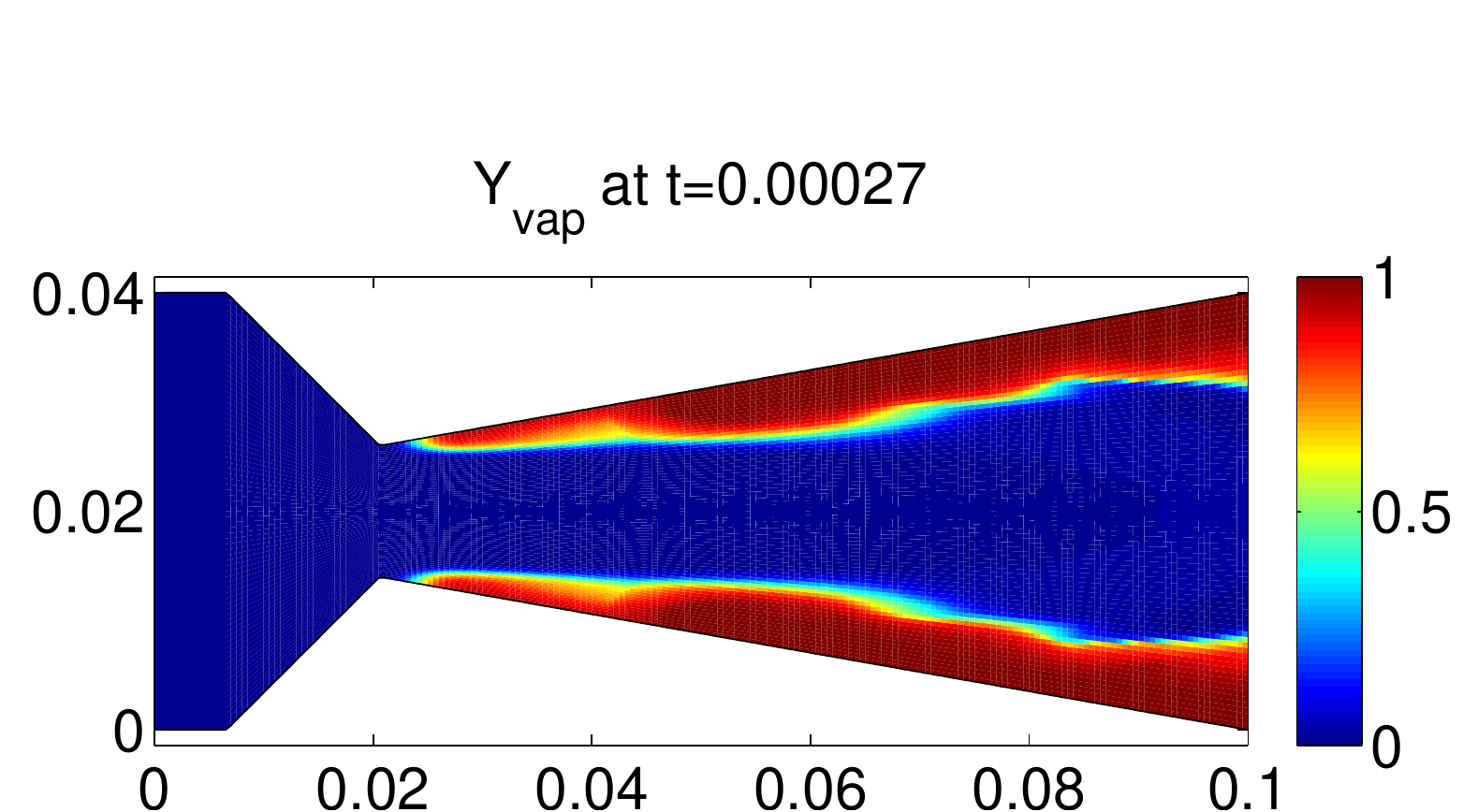}
\includegraphics[height=2.4cm,viewport=10 0 460 250,clip=]{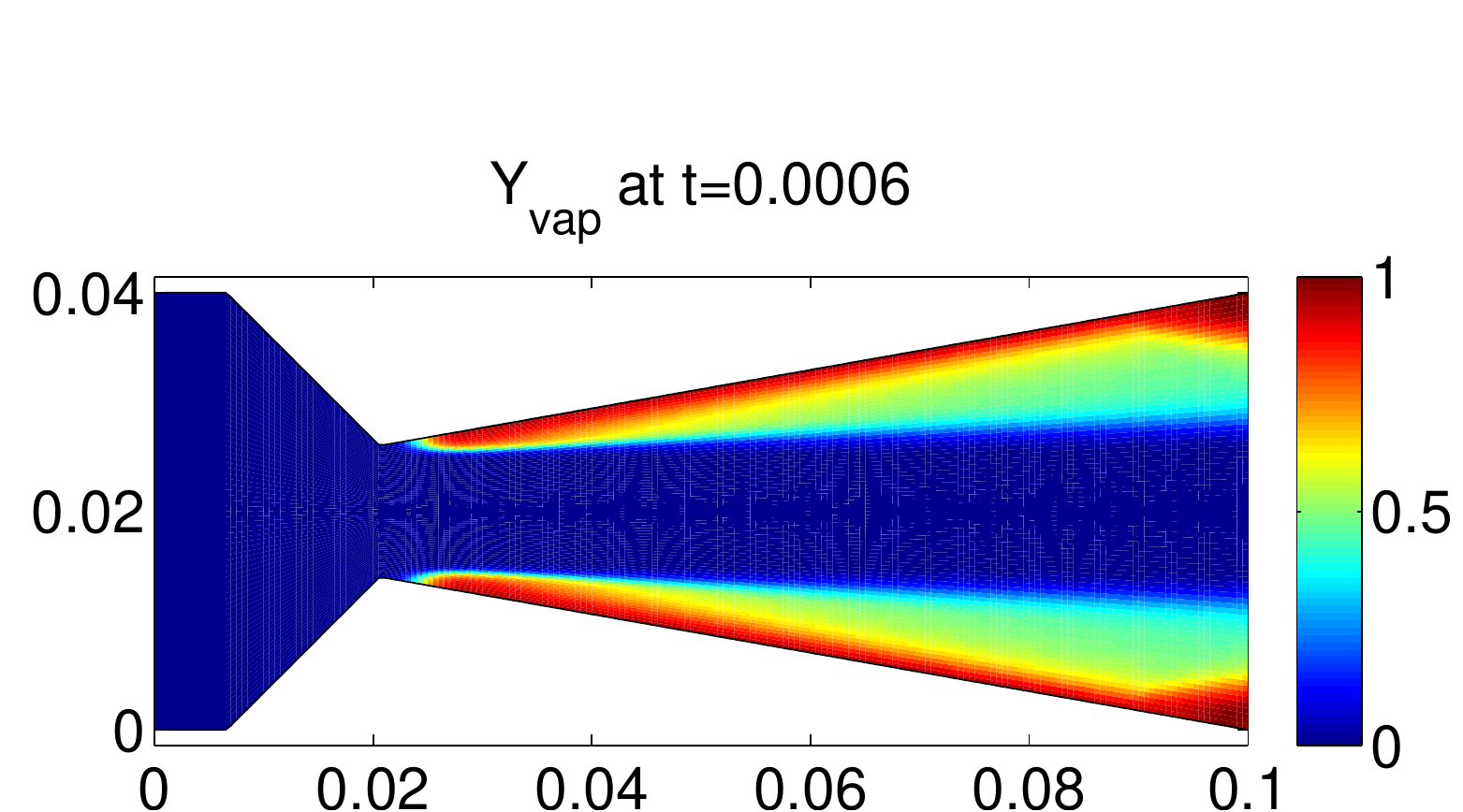}
\includegraphics[height=2.4cm,viewport=10 0 460 250,clip=]{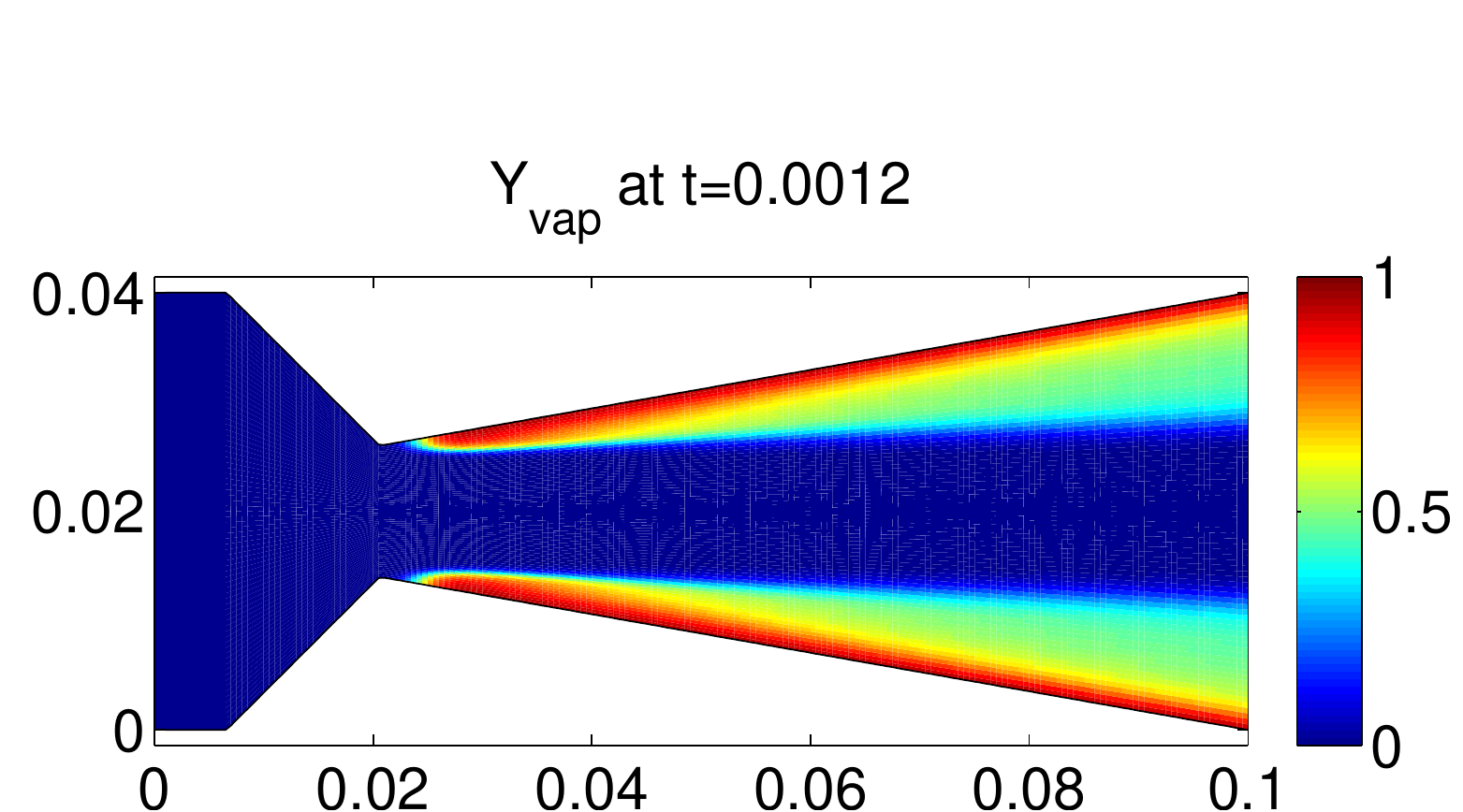}

\vspace{5mm}

{\small $\nu\rightarrow \infty$}

\includegraphics[height=2.4cm,viewport=10 0 460 250,clip=]{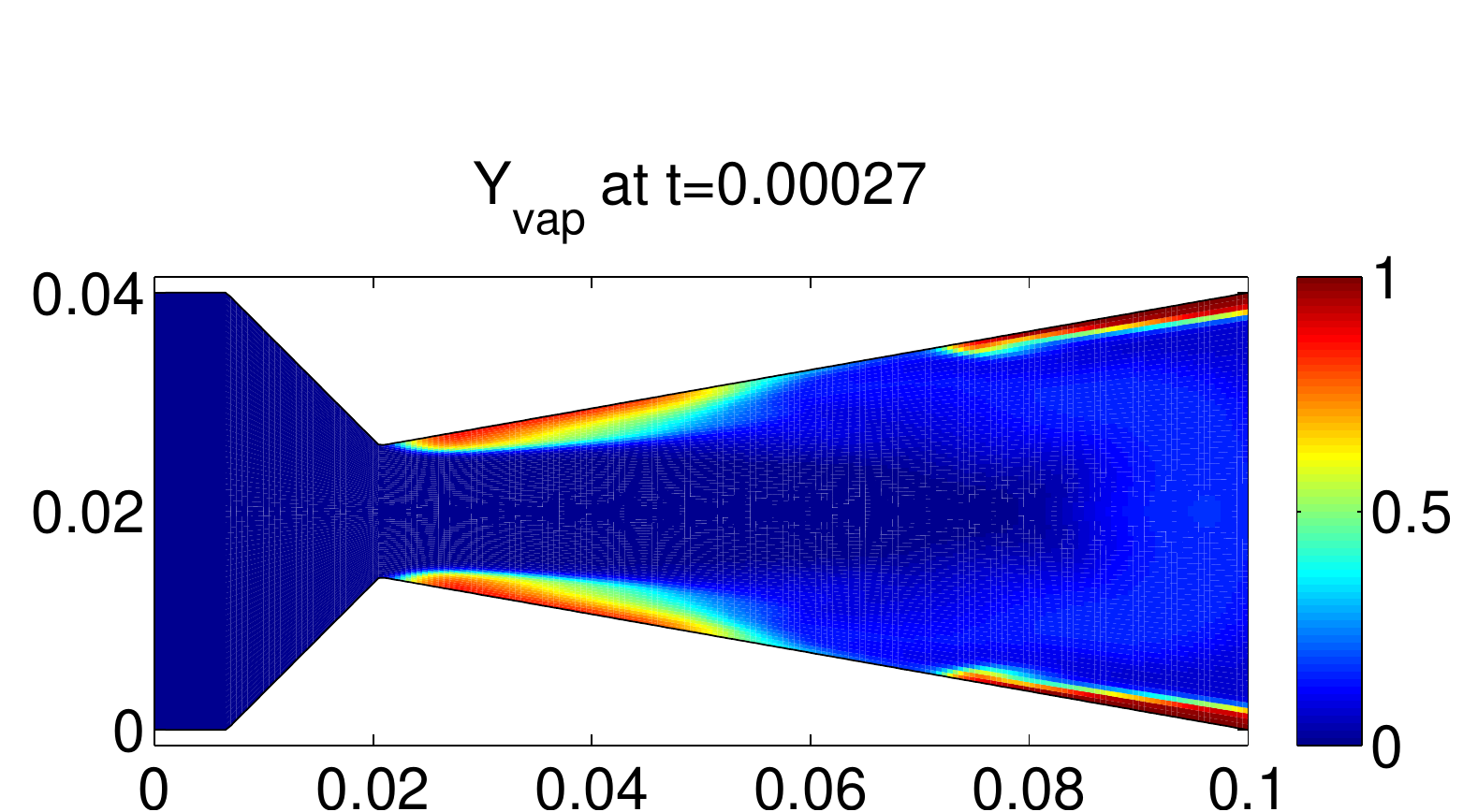}
\includegraphics[height=2.4cm,viewport=10 0 460 250,clip=]{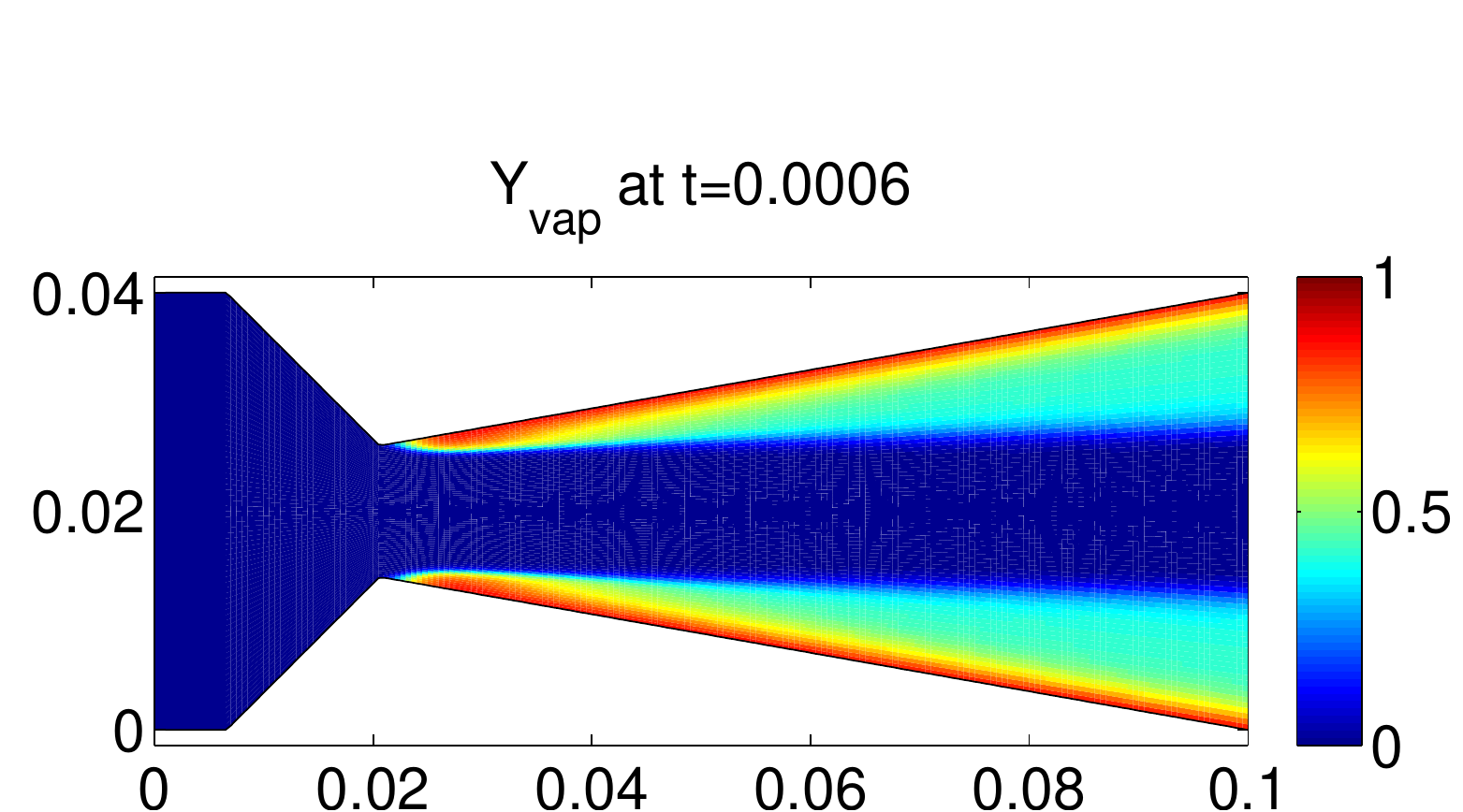}
\includegraphics[height=2.4cm,viewport=10 0 460 250,clip=]{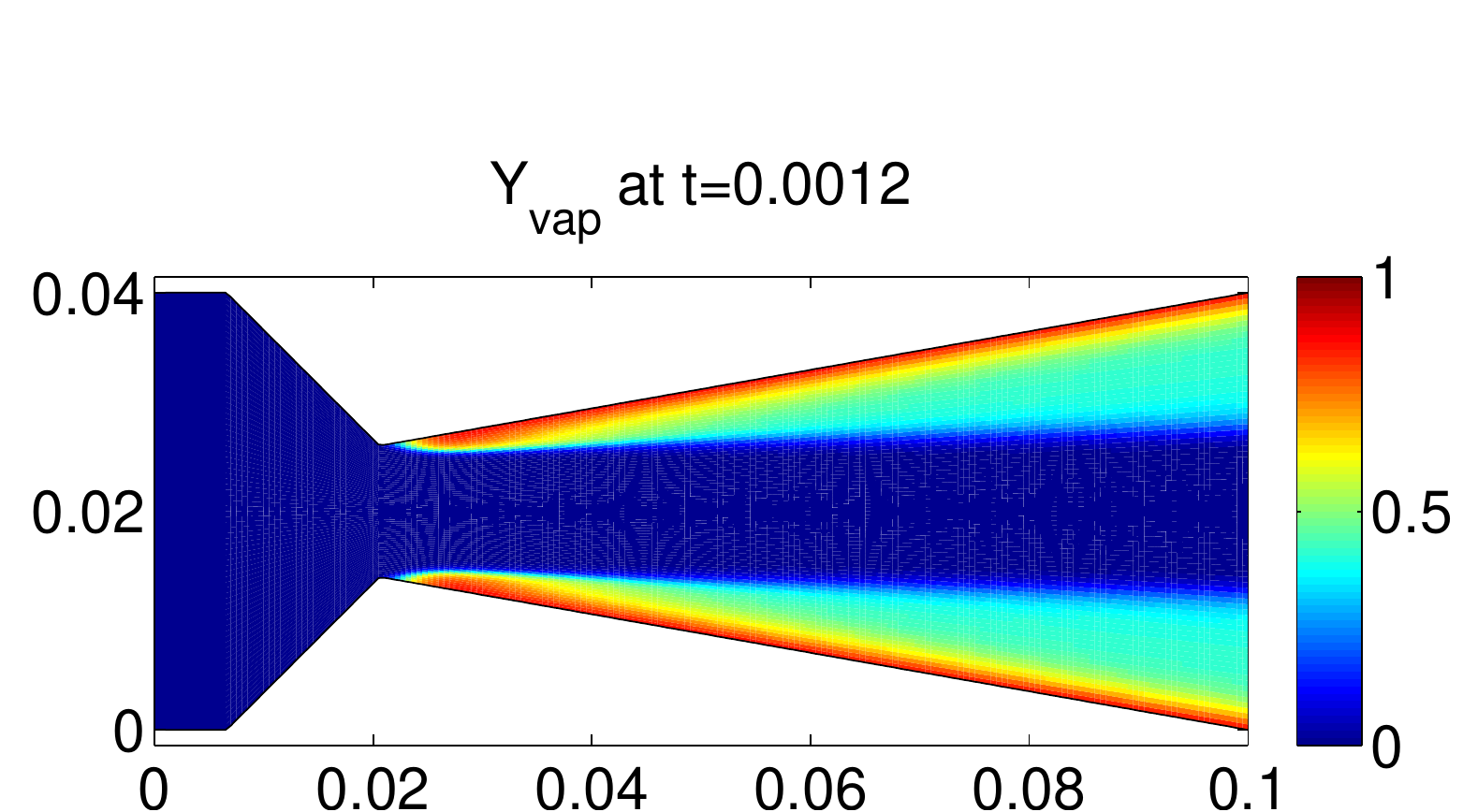}





\caption{High-pressure fuel injector experiment. Computed vapor mass fraction $Y_{\rm vap}$ at times
$t=0.00027,\,\,0.0006,\,\,0.0012\,{\rm s}$ (columns from left to right)
for $\nu=0$, $\nu=0.1$, $\nu=25$, $\nu\rightarrow \infty$ (rows from top to bottom), using a $200\times 80$ grid.}
\label{fig:fueldode2}
\end{figure}

Finally, we simulate a two-dimensional fuel injector. This test is also a variant of a test proposed
in \cite{saurel-PA}, which we solved in \cite{pelanti-shyue} with the stiffened gas equation of state
and instantaneous relaxation processes. 
We consider a nozzle
where liquid fuel (dodecane) is injected from a high-pressure tank to a chamber at atmospheric pressure. 
The nozzle has the shape shown in the plots of Figures~\ref{fig:fueldode1}-\ref{fig:fueldode2}, and it has a length 
of $10$~cm and a height of $4$~cm. The height of the throat is $1.2$~cm, and the
outer inclination angles of the converging and diverging chambers with respect to the horizontal direction are 
$45^\circ$ and $10^\circ$,
respectively. We set an initial discontinuity at $x = 0.8$~cm between 
a region of liquid dodecane at a pressure $p=10^8$~Pa~
and at temperature  $T=550$~K and a region of dodecane vapor at 
pressure $p=10^5$ and with phasic density $\rho_{\rm vap} = 5\,\,{\rm kg \cdot m^{-3}}$. 
At the initial time, a small amount of vapor  is present in the liquid 
with $\alpha_{\rm vap} =10^{-4}$, and a small amount of liquid is present
in the vapor with  $\alpha_{\rm liq} = 10^{-6}$.  
Phases are initially assumed in thermal equilibrium.
The dodecane liquid and vapor phases are modeled by the NASG EOS with
the parameters in Table~\ref{tab_dode}.
In this experiment we activate instantaneous heat and mass transfer
 at interfaces defined by
$\min(\alpha_{\rm vap},\alpha_{\rm liq}) > \epsilon$, with $\epsilon=0.9 \times 10^{-4}$.
We assume instantaneous thermal relaxation 
($\vartheta \rightarrow \infty$),
while we use  different  values of the chemical relaxation parameter $\nu$.
In Figure~\ref{fig:fueldode1} we plot results computed with instantaneous heat and mass transfer at a time
at which stationary conditions are approximately attained for the mixture density, 
the vapor volume fraction, the pressure, the vapor mass fraction, the velocity field,
and the liquid temperature. For these results we have used 400 $\times$ 160 grid cells
and CFL number = 0.4. In Figure~\ref{fig:fueldode2} we plot results at three different times 
for the vapor mass fraction computed by using four different values of the
chemical relaxation parameter $\nu$ [${\rm Pa\cdot kg^2/(s \cdot J^2)}$], $\nu=0$ (no mass transfer),
$\nu=0.1$, $\nu=25$, $\nu\rightarrow \infty$ (instantaneous mass transfer).
For the results in this Figure~\ref{fig:fueldode2} we have used $200 \times 80$ grid 
cells and CFL number = 0.4.
Note that we can compare results for the vapor mass fraction at $t=0.0006$~s obtained for
the test with instantaneous mass transfer with two different mesh sizes, see Figure~\ref{fig:fueldode1},
plot at the center-right, and Figure~\ref{fig:fueldode2}, plot at the center of the bottom row.
Overall the results of this numerical test show the capability of the numerical  model
to simulate mass transfer processes of arbitrary rate, from slow to very fast processes.

\section{Conclusions}

We have proposed new efficient numerical techniques to treat the mechanical, thermal, and chemical
relaxation source terms of the single-velocity
two-phase flow model (\ref{eq:2phasesys}) that we have presented in previous work \cite{pelanti-shyue}.
The new techniques are based on analytical semi-exact exponential solutions of the systems
of ordinary differential equations used to model the relaxation processes, and 
they have two significant features: the applicability to a general
equation of state, and the capability to describe
arbitrary-rate heat and mass transfer.
These relaxation procedures also ensure a mixture-energy-consistent scheme.
The procedures are simple, and for equations of state that can be written in the form
of  the Mie--Gr\"uneisen EOS they do not need iterative methods.
In general, for more complex equations of state, only the mixture energy equation (\ref{eq:mixencons2}) 
that defines implicitly the equilibrium mixture pressure might need
an iterative solution method.
The relaxation techniques for heat and mass transfer can robustly handle
both stiff instantaneous processes and non-stiff slow finite-rate relaxation processes.
This is an important improvement with respect to our previous work \cite{delor-laf-pelanti-JCP19,delor-laf-pelanti-NED}.
Let us also note that, by construction, the techniques that we have proposed for the six-equation model 
(\ref{eq:2phasesys}) can be used to treat relaxation terms
of the $p$-relaxed (\ref{eq:sisp}) and $pT$-relaxed (\ref{eq:sysT}) models, when one solves these models directly (see for instance the application to the $pT$-relaxed model in \cite{demou-boil}). 
Several numerical tests show the effectiveness of the new  relaxation techniques.
 We have first observed the good performance of the numerical model in problems with interfaces and strong
 shocks and complex equations of state. Then we have shown the ability of the methods to
 describe finite-rate mass transfer processes, which for instance is essential for predicting
  the occurrence of metastable superheated liquid in fast depressurization problems.
  The capability of approximating efficiently solutions to the $p$-relaxed and $pT$-relaxed models
  in the limit of instantaneous mechanical and thermal relaxation has been also  demonstrated numerically.

Concerning  future work, one objective is to couple the new relaxation
techniques with a  Table Look-up Method similar to the one that we have
developed in   \cite{delor-laf-pelanti-IJMF, delor-laf-pelanti-NED} to employ a very precise equation of state for water, 
the IAPWS Industrial Formulation 1997 for Water and Steam
\cite{iapws97}. Moreover, we plan to extend the 
proposed relaxation techniques to the three-phase flow model that we have presented in \cite{pelanti-shyue-3p},
in particular for applications to underwater explosion problems.

\section*{Acknowledgments}
 
The  author  was  supported by the French Government Directorate for Armament (Direction G\'en\'erale de l'Armement, DGA) 
under grant N.~2018.60.0071.00.470.75.01. 

\appendix

\section{Derivation of the $p$-relaxed   model}

\label{sec_redmod}

In this section we derive the $p$-relaxed model in (\ref{eq:sisp}) from the
two-phase model in (\ref{eq:2phasesys}). For simplicity, we shall consider the one-dimensional case $d=1$.
We follow in particular the technique of Murrone--Guillard \cite{gui-mur} to derive the 5-equation model 
from the seven-equation model (see also \cite{ch-le-liu}).
First, we write the system (\ref{eq:2phasesys}) in one dimension in terms of the vector of primitive variables
$w\in\mathbb{R}^{6}$ as:
\begin{subequations}
\label{eq:sysrel}
\begin{equation}
\dts w + A(w)\dxs w = \frac{1}{\tau}\Psi(w) + \Phi(w), 
\end{equation}
where $\tau\equiv \frac{\tilde{\mu}}{\mu}$ ($\tilde{\mu}$ is an arbitrary reference quantity
to non-dimensionalize $\mu$), and 
\begin{equation}
w=
\left[
\begin{array}{c}
\alpha_1\\
[1mm]
 \rho_1\\
 [1mm]
 \rho_2\\
 [1mm]
u\\
[1mm]
p_1\\
[1mm]
p_2
\end{array}
\right ],\,\,
A =\left[
\begin{array}{ccccccccccccc}
u & 0& 0 & 0   &0 & 0\\
[1mm]
0 & u & 0 & \rho_1  & 0 & 0 \\
[1mm]
0 & 0 &  u &   \rho_2  & 0 & 0\\
[1mm]
\frac{p_1-p_2}{\rho} &0 & 0 & u  & \frac{\alpha_1}{\rho} & \frac{\alpha_2}{\rho}\\
[1mm]
0 & 0 &  0& \rho_1c_1^2  & u & 0 \\
[1mm]
0 & 0 &  0&  \rho_2 c_2^2  & 0 & u 
\end{array}
\right],
\end{equation}
\begin{equation}
\label{eq:PsiPhi}
\Psi = \tilde{\mu}\left[
\begin{array}{c}
p_1-p_2\\
[1mm]
-\frac{\rho_1}{\alpha_1}(p_1-p_2)\\
[1mm]
\frac{\rho_2}{\alpha_2}(p_1-p_2)\\
[1mm]
0\\
[1mm]
-\frac{1}{\alpha_1}[\Gamma_1(\mathcal{E}_1  +\pint)+\chi_1 \rho_1](p_1-p_2)\\
[1mm] 
 \frac{1}{\alpha_2}[\Gamma_2(\mathcal{E}_2 +   \pint)+\chi_2 \rho_2](p_1-p_2)
\end{array}
\right],\,\,
\Phi = \left[
\begin{array}{c}
0\\
[1mm]
\frac{\mathcal{M}}{\alpha_1}\\
[1mm]
-\frac{\mathcal{M}}{\alpha_2}\\
[1mm]
0\\
[1mm]
\frac{\Gamma_1}{\alpha_1}\mathcal{Q}+\left(\Gamma_1\gint +\chi_1\right) \frac{\mM}{\alpha_1}\\
[1mm]
-\frac{\Gamma_2}{\alpha_2}\mathcal{Q}-\left(\Gamma_2\gint +\chi_2\right) \frac{\mM}{\alpha_2}
\end{array}
\right].
\end{equation}
\end{subequations}
We are interested in the behavior of the solutions of (\ref{eq:sysrel}) in the limit $\tau \rightarrow 0^+$
($\mu = \frac{1}{\tau}\rightarrow+\infty$). We expect that these solutions are close to
the set $\mathfrak{U} = \{w\in \mathbb{R}^{6};\Psi(w)=0\}$. We assume that the set of equations
$\Psi(w)=0$ defines a smooth manifold of dimension $L$ and that for any $ w\in \mathfrak{U}$ we know 
a parameterization $\Xi$ (the Maxwellian) from an open subset $\Omega$  of $\mathbb{R}^L$ on a neighborhood of $w$ in $\mathfrak{U}$. For any $v \in \Omega \subset \mathbb{R}^L$ the Jacobian matrix $d\Xi_{v}$ is a full rank matrix,
moreover, the column vectors of  $d\Xi_{v}$ form a basis of $\ker (\Psi'(\Xi(v)))$ \cite{gui-mur}.
Now let us define the matrix $C\in \mathbb{R}^{6 \times 6}$: 
\begin{equation}
\label{eq:Cmatr}
C = [d\Xi_v^1\ldots d\Xi_v^L\,V^1 \ldots V^{6-L}]
\end{equation}
where $d\Xi_v^1,\ldots, d\Xi_v^L$ are the column vectors of $d\Xi_{v}$ and $\{V^1,\ldots,V^{6-L}\}$
is a basis of the range of $\Psi'(\Xi(v))$. Based on the observations above, the matrix $C$ is invertible.
Let us now denote with $P$ the $L \times 6$ matrix composed of the first $L$ rows of the inverse $C^{-1}$.
We have also the following results (see \cite{gui-mur}): 
\begin{equation}
\label{eq:ident}
P \,d\Xi_v =\mathbb{I}_{L} \qeq P \, \Psi'(\Xi(v)) = 0,
\end{equation}
where $\mathbb{I}_L$ denotes the $L\times L$ identity matrix.
Now to obtain a reduced pressure equilibrium model we look for solutions in the form
$w=\Xi(v)+\tau z$, where $z$ is a small perturbation around the equilibrium state $\Xi(v)$.
Using this into the system (\ref{eq:sysrel}) we obtain
\begin{equation}
\dts (\Xi(v)) +A(\Xi(v))\dxs(\Xi(v)) -\Psi'(\Xi(v)) \,z=\Phi(\Xi(v)) +\mathcal{O}(\tau).
\end{equation}
Multiplying the above equation by $P$, by using (\ref{eq:ident}), and by neglecting terms of order $\tau$,
 we obtain the reduced model system:
\begin{equation}
\label{eq:redsys}
\dts v + P A(\Xi(v))d\Xi_v\dxs v= P \Phi(\Xi(v)).
\end{equation} 
In the limit of instantaneous pressure relaxation we have $p_1=p_2$,  hence
the vector of the variables of the reduced pressure-relaxed model is
\begin{equation}
v = [\alpha_1,\rho_1,\rho_2,u,p]\trasp \in \mathbb{R}^{5}.
\end{equation}   
Note that here $L=5$. The equilibrium state $\Xi(v)$ is defined by:
\begin{equation}
\label{eq:eqmaxw}
\Xi: v\rightarrow \Xi(v) = [\alpha_1, \rho_1,\rho_2,u,p,p]\trasp
\in \mathbb{R}^{6}.
\end{equation}  
The Jacobian $d\Xi_v \in \mathbb{R}^{6\times 5}$ of the Maxwellian  is:
\begin{equation}
d\Xi_v = \left[
\begin{array}{c|c}
\mathbb{I}_{4} & \begin{array}{c}
0 \\
\vdots\\
0
\end{array}
\\
\hline
\begin{array}{ccc}
0 & \ldots &0
\end{array}              & 1\\
\begin{array}{ccc}
0 & \ldots & 0
\end{array} & 1
\end{array}
\right].
\end{equation}
A basis  $V^1 \in \mathbb{R}^{6}$,  for the range of $\Psi'(\Xi(v))$ is found as 
\begin{equation}
V^1 = \left[
\begin{array}{c}
1 \\
[1mm]
-\frac{\rho_1}{\alpha_1}\\
[1mm]
\frac{\rho_2}{\alpha_2}\\
[1mm]
0\\
[1mm]
-\frac{\rho_1}{\alpha_1}c_1^2\\
[1mm]
\frac{\rho_2}{\alpha_2}c_2^2
\end{array}
\right]\,.
\end{equation}
Hence we can construct the matrix $C \in \mathbb{R}^{6 \times 6}$ (\ref{eq:Cmatr}), compute  the inverse
$C^{-1}$, and finally obtain the matrix $P \in \mathbb{R}^{5 \times 6}$ 
by taking the first $5$ rows of $C^{-1}$.
We find:
\begin{equation}
P=\left[
\begin{array}{c|c}
\mathbb{I}_{4} & \begin{array}{cc}
\frac{\alpha_1\alpha_2}{D}
& \frac{\alpha_1\alpha_2}{D} \\
[1mm]
-\frac{\rho_1\alpha_2}{D} &
 \frac{\rho_1\alpha_2}{D}\\
 [1mm]
 \frac{\rho_2\alpha_1}{D}
& -\frac{\rho_2\alpha_1}{D}\\
[1mm]
 0 & 0
\end{array}
\\
\hline \\
\begin{array}{ccc}
0 & \ldots & 0
\end{array} & \begin{array}{cc}
\frac{\rho_2 c_2^2\alpha_1}{D}& \frac{\rho_1 c_1^2\alpha_2}{D}
\end{array}
\end{array}
\right],
\end{equation}
where $D$ is given in (\ref{eq:Dpar}).
Finally, the reduced $p$-relaxed multiphase flow model in (\ref{eq:sisp}) is obtained
from (\ref{eq:redsys}) by using the above expression of the matrix $P$ and by evaluating the matrix
$A$ and the source term $\Phi$ in the equilibrium state $\Xi(v)$ in (\ref{eq:eqmaxw}).
Let us also note that we use the relations $\chi_k = c_k^2- \Gamma_k h_k$ in the entries
of $\Phi$ in (\ref{eq:PsiPhi}).

\section{Source terms of the $pT$-relaxed  model}

\label{sec:source4eq}

We derive here the expressions (\ref{eq:sourcep2}) appearing in the mass transfer source terms
of the four-equation $pT$-relaxed model (\ref{eq:sysT}) starting from the ordinary differential equations
 obtained from (\ref{eq:sysT})
for  the partial densities  and the mixture internal energy: 
\begin{subequations}
\label{eq:grelsys1}
\begin{eqnarray}
&&\dts (\alpha_1\rho_1)  =\mM\,,\\
[2mm]
&&\dts (\alpha_2\rho_2)  =-\mM\,,\\
[2mm]
&&\dts \Ei  =0\,.
\end{eqnarray}
\end{subequations}
Now we determine the source terms corresponding to the equations for the
volume fraction $\alpha_1$, the equilibrium temperature $T$, and the 
equilibrium pressure $p$. To this aim, we write the transformation matrix
$\frac{d \tilde{q}}{d w}$, where 
\begin{equation}
\tilde{q} =\left[
\begin{array}{c}
\alpha_1\rho_1\\
[0.5mm]
\alpha_2 \rho_2\\
[0.5mm]
\Ei
\end{array}
\right] = 
\left[
\begin{array}{c}
\alpha_1\rho_1(p,T)\\
[0.5mm]
(1-\alpha_1) \rho_2(p,T)\\
[0.5mm]
\alpha_1\Ei_1(p,T)+(1-\alpha_1)\Ei_2(p,T)
\end{array}
\right], \qqeqq
w = \left[
\begin{array}{c}
\alpha_1\\
[0.5mm]
T\\
[0.5mm]
p
\end{array}
\right].
\end{equation}
We have:
\begin{equation}
\frac{d \tilde{q}}{d w} =
\left[
\begin{array}{ccc}
\rho_1 & \alpha_1 \phider_1 & \alpha_1 \zetader_1\\
 -\rho_2 & \alpha_2 \phider_2 & \alpha_2 \zetader_2\\
\Ei_1-\Ei_2 & C_{\Ei p} & C_{\Ei T}
\end{array}
\right]
\end{equation} 
where $\phider_k$ and $\zetader_k$ are the derivatives defined in (\ref{eq:phizeta})
(with $T_k=T$, $p_k=p$, $k=1,2$) and
\begin{subequations}
\begin{eqnarray}
&&C_{\Ei p} = \alpha_1 \left(\frac{\partial \Ei_1}{\partial T}\right)_p +
\alpha_2 \left(\frac{\partial \Ei_2}{\partial T}\right)_p \,,\\
[2mm]
&&C_{\Ei T} = \alpha_1 \left(\frac{\partial \Ei_1}{\partial p}\right)_T+
\alpha_2 \left(\frac{\partial \Ei_2}{\partial p}\right)_T\,.
\end{eqnarray} 
\end{subequations}
The system of ordinary equations for $w=[\alpha_1,T,p]\trasp$
is then obtained as 
\begin{equation}
\label{eq:sysrelg2}
\dts w = \left(\frac{d \tilde{q}}{d w}\right)^{-1}\left[
\begin{array}{c}
\mM\\
-\mM\\
0
\end{array}
\right] = \mM\left[
\begin{array}{c}
\mathcal{S}_{\alpha}\\
\mathcal{S}_T\\
\mathcal{S}_p
\end{array}
\right],
\end{equation}
where
\begin{subequations}
\label{eq:sourcepT}
\begin{eqnarray}
&&\mathcal{S}_{\alpha} = \frac{1}{D_T}[C_{\Ei T}(\alpha_1 \phider_1 + \alpha_2 \phider_2)-
C_{\Ei p} (\alpha_1 \zetader_1+\alpha_2 \zetader_2)],\\
[1mm]
&&\mathcal{S}_{T}= \frac{1}{D_T}[C_{\Ei T} (\rho_2-\rho_1)+
(\Ei_1-\Ei_2)(\alpha_1 \zetader_1+\alpha_2 \zetader_2)],\\
[1mm]
&&\mathcal{S}_{p}=\frac{1}{D_T}[C_{\Ei p}(\rho_1-\rho_2)-
(\Ei_1-\Ei_2)(\alpha_1 \phider_1 + \alpha_2 \phider_2)],
\end{eqnarray}
with $D_T$ given in (\ref{eq:DTsource}). 
\end{subequations}
Note that we can write the derivatives of $\Ei_k(p_k,T_k)$ appearing in the expressions above as:
\begin{equation}
\left(\frac{\partial \Ei_k}{\partial T_k}\right)_{p_k} = -\frac{\chi_k}{\Gamma_k} \phider_k\,,
\quad \quad\left(\frac{\partial \Ei_k}{\partial p_k}\right)_{T_k} =
\frac{1}{\Gamma_k}\left(1-\chi_k \zetader_k  \right)\,, \quad k=1,2\,.
\end{equation}
Using this, together with $\Ei_1-\Ei_2 = \rho_1 h_1 -\rho_2 h_2$
and $h_k = \frac{c_k^2-\chi_k}{\Gamma_k}$, we can rewrite the numerators of 
(\ref{eq:sourcepT}) and we obtain the expressions reported in (\ref{eq:sourcep2}).
Let us remark that the derivation illustrated above can be extended to the 
case of constant temperature difference $T_2-T_1=\Delta T$ by considering
 the  variables associated to the phase $k$ as functions of $p$ and
$T_k$ and by taking for instance $w=[\alpha_1,T_1, p]\trasp$,
with the constraint $T_2=T_1+\Delta T$, $\Delta T$ = constant.
Let us finally note that the derivation of the homogeneous equations for $\alpha_1$, $p$ and $T$ of the 
four-equation $pT$-relaxed model (\ref{eq:sourcep2}) from the seven-equation Saurel--Abgrall model~\cite{sa-ab:multi}
in the limit of instantaneous velocity, pressure, and temperature equilibrium has been presented 
in \cite{demou-boil}. 

\section{Pressure invariance at interfaces}

\label{sec:pressinv}

It is well known that finite volume conservative schemes for compressible flows  may produce spurious pressure oscillations at contact interfaces, as first investigated in \cite{abgrall:mc}. 
This problem can be easily observed
for multi-component flow models, however it appears also
when computing single-component flows when   non-linear equations of state are used.
Indeed this issue is a consequence of the cell-based description
of the discrete solution, together with the choice of the 
conserved variables as principal variables, since the pressure derived
from the cell-averaged conserved quantities  might
differ from the uniform pressure value across contact discontinuities
(see e.g.\ discussion in \cite{mpthesis}). 
Typically the choice of pressure laws linear in the density and the internal energy 
per unit volume allows one to avoid difficulties. For more complex equations 
of state different strategies can be devised, for instance hybrid conservative/non-conservative methods
or methods that introduce additional variables to be used in the pressure updating
 \cite{karni:mc,abgrall:mc,ab-ka:multifl,shyue:multifluid,shyue:vdWaals,shyue:mixture,
 pelanti:hyp02,toro:nonlineos}.
Here we show that the pressure relaxation procedure presented in Section~\ref{sec:relax-step} allows us to ensure velocity 
and pressure invariance at material interfaces at least when the stiffened gas equation of state
is used (which is linear in $\rho$ and $\Ei$). Hence we consider here for each phase the pressure law (\ref{miegrun_eos}) with
constant parameters $\Gamma(\rho)\equiv \bar{\Gamma}$, 
$\varepsilon_r(\rho) \equiv \bar{\varepsilon}_r(\rho)$,
$p_r(\rho)\equiv \bar{p}_r$:
\begin{equation}
\label{eq:sgeos}
p_k(\rho_k, \mathcal{E}_k) = \bar{\Gamma}_k(\mathcal{E}_k - \rho_k \bar{\varepsilon}_{rk})
+\bar{p}_{rk}\,.
\end{equation}
Following \cite{abgrall:mc}, let us consider an isolated  material interface moving in a
flow with uniform velocity $\bar{u}$ and uniform pressure $\bar{p}$. 
For simplicity we assume $\bar{u}>0$, but the proof below can be analogously written
for $\bar{u}<0$. We consider the one-dimensional case along the $x$ direction. 

\noindent

\textit{Proposition.} If at time level $n$ we have $u_i^n=\bar{u}$
and $p_i^n=\bar{p}$, $\forall i$, then the computation by the first-order numerical scheme with
instantaneous pressure relaxation (\ref{eq_wp_1d}), (\ref{eq:alpha1eqp}), (\ref{eq:mixencons2p}) 
at time level $n+1$ gives 
  $u_i^{n+1}=\bar{u}$ and $p_i^{n+1}=\bar{p}$, when the linear equation of state (\ref{eq:sgeos})
  is used for each phase.


\vspace{1mm}

\noindent
  
\textit{Proof.} (i) Solution of the homogeneous system by the wave  propagation scheme (\ref{eq_wp_1d}).
The hypothesis   $u_i^n=\bar{u}$
and $p_i^n=\bar{p}$ implies that for each Riemann problem at the interface $i+1/2$
between the cells $i$ and $i+1$ the first and third HLLC waves are $\mathcal{W}^1_{i+1/2}=
\mathcal{W}^3_{i+1/2}=0$, based on (\ref{eq:mstate}). Hence the HLLC Riemann solution structure
at  $i+1/2$ consists of 
a single 2-wave $\mathcal{W}_{i+1/2}^2$ moving at speed $s_{i+1/2}^2=S^\star_{i+1/2} = \bar{u}$
(based on (\ref{Sspeed})). 
Hence the updating formula (\ref{eq_wp_1d}) becomes (omitting here second-order corrections):
\begin{equation}
Q_i^{n+1} =  Q_i^{n} -\bar{u}\frac{\Delta t}{\Delta x}(Q_i^n-Q^{n}_{i-1})=
(1-\xi_c)Q_i^{n} +\xi_c Q^{n}_{i-1}\,,\quad \xi_c \equiv \bar{u}\frac{\Delta t}{\Delta x}\,. 
\end{equation}
Therefore, first we easily verify
\begin{equation}
u_{i}^{n+1} = \frac{(\rho u)_{i}^{n+1}}{\rho_i^{n+1}} = \bar{u}\,,\quad \forall i\,.
\end{equation}
Note that this results for the velocity invariance holds in general for any pressure law.
Then we compute the phasic pressures of the homogeneous system solution step by using the equation of state above (\ref{eq:sgeos}):
\begin{equation}
p_{ik}^{n+1}  = \frac{1}{\alpha_{ik}^{n+1}}\left(
\bar{\Gamma}_k((\alpha_k\mathcal{E}_k)_{i}^{n+1} - (\alpha_k\rho_k)_i^{n+1}
 \bar{\varepsilon}_{rk})\right)
+\bar{p}_{rk}= \bar{p}\,,\quad \forall i, \,\,k=1,2\,.
\end{equation}

\noindent

(ii) Pressure relaxation step and pressure update.
Since the solution of the homogeneous system gives $(p_1^0)_i = (p_2^0)_i =\bar{p}$ 
for each cell $i$ of the computational domain,
the relaxed volume fraction $\alpha_1^*$ (\ref{eq:alpha1eqp}) computed in the mechanical
relaxation step is $(\alpha_1^*)_i=(\alpha_1^0)_i = (1-\xi_c)\alpha_{1,i}^n+\xi_c\alpha_{1\,i-1}^n$.
Finally, the updated mixture equilibrium pressure is, based on (\ref{eq:mixencons2p})
with (\ref{eq:sgeos}):
 \begin{subequations}
 \begin{eqnarray}
p_i^{n+1}\!= \!p^*_i &\!\!\!=\!\!\!& \frac{\Ei^0_i-
\left((\alpha_1\rho_1)^0_i \bar{\varepsilon}_{r1} + (\alpha_2\rho_2)^0_i\bar{\varepsilon}_{r2}\right)
+\left(\frac{\alpha_{1,i}^0 \bar{p}_{r1}}{\bar{\Gamma}_1} + 
\frac{\alpha_{2,i}^0 \bar{p}_{r2}}{\bar{\Gamma}_2}\right)}{
\frac{\alpha_{1,i}^0}{\bar{\Gamma}_1} + \frac{\alpha_{2,i}^0}{\bar{\Gamma}_2}}\\
&\!\!\!=\!\!\!& \frac{\frac{1}{\bar{\Gamma_1}}\left((1-\xi_c)\alpha_{1,i}^n \bar{p}+\xi_c\alpha_{1,i-1}^n\bar{p}\right)
+\frac{1}{\bar{\Gamma_2}}\left((1-\xi_c)\alpha_{2,i}^n\bar{p}+\xi_c\alpha_{2,i-1}^n\bar{p}\right)}
{\frac{1}{\bar{\Gamma_1}}\left((1-\xi_c)\alpha_{1,i}^n+\xi_c\alpha_{1,i-1}^n\right)
+\frac{1}{\bar{\Gamma_2}}\left((1-\xi_c)\alpha_{2,i}^n+\xi_c\alpha_{2,i-1}^n\right)} = \bar{p}\,,
\,\, \forall i.
\end{eqnarray}
\end{subequations}
\begin{flushright}
\qed
\end{flushright}
Although the pressure invariance is proven here only for a linear equation of state, 
we have observed  numerically by performing numerous tests that no oscillations appear around material interfaces for more general  nonlinear pressure laws of the form (\ref{miegrun_eos}).

\bibliographystyle{plain}
\bibliography{main}
\end{document}